%% file: main.tex
\documentclass[3p, times]{elsarticle}

\usepackage{lineno,hyperref}
\usepackage{amsmath}
\usepackage{mathtools}
\DeclarePairedDelimiter{\abs}{\lvert}{\rvert}
\usepackage{bm}
\usepackage{amssymb}
\usepackage{amsthm}
\usepackage{array}
\usepackage{enumitem}
\usepackage{graphicx}
\usepackage[ruled,vlined,linesnumbered]{algorithm2e}
\usepackage{todonotes}
\usepackage{optidef}
\usepackage{caption}

\usepackage{pgfplots}
\usepackage{pgfkeys}
    \newenvironment{customlegend}[1][]{%
        \begingroup
        \csname pgfplots@init@cleared@structures\endcsname
        \pgfplotsset{#1}%
    }{%
        \csname pgfplots@createlegend\endcsname
        \endgroup
    }%
    \def\addlegendimage{\csname pgfplots@addlegendimage\endcsname}
\usepackage{subfigure}

\pgfplotsset{compat=newest}
\usepgfplotslibrary{external}
\usepgfplotslibrary{patchplots}
\usetikzlibrary{shapes,arrows}
\usetikzlibrary{positioning,calc}
\usetikzlibrary{decorations.pathreplacing,decorations.markings,math,arrows.meta}
\usetikzlibrary{spy}
\usetikzlibrary{plotmarks}

\input{math}
\input{figures/formats}

\modulolinenumbers[5]

\journal{Computer Methods in Applied Mechanics and Engineering}

\begin{document}

\begin{frontmatter}

\title{Moment fitted cut spectral elements for explicit analysis of guided wave propagation}

%% Group authors per affiliation:
\author[1]{Sergio Nicoli\corref{cor1}}
\ead{nicoli@ibk.baug.ethz.ch}
\author[2]{Konstantinos Agathos}
\ead{K.Agathos@exeter.ac.uk}
\author[1]{Eleni Chatzi}
\ead{chatzi@ibk.baug.ethz.ch}

\address[1]{Department of Civil, Environmental, and Geomatic Engineering,
ETH Z\"{u}rich, Stefano-Franscini-Platz 5,
CH-8093 Z\"{u}rich, Switzerland}
\address[2]{ College of Engineering, Mathematics and Physical Science,
            Exeter University,
            Exeter, UK}

\cortext[cor1]{Corresponding author}
%% or include affiliations in footnotes:
%\author[mymainaddress,mysecondaryaddress]{Elsevier Inc}
%\ead[url]{www.elsevier.com}

%\author[mysecondaryaddress]{Global Customer Service\corref{mycorrespondingauthor}}
%\cortext[mycorrespondingauthor]{Corresponding author}
%\ead{support@elsevier.com}

%\address[mymainaddress]{1600 John F Kennedy Boulevard, Philadelphia}
%\address[mysecondaryaddress]{360 Park Avenue South, New York}

\begin{abstract}
In this work, a method for the simulation of guided wave propagation in solids defined by implicit surfaces is presented. The method employs structured grids of spectral elements in combination to a fictitious domain approach to represent complex geometrical features through singed distance functions. A novel approach, based on moment fitting, is introduced to restore the diagonal mass matrix property in elements intersected by interfaces, thus enabling the use of explicit time integrators. Since this approach can lead to significantly decreased critical time steps for intersected elements, a ``leap-frog'' algorithm is employed to locally comply with this condition, thus introducing only a small computational overhead. 
The resulting method is tested through a series of numerical examples of increasing complexity, where it is shown that it offers increased accuracy compared to other similar approaches. Due to these improvements, components of interest for SHM-related tasks can be effectively discretized, while maintaining a performance comparable or only slightly worse than the standard spectral element method.
\end{abstract}

\end{frontmatter}

%\linenumbers
\input{1_introduction}

\input{2_problemStatement}
\input{3_Method}
\input{4_1_example2D}
\input{4_2_willberg}
\input{4_3_experiment}
\input{5_conclusions}

\bibliography{refs.bib}

\end{document}

%% file: math.tex
\newcommand{\bsigma}{\mbox{\boldmath$\sigma$}}

\newcommand{\bu}{\mbox{\boldmath$u$}}
\newcommand{\bv}{\mbox{\boldmath$v$}}

\newcommand{\bepsilon}{\mbox{\boldmath$\epsilon$}}

\newcommand{\surfaceLoad}{\mbox{\boldmath$p_s$}}
\newcommand{\interfaceLoad}{\mbox{\boldmath$p_c$}}

\newcommand{\globalVec}{\mbox{\boldmath$x$}}
\newcommand{\localVec}{\mbox{\boldmath$\xi$}}

\newcommand{\dt}{\mbox{$\Delta t$}}
\newcommand{\dtcrit}{\mbox{$\Delta t_c$}}

%% file: figures/formats.tex
\newcommand\figureFontSize{8}
\newcommand\bigFigureFontSize{6}
\newcommand\plotLineWidth{0.75}

\newcommand\explicitMarker{}

\newcommand\conformMarker{triangle}

\newcommand\uncutColor{black}
\newcommand\uncutColorb{gray}

\newcommand\cutConsistentColor{black}
\newcommand\cutHRZColor{red}

\newcommand\cutOptimizedColor{blue}
\newcommand\cutOptimizedColorb{green}

\newcolumntype{C}[1]{>{\centering\arraybackslash}p{#1}}
\newcolumntype{L}{>{$}l<{$}} 

%% file: 1_introduction.tex
\section{Introduction}
%\subsection{Context}
A grand challenge of modern engineering lies in stewarding, i.e., managing and maintaining, critical infrastructure, which comprises a mix of existing and ageing, as well as new and ever-complex structures. In understanding the condition and capacity of existing structures, Structural Health Monitoring (SHM) offers a set of strategies aimed at the continuous supervision of structures, targeting detection of damage onset, its localization and assessment, and the estimation of a structure's remaining life \cite{rytter1993vibrational, worden2015structural}. Vibration-based condition monitoring installations have proven extended capabilities in detecting global damage occurrence, which can affect the dynamic/modal properties of a system \cite{avci2021review, ou2017vibration, an2019recent,limongelli}. However, such methods display reduced efficacy in the case of more local damage effects, even for the case of the more promising alternative of strain-based measurements \cite{laflamme2016damage, anastasopoulos2018damage}. As a more targeted solution for discovery and localization of flaws within structures, non-destructive evaluation techniques (NDE) are employed, typically in the form of periodic inspections. 

Among NDE procedures, Guided Waves (GW) show promising capabilities in accurately predicting the location of possible damage within a medium. A common means to this end, piezo-electric sensors are employed \cite{An2014,Nienwenhui2005,GIURGIUTIU2008239,CANTEROCHINCHILLA2019192}, enabling to both, generate elastic ultrasonic waves and record their propagation in terms of an electric signal. Damage can then be diagnosed either in a purely data-driven manner, or using a model based approach. As far as the former approach is concerned, a number of works have relied on interpretation of characteristics of the propagating waves - such as non-linear, multimodal behavior, scattering and energy leakage of guided waves - for the detection and sizing of flaws, including delamination-type defects \cite{samaitis2020assessment,ramadas2011characterisation, santos2005leaky, sohn2011delamination, zhao2019detection}. In this work, we focus on the latter and, thus, rely on use of a model that is able to match a virtual representation of the monitored structure \cite{lee2003modelling1, lee2003modelling2, douglass2020model}. 

%\subsection{Lamb waves}
%In media bounded by two equidistant surfaces, repeated reflections of a wave along the cross-section lead to longitudinal propagation, hence the designation as it being "guided" by the structure. When considering plate or shell structures (i.e. with stress-free surface boundaries) with normal loading, Lamb waves \cite{lamb1917waves} arise form the interaction of shear and compression waves reflected at the boundaries. Due do these interactions, these waves are highly dispersive (ref) and can emerge in a multitude of possible modes, which can be classified as symmetric or anti-symmetric with respect to the particle's motion observed along the cross section \cite{ostachowicz2011guided}.
The effectiveness of using GW, and Lamb waves in particular \cite{lamb1917waves, kessler2002damage}, for NDE procedures stems from some of their physical properties. GW experience only a small amplitude attenuation over distance, which reduces the size of the sensor network and its energy consumption, and further allows to evaluate regions that are inaccessible to the inspector \cite{giurgiutiu2007structural}. Additionally, the use of short wavelengths excites modes interacting with small, localized, features, so that even minor damage can be detected, and a wide range of modes can be used to classify the faults \cite{su2009identification}.

Nonetheless, the numerical modeling of these phenomena poses several challenges. To represent high frequency modes, fine temporal and spatial discretizations are necessary, resulting in large models and a large number of time integration steps. In modeling a structure of interest, conformal meshing of complex components (and/or localized damage) is often employed, which requires intensive human intervention and can impose a drastic reduction in the permissible element size due to the need to conform with small details. This in turn leads in reduction of the critical time step prescribed for the stability of explicit solvers \cite{courant1967partial}, thus further increasing the cost of time integration. In the context of damage detection, these costs are multiplied by repeated evaluations of a model (inverse solution), while the necessity of automatically updating the damage configuration precludes the exclusive use of traditional meshing techniques.

These and other limitations, such as the recurrent concern with mesh quality, have emerged in various fields of computational mechanics dealing with complex and/or evolving geometries, and led to the development of the eXtended, or Generalized, FEM (XFEM/GFEM) \cite{moes1999finite, strouboulis2000design}, as well as fictitious domain methods such as the Finite Cell Method (FCM) \cite{parvizian2007finite, duster2008finite}, and the CutFEM \cite{burman2015cutfem, claus2018stable}. To effectively tackle the aforementioned challenges, we seek to combine such approaches, which provide geometrical descriptions that are independent of the underlying mesh, in environments that are effective for the analysis of GW. In this context, two important requisites are the availability of high order Ansatz functions, effective in the modeling of high frequency modes; and mass matrix diagonalization, which enables use of highly efficient explicit solvers. Mass lumping techniques for the XFEM \cite{menouillard2006efficient, elguedj2009explicit} were used in explicit simulations of dynamic crack propagation \cite{menouillard2008mass} and combined with the Spectral Element Method in the time domain (SEM) \cite{liu2011xfem}. 

More recently, variationally consistent lumping \cite{schweitzer2013variational} has been proposed for the global-local GFEM \cite{sanchez2021high, geelen2021scale}. The FCM was also applied to dynamic analysis by Duczek et al. \cite{duczek2014higher}, who proposed the Spectral Cell Method (SCM) by combining a fictitious domain approach with the SEM. This was successfully applied to simulate Lamb waves on 2 and 3D aluminium plates with holes \cite{duczek2014numerical} and piezo-electric sensors for SHM \cite{duczek2015finite}. Among other high order methods suitable for dynamic analysis \cite{hughes2005iso, cottrell2009iso, duster2001p, duczek2019critical, gravenkamp2020mass}, the SEM is often one of the preferred approaches \cite{PATERA1984468, ostachowicz2011guided, zak2011certain}, since use of Gauss-Lobatto-Legendre (GLL) integration points delivers a variationally consistent diagonal mass matrix, without incurring loss of accuracy \cite{ostachowicz2011guided, schulte2010spectral, lonkar2014modeling} or the need for additional lumping procedures. However, when decoupled geometrical descriptions as in the SCM or the XFEM are employed, special integration rules are applied for elements intersected by a boundary, thus eliminating the diagonal property of the mass matrix. To recover this quality, Joulaian et al. \cite{joulaian2014finite} proposed to perform HRZ (Hinton, Rock, and Zienkiewicz) lumping \cite{hinton1976note}, a solution later applied by Giraldo and Restrepo in earthquake modeling \cite{giraldo2017spectral} and also adopted by Mossaiby et al. \cite{mossaiby2019spectral} in a GPU implementation of the SCM. As hypothesized in \cite{joulaian2014finite} and confirmed in \cite{duczek2019mass}, the lumping procedure might introduce some error which negatively affects the convergence of SE, although it guarantees positiveness of the resulting mass coefficients. 

In this paper, we propose an improved mass lumping method for cut Spectral Elements (SE). It can be summarized as using element partitions to evaluate integrals of the polynomial basis for the moment fitting equations, so that integration weights within a nodal quadrature rule can be determined to account for the cut configuration of an element, thus preserving a diagonal mass matrix. These moment fitting equations are not solved directly; instead, a quadratic programming problem is derived, allowing to guarantee positiveness of the weights (and thus of the mass coefficients) through appropriate constraints. Like in other, similar methods, the decay of critical time step for cut elements is of concern, since it can compromise the performance of time integration. In this contribution, we address this problem by means of a frog-leap algorithm, which enables to efficiently tailor different time integration steps to intact and the cut elements.

The remainder of this work is organized as follows. The elastodynamics problem is defined in \autoref{sec:problemStatement}, followed by a brief review of the SEM in \autoref{sec:SEM}. In \autoref{sec:LSM} we use element partitions \cite{fries2017higher} conforming with mesh-independent implicit interfaces \cite{osher1988fronts} to accurately introduce voids in the domain. The novel method is then introduced in \autoref{sec:momentFitting} and its effect on the critical time step is studied and compared to the available lumping strategies. Considerations regarding time integration and a review of the leap-frog solver by Diaz, Grote, et al. \cite{diaz2009energy, grote2013explicit} are offered in \autoref{sec:leapfrog}. In \autoref{sec:ex1} the performance of the proposed approach is assessed on th 2D benchmark of a cut beam and again compared against available alternative schemes. In \autoref{sec:ex2} the problem of a 3D plate with a conic hole by Willberg \cite{willberg2012comparison} and Duczek \cite{duczek2014numerical} is adopted to benchmark the novel method in comparison with the SEM. A more realistic example is then offered in \autoref{sec:ex3} in preliminary studies of an aluminum specimen due for experimental testing. Based on the outcome of these analyses, concluding remarks are formulated in \autoref{sec:conclusions}.

%% file: 2_problemStatement.tex
\section{Problem Statement}\label{sec:problemStatement}

\begin{figure}[h]
    \centering
	\includegraphics[width=.55\linewidth]{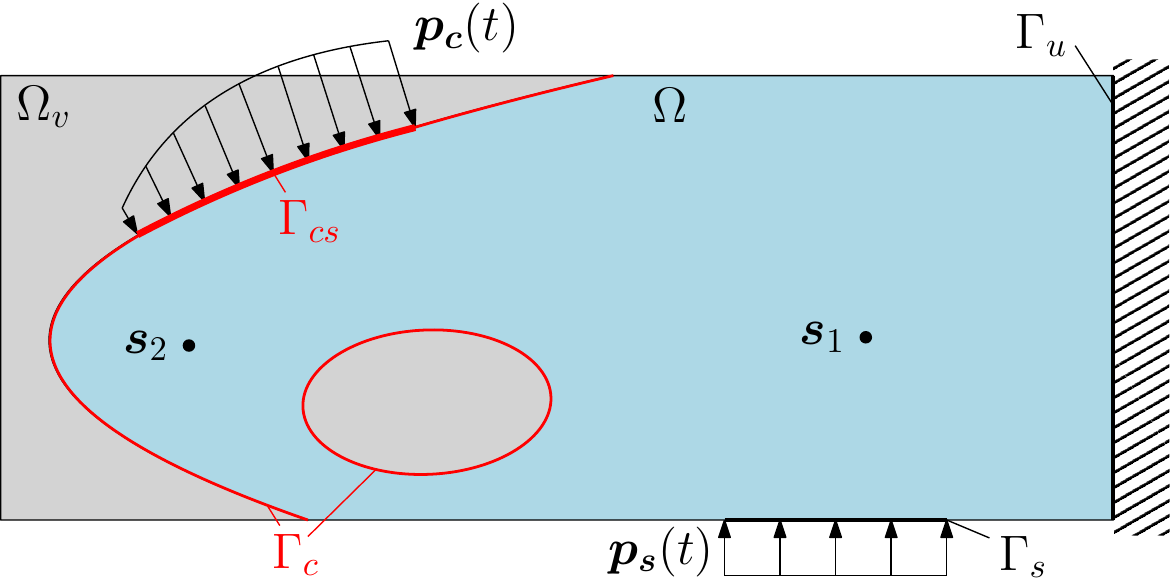}
	\caption{Embedding domain $\Omega_{tot}$ subdivided into the domain of interest $\Omega$ and the void domain $\Omega_{v}$ via the mesh-independent interface $\Gamma_c$. $\mathbf{s}_{1,2}$ denote sensor locations, while actuators are modeled by the time-dependent surface tractions $\surfaceLoad(t)$ and $\interfaceLoad(t)$.}
	\label{fig:problemStatement}
\end{figure}

Let us consider a 2D or 3D domain, denoted by $\Omega$, for which the solution of the elastodynamics problem is sought. To ease discretization, $\Omega$ is complemented with a void domain $ \Omega_v$, resulting in the domain $\Omega_{tot} = \Omega \cup \Omega_{v}$, which in the ideal case can be represented by a structured Cartesian mesh. In the FCM/SCM literature, $\Omega_{tot}$ is called extended or embedding domain, $\Omega_{v}$ is the fictitious domain \cite{parvizian2007finite, duster2008finite, duczek2014numerical}, and $\Omega$ is the physical domain or domain of interest. It can now be expressed as:
\begin{equation}
     \Omega = \Omega_{tot} \setminus \Omega_{v}
\end{equation}
In \autoref{fig:problemStatement}, $\Omega_{v}$ and $\Omega$ are schematically represented. Importantly, their common boundary within $\Omega_{tot}$ is the ``cut'' interface 
$\Gamma_{c}$, which, in practice, represents the boundaries of $\Omega$ that are defined independently from the mesh. 
Essential boundary conditions of the form $\bu(t) = \bar{\bu}$ are applied on the mesh-conforming boundaries of $\Omega$ denoted by $\Gamma_u$. Surface tractions $\surfaceLoad(t)$ and $\interfaceLoad(t)$ act on mesh conforming, as well as cut, boundaries, denoted $\Gamma_s$ and $\Gamma_{cs}$, respectively. In this context, such loads are meant to model excitation of the structure by means of PZT actuators. They are accordingly complemented by the definition of $n$ sensor locations $\mathbf{s}_i, \ i = 1,...,n$ at which the numerical solution is of great practical interest. Modeling of the piezo-electric effect and optimization of the sensor layout are important aspects of this technology, however they are beyond the scope of this contribution.

\noindent Based on these definitions, the weak from of the linear elastodynamics problem can be expressed as:
\begin{equation}
\label{eq:weak_form}
	\int_{\Omega} \rho \ddot{\bu}(t)\cdot{\bv} \ d\Omega + \int_{\Omega} \bsigma (\bu(t)):\bepsilon (\bv)\ d\Omega = \int_{\Gamma_s} \surfaceLoad(t)\cdot\bv \ d\Gamma_s + \int_{\Gamma_{cs}} \interfaceLoad(t) \cdot\bv \ d\Gamma_{cs}
\end{equation}
where $\rho$ is the material density, $\bsigma$ is the Cauchy stress tensor and $\bepsilon$ the linear strain. For the trial function $\bv$ holds: 
\begin{equation}
	\mathcal{V}^0 = \left\{\bv|\bv \in \left(H^1\left(\Omega\right)\right)^d,\bv = 0  \text{ on }  \Gamma_u \right\}
\end{equation}
where $d$ is the number of spatial dimensions, while $\bu(t)$ represents the displacement solution at time $t$:
\begin{equation}
	\mathcal{U}_t = \left\{\bu(t)|\bu(t) \in \left(H^1\left(\Omega\right)\right)^d,\bu(t) = \bar{\bu}  \text{ on }  \Gamma_u \right\}
\end{equation}
To obtain the solution $\bu(t)$, \autoref{eq:weak_form} must be discretized. As opposed to mesh-conforming discretization methods, the presence of the interface $\Gamma_c$ has important implications in both, space and time discretizations, which will be addressed and discussed in the following section.

%% file: 3_Method.tex
\section{Moment Fitting for Cut Spectral Elements}\label{sec:method}
In the following, we present our mass matrix lumping approach, which builds on the concepts of the SEM (\autoref{sec:SEM}) and element partitioning techniques (\autoref{sec:LSM}). While the former delivers the approximation space used to discretize the continuum equations (\autoref{eq:weak_form}), the latter is used to generate an integration rule for elements traversed by the boundary $\Gamma_c$. These two components are then merged in the novel moment fitting procedure, presented in \autoref{sec:momentFitting}. Aspects concerning explicit time integration are further addressed in \autoref{sec:leapfrog}, where a frog-leap solver is adopted to accelerate the solution for domains modeled with the new method.
%%%%%%%%%%%%%%%%%%%%%%%%%%%%%%%%%%%%%%%%%%%%%%%%%%%%%%%%%%%%%%%%%%%%%%%%%%%%%%%%%%%%%%%%%%%%%%%%
\subsection{The Spectral Element Method}\label{sec:SEM}
\label{SEM}
In the SEM \cite{PATERA1984468}, the domain is discretized as in traditional FEM, although particular consideration is given to the nodal configuration. Besides the Chebyshev nodal distribution \cite{dauksher1997accuracy, dauksher2000solution} used at inception, GLL quadrature points have been extensively used, especially in dynamic analysis \cite{komatitsch2000simulation, komatitsch2002spectral, kudela2007wave, kudela2007modelling, zak2011certain}. Both approaches crucially enable the use of high order polynomials by overcoming the  Runge phenomenon (see, e.g. \cite[Chapter~4.2]{boyd2001chebyshev} and \cite[Chapter~3]{pozrikidis2005intr}): However, GLL-SE additionally enable the formulation of variationally consistent diagonal mass matrices by exploiting the orthogonality of the shape functions and performing integration at the element nodes. This strategy is referred as to Lumping by nodal quadrature \cite{duczek2019critical} or by integration \cite{FRIED1975461}. For the case of GLL-SE, it preserves the optimal convergence properties of the method and guarantees positiveness of the mass coefficients \cite{zak2011certain, tschoke2018numerical, duczek2019mass}, hence the designation ``optimal lumping'' \cite{cook2007concepts}. In what follows, we offer a summarized view of the SEM applied in this contribution, as well as in previous instances of the SCM \cite{duczek2015finite}. The interested reader is referred to \cite{pozrikidis2005intr, ostachowicz2011guided, canuto2012spectral} for more detailed derivations.

\subsubsection{Shape Functions}
Consider a one-dimensional SE of order $p$. According to the GLL nodal configuration, the locations of its nodes $\xi_i \text{, with } i \in \{1,2,..., p+1\}$ are given in the local (reference) coordinate system $ \xi \ \in \ [-1, 1] $ as:
\begin{equation}
    \left(1-\xi^2\right) L_{p-1}(\xi) = 0
\end{equation}
i.e. by the vertices $\{-1, 1 \}$ and roots of the Lobatto polynomial $ L_{p-1}$ of order $p-1$, which consists in the first derivative of the Legendre polynomial $P_p$ of order $p$:
\begin{equation}
    L_{p-1}(\xi) = \frac{\delta P_p(\xi)}{\delta \xi}
\end{equation}
The element shape functions $N_{p,i}(\xi)$ are then defined by Lagrangian interpolations supported at the nodes $\xi_i$:
\begin{equation}
    \label{eq:shapeFuncs}
    N_{p,i}(\xi) = \prod_{j=1, j \neq i}^{p + 1}\frac{\xi - \xi_j^p} {\xi_i^p - \xi_j^p}.
\end{equation}
Elements in higher dimensions can elegantly be constructed by taking the sparse product(s) of the shape functions of the one-dimensional system, i.e.:
\begin{equation}
\label{eq:2d3dshapeFunc}
\begin{split}
    \mathcal{N}_{p,q}(\localVec) &= \left \{ \left \{  N_{p,1}(\xi), N_{p,2}(\xi),..., N_{p,p+ 1}(\xi) \right \} \times \left\{  N_{q,1}(\eta), N_{q,2}(\eta),..., N_{q,q + 1}(\eta) \right \}   \right \}\\
    \mathcal{N}_{p,q,r}(\localVec) &= \left \{ \left \{  N_{p,q,1}(\xi, \eta), N_{p,q,2}(\xi, \eta),..., N_{p,q,(p+ 1)(q+1)}(\xi, \eta) \right \} \times \left \{  N_{r,1}(\zeta), N_{r,2}(\zeta),..., N_{r,r + 1}(\zeta) \right \}   \right \}
\end{split}
\end{equation}
In eq. \ref{eq:2d3dshapeFunc}, $\mathcal{N}_{p,q} $ represents the group of shape functions for a quadrilateral element of orders $p$ and $q$ in the respective local coordinates $\localVec = [ \xi, \eta ]^T$. For a hexahedral element with $\localVec = [ \xi, \eta,\zeta ]^T$, the shape functions are similarly obtained as the sparse product of $\mathcal{N}_{p,q}$ with the interpolants of order $r$ in the third local dimension $\zeta$. These distinctions enable the construction of hybrid elements (in the sense of polynomial degree), which is important in the modeling of GW, due to the fact that the spatial discretization must be carefully tailored to the expected wave modes \cite{willberg2012comparison}.

\subsubsection{Discretized equilibrium equations}
In this section, the subscripts $p,q,r$ are omitted to reduce clutter. In the reference system of a generic element with $n$ nodes and shape functions $N_i(\localVec), \ i = 1, ..., n$, the unknown displacement field $ \mathbf{u}(\localVec, t) $ at time $t$ is interpolated from the nodal displacements $\mathbf{u}_i(t) $: 
\begin{equation}
    \label{eq:Nf}
  \mathbf{u}(\localVec, t)  = \sum_{i = 1}^n N_i(\localVec) \ \mathbf{u}_i(t)
  = \mathbf{N}(\localVec) \ \mathbf{u}_e(t)
\end{equation}
Eq. \ref{eq:Nf} can be conveniently written in matrix form by distributing the shape functions in the matrix $\mathbf{N}(\localVec)$ to match the $d$ Degrees Of Freedom (DOFs) of the respective node $i$ within the element's displacement vector $\mathbf{u}_e(t)$:
\begin{equation}
    \label{eq:Nf2}
    \mathbf{N}(\localVec) = 
    \begin{bmatrix}
        N_1 \mathbb{I}_d & N_2 \mathbb{I}_d & ... & N_n \mathbb{I}_d
    \end{bmatrix}
\end{equation}

\noindent where $\mathbb{I}_d$ is the $d \times d$ unit matrix.

Under the assumption of eq. \ref{eq:Nf}, and after application of Hook's constitutive law, \autoref{eq:weak_form} can be discretized with respect to the displacements and the excitation at the nodal DOFs, which are collected in the system vectors $\mathbf{u}_{s}(t)$ and $\mathbf{f}_{s}(t) $, respectively:
\begin{equation}
\label{eq:system}
 \mathbf{M} \ \ddot{\mathbf{u}}_{s}(t) + \mathbf{K} \ \mathbf{u}_{s}(t) = \mathbf{f}_{s}(t).
\end{equation}
The mass matrix $\mathbf{M}$ and the stiffness matrix $\mathbf{K}$ are assembled form the respective element contributions $\mathbf{M}_e, \ \mathbf{K}_e$:
\begin{equation} \label{eq:elementMass}
    \mathbf{M}_e = \int_{\Omega_e } \rho {\mathbf{N}^T \mathbf{N}\ }{d\Omega_e}
\end{equation}
\begin{equation} \label{eq:elementStiff}
    \mathbf{K}_e = \int_{\Omega_e } {\mathbf{B}^T \mathbf{D} \mathbf{B}\ }{d\Omega_e}
\end{equation}
where $\mathbf{B}$ is the matrix of strain coefficients and $\mathbf{D}$ is Hooke's tensor. The time-dependent element force vector $\mathbf{f}_e(t)$ results from integration of the interpolated nodal values for surface tractions acting on mesh conforming ($\Gamma_s$) as well as non-conforming ($\Gamma_{cs}$) boundaries:
\begin{equation}
    \mathbf{f}_e(t) = \int_{\Gamma_s } {\mathbf{N}^T \mathbf{p}_{s}(t) \ }{d\Gamma_s} + \int_{\Gamma_{cs} } {\mathbf{N}^T \mathbf{p}_{c}(t) \ }{d\Gamma_{cs}} 
    \label{eq:loads}
\end{equation}
In this notation, $\Omega_e$ represents the portion of the physical domain $\Omega$ contained by the element, and shall not be confused with the embedding domain. If an element is fully in the void (i.e. $\Omega_e = \emptyset$), the number of nodes and DOFs of the model can be reduced. In our implementation, sections of the elements belonging to $\Omega_v$ are discarded. We should note that, in many instances of the FCM/SCM, the void domain is instead considered by penalizing its integration by a numerical tolerance factor (typically $\alpha = 10^{-5}, ..., 10^{-10}$), which offers a measure to prevent bad conditioning of the system matrices. This problem, however, is less pronounced in an explicit dynamics application, since no inversion of the stiffness matrix occurs (either by factorization or iteration) and the mass matrix is diagonal. 
%%%%%%%%%%%%%%%%%%%%%%%%%%%%%%%%%%%%%%%%%%%%%%%%%%%%%%%%%%%%%%%%%%%%%%%%%%%%%%%%%%%%%%%%%%%%%%%%
\subsection{Element partitioning}\label{sec:LSM}
\begin{figure}[h]
    \centering
	\includegraphics[width=.95\linewidth]{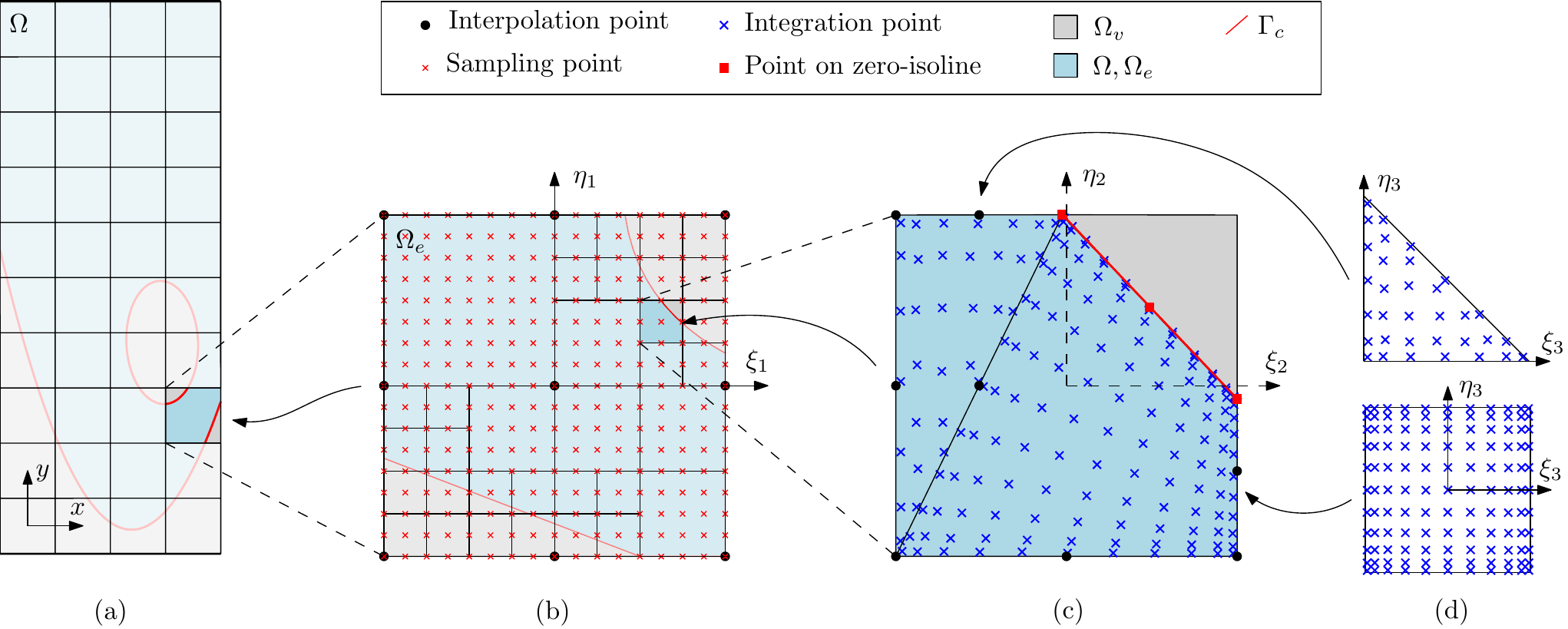}
	\caption{(a) Discretization of the embedding domain $\Omega_{tot}$ with a structured SE mesh. (b) The level set values are sampled within a SE to construct a local quadtree (respectively, octree) mesh. (c) Boundary-conforming element partitions are generated for the cut sub-elements of the quadtree. (d) To represent the physical element domain $\Omega_e$, integration points are mapped from the reference system of the element partitions $( \xi_3, \eta_3 ) $ to the one of the initial SE $( \xi_1, \eta_1 ) $.
	}
	\label{fig:LSM}
\end{figure}

With the level set method (LSM), an interface or boundary can be implicitly represented by the zero iso-surface of a signed distance function $\Phi (\globalVec )$ \cite{osher2006level}. A generic point of the domain $\Omega_{tot}$ can thus be classified as belonging to $\Omega_{v}$ (with $\Phi(\globalVec) < 0 $), to  $\Omega$ (with $\Phi(\globalVec) > 0 $) or to the boundary $ \Gamma_c $ (with $\Phi(\globalVec) = 0 $).
Since the Gauss integration rule assumes smoothness of the integrand, elements traversed by a discontinuity must be handled accordingly. Generally speaking, the main challenges in this context involve accurately approximating the boundary and limiting the escalation in the number of integration points, while ensuring the accuracy of the resulting integration rule. This often requires the use of complex data structures. In the XFEM/GFEM and SCM/FCM communities, different strategies have been proposed to tackle these issues (e.g. \cite{abedian2013performance, duczek2015efficient, fries2017higher, chin2019modeling}). In this work, we use quadtree (in 2D) and octree (in 3D) meshes in combination with boundary-conforming element partitions. The procedure is summarized in \autoref{fig:LSM} for the exemplary 2D domain introduced in \autoref{fig:problemStatement}, while we refer to the aforementioned literature for an in-depth view of these methods and alternative strategies. An integration rule in the reference system $( \xi_1, \eta_1 ) $ is sought for the physical portion $\Omega_e$ of the element highlighted in Figure \ref{fig:LSM}(a). In Figure \ref{fig:LSM}(b), the level set function $\Phi(\globalVec)$ is sampled over a fine grid of nodes, thus enabling the creation of a local, hierarchical mesh of \textit{sub-elements}. \autoref{fig:LSM}(c) shows that, when a sub-element is intersected, boundary-conforming element partitions are deployed. To this end, nodes on the boundary $\Gamma_c$ (i.e. on the zero iso-line of $\Phi(\globalVec)$) are generated by means of a Newton-Raphson algorithm \cite{fries2017higher} and the standard Lagrangian interpolation. By virtue of the quad/octree partitioning, possible cut configurations can be reduced to few, fundamental, topological cases and the distortion of the interface relative to the system $( \xi_2, \eta_2 )$ is diminished, thus improving accuracy. In figures \ref{fig:LSM}(c) and \ref{fig:LSM}(d), the relevant partitions are locally meshed with quadrilateral and triangular elements (in 2D) or tetrahedral elements (in 3D), and used to generate an integration rule in the initial reference system.
%%%%%%%%%%%%%%%%%%%%%%%%%%%%%%%%%%%%%%%%%%%%%%%%%%%%%%%%%%%%%%%%%%%%%%%%%%%%%%%%%%%%%%%%%%%%%%%%
\subsection{Moment Fitting}\label{sec:momentFitting}
%The moment fitting procedure introduced in this section aims at enabling
The element partitioning procedure described in the previous section, as well as other alternatives from the literature, typically lead to a large number of integration points, whose locations depend on the exact way in which each element is intersected by the interface. As a result, the optimal lumping property, associated with GLL points is lost, and non-diagonal mass matrices are produced, rendering the approach unusable in an explicit dynamics context. To overcome this limitation, Joulaian et al.~\cite{joulaian2014finite} and Duczek et al.~\cite{duczek2014numerical} have proposed techniques, also commonly used in FE analysis, to lump the resulting mass matrices, while preserving some desired properties, such as the total mass of the element. Herein, we introduce a novel technique, aiming at minimising the errors introduced by lumping, while allowing the imposition of some physical constraints, such as the aforementioned mass conservation.

As a starting point for our approach, we consider the construction of a rule for integrating polynomial functions up to a certain degree, defined in the reference system of an element. For clarity, it is pointed out that $\Omega_e$ represents only the physical portion of an element, for which integration rules can be derived as in the previous section. Constructing a new rule consists of determining a set of $n$ points, in terms of their coordinates in the reference system of the element $ \localVec_i $, and a set of weights $ w_i $, with $i = 1,2,...,n$, such that:
\begin{equation}
\label{eq:fitDef}
    \int_{\Omega_e} {f(\localVec) \ }{d\Omega_e} = \sum_{i = 1}^n {f(\localVec_i) w_i}
\end{equation}
\noindent where $f(\localVec)$ is a polynomial to be integrated.
%The choice of integration points $ \localVec_i $ falls on the roots of orthogonal polynomials \cite[Chapter~4.5]{flannery1992num}.
Since $f(\localVec)$ can be decomposed into a set of monomials $ g_i (\localVec) $, with $ i = 1,2,..., m $ such that $f( \localVec ) \in span \left\{g_1, g_2, ..., g_m \right \}$, the problem can be expressed as \cite[Chapter~4.5]{flannery1992num}:
\begin{equation}
\label{eq:fitLinear}
\renewcommand*{\arraystretch}{1.3}
   \begin{bmatrix}
        g_1(\localVec_1) & g_1(\localVec_2) & \dots & g_1(\localVec_n)\\
        g_2(\localVec_1) & g_2(\localVec_2) & &\vdots \\
        \vdots & &\ddots &\vdots\\
        g_m(\localVec_1) & \dots &\dots & g_m(\localVec_n)\\
    \end{bmatrix}
\renewcommand*{\arraystretch}{1.4}
    \begin{bmatrix}
        w_{1} \\
        w_{2} \\
        \vdots \\
        w_{n} \\
    \end{bmatrix}
    = 
    \begin{bmatrix}
        \int_{\Omega_e}{g_1(\localVec) \ }{d\Omega_e} \\
       \int_{\Omega_e}{g_2(\localVec) \ }{d\Omega_e}\\
        \vdots \\
        \int_{\Omega_e}{g_m(\localVec) \ }{d\Omega_e}\\
    \end{bmatrix}
\end{equation}
which is nonlinear with respect to $\localVec_i$ and linear in $w_i$. In essence, our approach consists in setting the nodes $\localVec_i$ as the original GLL nodes of the element and evaluating the right-hand side of \autoref{eq:fitLinear} with the quadrature rule obtained in \autoref{sec:LSM}. With this strategy, the main intent is to preserve the properties stemming from nodal quadrature, namely an efficient integration of the weak form and - most importantly - a diagonal mass matrix, while effectively accounting for the reduced volume and the cut configuration of the element by enforcing \autoref{eq:fitDef}. Moreover, the moment fitting problem is now reduced to the determination of the weights $w_i$ and can be expressed as a linear system of the form:
\begin{equation}\label{eq:fitLinearShort}
    \mathbf{A} \mathbf{w} = \mathbf{b} 
\end{equation}
An inspection of $\mathbf{A}$ highlights that the number of monomials is limited by the number of integration points, the system being overdetermined for $m > n$. It is important to mention that zero or negative weights might arise by solving \autoref{eq:fitLinear} for a cut element. This is very problematic since it leads to zero or negative diagonal coefficients in the integration of the mass matrix, which, in turn, cause explicit solvers to diverge. While the standard GLL integration weights are guaranteed to be positive, this property cannot be imposed directly, since, for the SE at hand, the number of basis monomials equals the number of nodes, i.e. $m = n = (p + 1)(q + 1)(r + 1)$ and, thus, matrix $\mathbf{A}$ is full rank. To overcome this, we relax the requirement that eq. \ref{eq:fitLinearShort} should be satisfied exactly by allowing for a nonzero residual:
\begin{equation} \label{eq:residual}
        \mathbf{r} = \mathbf{A} \mathbf{w} - \mathbf{b}.
\end{equation}
Then, boundary conditions can be applied, and weights are obtained by minimising some norm of this residual. In the present case, we minimise its $L_2$ norm $\sqrt{(\mathbf{A} \mathbf{w}-\mathbf{b})^T (\mathbf{A} \mathbf{w}-\mathbf{b})}$, and enforce positive weights, as well as mass conservation, leading to the following constrained quadratic programming problem:
\begin{mini!}|l|[3]
{\mathbf{w}}{\frac{1}{2} \mathbf{w}^T \bar{\mathbf{A}} \mathbf{w} - \mathbf{w}^T \bar{\mathbf{b}}}
{}{}
\label{eq:fitQuad}
\addConstraint{w_i \geq w_{min} \ \ \forall i \in [1,  n]}\label{eq:pos_weights}
\addConstraint{\sum_{i = 1}^n{w_i} =  \int_{\Omega_e}{}{d\Omega_e}}{}\label{eq:mass_conservation}
\end{mini!}
where $\bar{\mathbf{A}} = \mathbf{A}^T \mathbf{A} $ and $\bar{\mathbf{b}} =  \mathbf{A}^T \mathbf{b} $. The addition of Constraint \ref{eq:mass_conservation} is due to the fact that mass, conserved by default in \autoref{eq:fitLinear} by $ g_1(\localVec) \equiv 1$, is not necessarily preserved in \autoref{eq:residual}. The value $w_{min} $ corresponds to the smallest allowed weight, which directly determines the minimum magnitude of mass coefficients. To adapt the optimization procedure to the size of $\Omega_e$, we propose the following:

\begin{equation} \label{eq:wminLinear}
 w_{min} = \epsilon \ v_e \ w_{std}
\end{equation}
where $w_{std}$ is the smallest standard GLL weight and $v_e$ is the element's physical volume ratio:
\begin{equation}
    v_e = \frac{\int\limits_{\Omega_e}{d\Omega_e}}{\int\limits_{\Omega_{e,tot}}{d\Omega_{e,tot}}},
\end{equation}
with respect to the full (embedding) element $\Omega_{e,tot}$. Parameter $\epsilon \in \ (0,1]$ is a numerical factor that can be used to further tune the optimization and should be set while keeping two conflicting requirements in mind. On one hand, a small value will reduce the constraints on \autoref{eq:fitQuad}, leading to a better optimization and thus a smaller lumping error. On the other hand, this will lead to small diagonal entries in the mass matrix and, consequently, very high eigenvalues in the system. For explicit time integrators, whose critical time step is determined by the CFL condition, this translates to a very small critical time step, which, in turn, results in increased computational effort. In the following subsection, these effects are studied in greater detail, enabling to propose appropriate values for this parameter.
%the 0.1 threshold used will be justified and appropriate values for $\epsilon$ will be established by investigating its influence on the critical time step.

\subsection{Time integration}\label{sec:leapfrog}

\begin{figure}
    \centering
	\includegraphics[width=.3\linewidth]{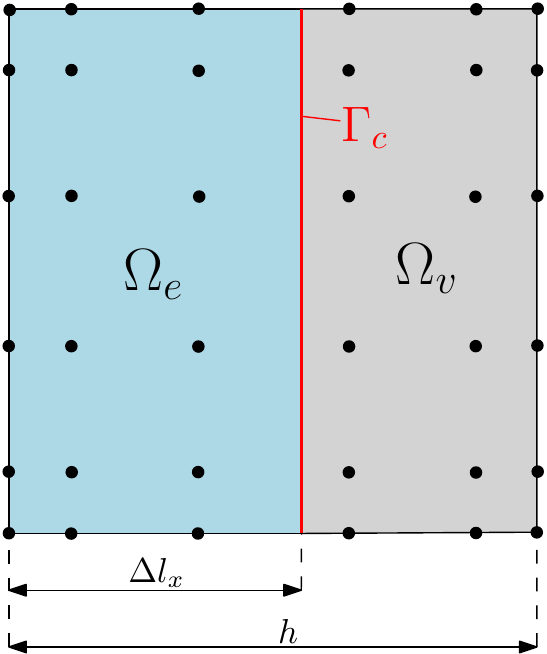}
	\caption{Unit square spectral element of order $p = 5$ intersected by a straight interface.}
	\label{fig:unit_square_cut}
\end{figure}

To illustrate the effect of the proposed moment fitting approach on the critical time step, we consider the case of a unit square element of varying order, intersected by a straight interface, as illustrated in \autoref{fig:unit_square_cut}. The critical time step corresponding to this element, can be computed as:
\begin{equation} \label{eq:critDt}
    \dt_{e} = \frac{2}{\omega_{max}}
\end{equation}
with $\omega_{max}$ being the highest eigenvalue of the element, obtained from the solution of the generalised eigenvalue problem:
\begin{equation}\label{eq:eigenvalProblem}
    \det \left( \bm{K}_e - \omega_e^2 \bm{M}_e \right) = 0
\end{equation}

\begin{figure}
\centering
\subfigure [$p = 4$] {
    \input{figures/time_integration/dtcrit_2d_sp4.tikz}
}
\subfigure [$p = 5$] {
    \input{figures/time_integration/dtcrit_2d_sp5.tikz}

}
\subfigure [$p = 6$] {
    \input{figures/time_integration/dtcrit_2d_sp6.tikz}
} 
\subfigure [$p = 7$] {
    \input{figures/time_integration/dtcrit_2d_sp7.tikz}
}

\subfigure{
    \centering
    \input{figures/time_integration/legend.tikz}
}
\caption{Critical time step ratios of SE of type $\mathcal{N}_{p,p}$ for different cutting fractions $ \Delta l_x $  and polynomial orders $p$. Element nodes are marked by the values at the abscissa. For the fitted SCM, \autoref{eq:wminLinear} has been applied.}
\label{fig:element_critical_dt}
\end{figure}
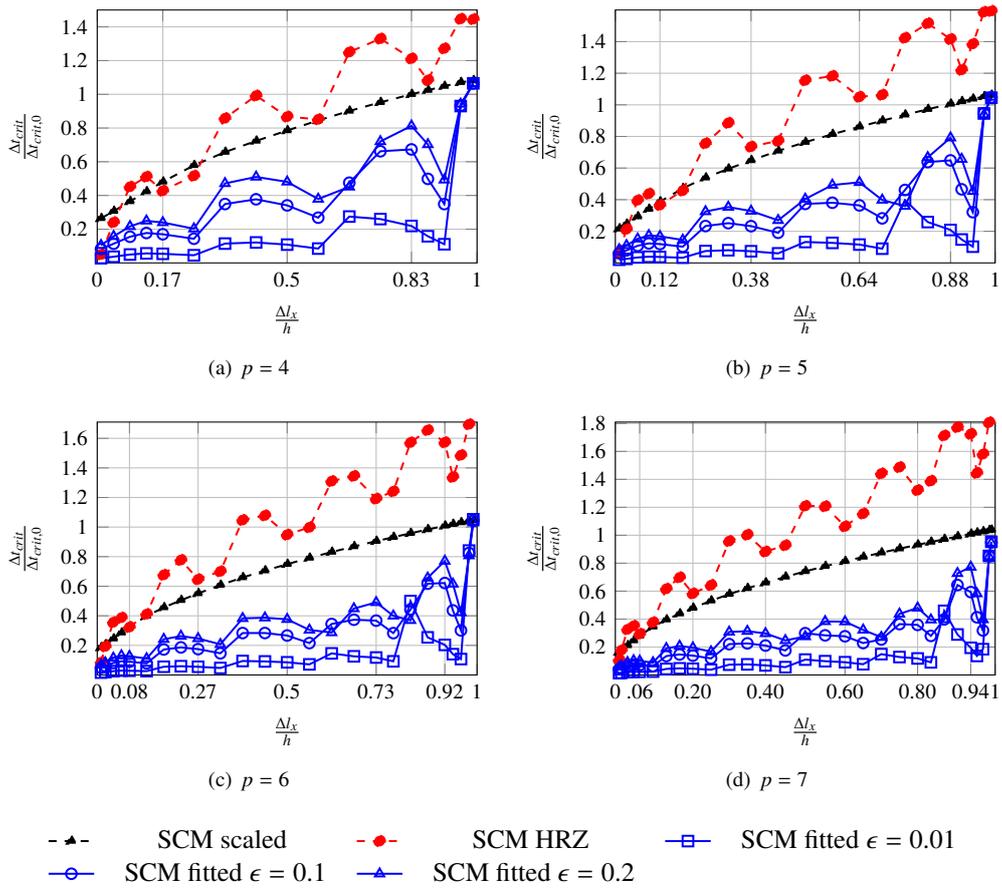
In \autoref{fig:element_critical_dt}, the ratio between the critical time steps of the cut ($\Delta t_{crit}$) and the initial element ($\Delta t_{crit, 0}$) is reported for different locations of the interface relative to the element size ($\frac{\Delta l_x}{h}$) and element orders ($p=4,5,6,7$). For comparison, the same results are obtained using the proposed approach (``SCM fitted''), as well as the lumping techniques ``1'' and ``2'' by Joulaian et al. \cite{joulaian2014finite}, which we label ``SCM scaled'' and ``SCM HRZ'', respectively. For the HRZ and the fitted methods, which include integration data from the cut configuration, oscillations of the results seem to be somewhat related to the location of the cut with respect to the element nodes, which are indicated by the marks on the abscissa of each graph. Most importantly though, the proposed approach results in the smallest critical time steps for all interface locations, with the difference in several cases being more than an order of magnitude. The critical time step for an assembly of elements is determined by the smallest critical time step among all individual elements:
\begin{equation}\label{eq:CFLcondition}
\dtcrit = \min_e \left\{ \dt_e \right\}
\end{equation}
Based on the above and on \autoref{fig:element_critical_dt}, the presence of a single intersected element with a small volume ratio will lead to a considerably reduced time step and a correspondingly increased computational effort. Nevertheless, in practice, this limitation becomes significant only for a small number of elements, for which $v_e$ is very small. Therefore, a first measure to alleviate this problem consists in replacing \autoref{eq:wminLinear} with:
\begin{equation} \label{eq:wminBilinear}
     w_{min} = \left\{ \begin{array}{ll}
        \epsilon \ v_e \ w_{std} & \mbox{if $v_e \geq 0.1$}\\
        v_e \ w_{std} & \mbox{if $v_e < 0.1$}.\end{array} \right.
\end{equation}
In the above, parameter $\epsilon$ is set to unity for elements with a volume ratio below a threshold of $10 \%$, which is chosen as to delimit the most problematic regions of \autoref{fig:element_critical_dt} without affecting accuracy in the remainder of the domain. This stronger restriction on the optimization problem is meant to lessen the emergence of very small mass coefficients as small values for $v_e$ are encountered. It is interesting to note that in this case, $w_{min}$ matches the corresponding smallest mass coefficient obtained in the scaled procedure by Joulaian et al.. For the remainder of the domain, a factor $\epsilon = 0.1$ is chosen, which in numerical investigations has been shown to provide a good compromise between the accuracy of mass lumping and performance of time integration. In \autoref{fig:element_critical_dt} it can be observed that this choice effectively reduces time step decay with respect to the option $\epsilon = 0.01$, while an increase to $\epsilon = 0.2$ would lead to only marginal improvements, but at the cost of higher lumping errors.

A third, more effective, measure, consists in the use of a ``leap-frog''~\cite{diaz2009energy} time stepping algorithm that allows to locally reduce the time step for intersected elements, while maintaining higher values for the remainder of the domain. As will be further investigated in \autoref{sec:numerical examples}, if a small fraction of the total number of elements is intersected, this approach introduces only a small computational overhead, leading to efficient solutions. For a review of the method it is convenient to re-write \autoref{eq:system} in the following form:
\begin{equation}
\label{eq:system_2}
 \ddot{\bm{z}}(t) + \bm{A} \ \bm{z}(t) = \bm{r}\left(t\right)
\end{equation}

\noindent where $\bm{z}\left(t\right) = \bm{M}^{\frac{1}{2}} \bm{u}_s\left( t \right)$, $\ddot{\bm{z}}\left(t\right) = \bm{M}^{\frac{1}{2}} \ddot{\bm{u}}_s\left( t \right)$, $\bm{r}\left(t\right) = \bm{M}^{-\frac{1}{2}} \bm{f}_s\left( t \right)$, $\bm{A} = \bm{M}^{-\frac{1}{2}} \bm{K} \bm{M}^{-\frac{1}{2}}$. Since $\bm{M}$ is a diagonal matrix, all these operations can be performed efficiently.

Assuming that $\bm{r},\bm{z} \in C^2$, the following holds for the analytical solution of \autoref{eq:system_2}:

\begin{equation}\label{eq:analytical_solution_integral}
    \bm{z}\left( t + \Delta t \right) + 2 \bm{z}\left( t \right) + \bm{z}\left( t - \Delta t \right) = \Delta t ^2 \int\limits_{-1}^{1} \left( 1 - \abs{\theta} \right) \left[ \bm{r}\left(t + \theta \Delta t \right) +  \bm{A} \bm{z}\left(t + \theta \Delta t \right)\right] d\theta
\end{equation}

\noindent where $\Delta t$ is some interval used as a time step. Approximating $\bm{r}$ and $\bm{z}$ in the above equation with their values at $t$ and denoting $t_n = n \Delta t$, $\bm{z}_n=\bm{z}\left(t_n\right)$, $\bm{r}_n=\bm{r}\left(t_n\right)$, the standard second order leap-frog scheme can be obtained:

\begin{equation}\label{eq:2nd_order_lf}
    \bm{z}_{n+1} - 2 \bm{z}_n + \bm{z}_{n-1} = \Delta t^2 \left( \bm{r}_n - \bm{A} \bm{z}_n \right)
\end{equation}

\noindent With known $\bm{z}_n$ and $\bm{z}_{n-1}$ it can be solved to yield $\bm{z}_{n+1}$. Then, Diaz and Grote~\cite{diaz2009energy, grote2013explicit} decompose $\bm{z}$ and $\bm{r}$ into a coarse and fine part as follows:

\begin{align}\label{eq:z_r_decomposition}
    \bm{z}\left(t\right) = \underbrace{\left(\bm{I}-\bm{P} \right)\bm{z}\left(t\right)}_{\bm{z}^{coarse}\left(t\right)} + \underbrace{\bm{P}\bm{z}\left(t\right)}_{\bm{z}^{fine}\left(t\right)} \\ \nonumber
    \bm{r}\left(t\right) = \underbrace{\left(\bm{I}-\bm{P} \right)\bm{r}\left(t\right)}_{\bm{r}^{coarse}\left(t\right)} + \underbrace{\bm{P}\bm{r}\left(t\right)}_{\bm{r}^{fine}\left(t\right)}
\end{align}

\noindent where $\bm{I}$ is a unit matrix and $\bm{P}$ is a diagonal selection matrix. Diagonal entries of $\bm{P}$ assume a value of either zero or one, allowing to select DOFs for which a reduced time step is to be used. Substituting Equation~\eqref{eq:z_r_decomposition} into \eqref{eq:analytical_solution_integral} and assuming the coarse part of the solution to remain constant during a time step, we obtain:

\begin{align}\label{eq:analytical_integrand_coarse_fine}
    & \bm{z}\left( t + \Delta t \right) + 2 \bm{z}\left( t \right) + \bm{z}\left( t - \Delta t \right) =   \\ \nonumber
    & \Delta t^2 \int\limits_{-1}^{1} \left( 1 - \abs{\theta} \right) \left[ \left(\bm{I}-\bm{P} \right) \bm{r}\left(t + \theta \Delta t \right) + \bm{A} \left(\bm{I}-\bm{P} \right) \bm{z}\left(t + \theta \Delta t \right) + \bm{P} \bm{r}\left(t + \theta \Delta t \right) + \bm{A} \bm{P} \bm{z}\left(t + \theta \Delta t \right) \right] d\theta
\end{align}

\noindent The integrand in the above equation is approximated by:

\begin{align}
    \left[ \left(\bm{I}-\bm{P} \right) \bm{r}\left(t + \theta \Delta t \right) + \bm{A} \left(\bm{I}-\bm{P} \right) \bm{z}\left(t + \theta \Delta t \right) + \bm{P} \bm{r}\left(t + \theta \Delta t \right) + \bm{A} \bm{P} \bm{z}\left(t + \theta \Delta t \right) \right] \approx \\ \nonumber  \left(\bm{I}-\bm{P} \right) \bm{r}\left(t \right) + \bm{A} \left(\bm{I}-\bm{P} \right) \bm{z}\left(t \right) 
    +   \bm{P} \bm{r}\left(t + \theta \Delta t \right) + \bm{A} \bm{P} \Tilde{\bm{z}}\left(t + \theta \Delta t \right) 
\end{align}

\noindent where $\Tilde{\bm{z}}$ is the solution of the equation:

\begin{align} \label{eq:z_tilde_equation}
\ddot{\Tilde{\bm{z}}} \left( \tau \right) = \left(\bm{I}-\bm{P} \right) \bm{r}\left(t \right) + \bm{A} \left(\bm{I}-\bm{P} \right) \bm{z}\left(t  \right) + \bm{P} \bm{r}\left(t + \tau \right) + \bm{A} \bm{P} \Tilde{\bm{z}}\left( \tau \right) \\ \nonumber
\Tilde{\bm{z}}\left( 0 \right) = \bm{z} \left( t \right), \ \dot{\Tilde{\bm{z}}} \left( 0 \right) = \bm{\nu}
\end{align}

\noindent where $t$ is considered fixed and $\bm{\nu}$ is the initial value of the derivative of $\Tilde{\bm{z}}$. It can be shown that \cite{grote2013explicit}:

\begin{equation}
    \bm{z} \left( t + \Delta t \right) + \bm{z} \left( t - \Delta t \right) \approx \Tilde{\bm{z}} \left( \Delta t \right) + \Tilde{\bm{z}} \left( - \Delta t \right)
\end{equation}

\noindent If a new variable is defined as:

\begin{equation}
    \bm{q}\left( \tau \right) = \Tilde{\bm{z}} \left( \tau \right) + \Tilde{\bm{z}} \left( - \tau \right)
\end{equation}

\noindent then $\bm{z}$ at time $t + \Delta t$ can be approximated as:

\begin{equation}
    \bm{z}\left( t + \Delta t \right) \approx \bm{q}\left( \Delta t \right) - \bm{z}\left( t - \Delta t \right)
\end{equation}

\noindent while $\bm{q}$ can be obtained as the solution of equation:

\begin{align} \label{eq:q_equation}
\ddot{\bm{q}} \left( \tau \right) = 2 \left[ \left(\bm{I}-\bm{P} \right) \bm{r}\left(t \right) + \bm{A} \left(\bm{I}-\bm{P} \right) \bm{z}\left(t  \right) \right] + \bm{P} \left[ \bm{r}\left(t + \tau \right) + \bm{r}\left(t - \tau \right) \right] + \bm{A} \bm{P} \bm{q} \left( \tau \right) \\ \nonumber
\bm{q}\left( 0 \right) = 2 \bm{z} \left( t \right), \ \dot{\bm{q}} \left( 0 \right) = \bm{0}
\end{align}

\noindent The above equation, in contrast to Equation~\eqref{eq:z_tilde_equation}, does not depend on the choice of initial value for the derivative and can be solved using a leap-frog algorithm and a fraction of the time step used for the coarse solution, as summarised in \autoref{alg:leapfrog}.

\begin{algorithm}
\SetAlgoLined
\KwData{$\bm{A}$, $\bm{P}$, $\bm{r}\left( t \right)$, $\mathbf{z}_{n}$, $\mathbf{z}_{n-1}$, $t_n$,$\Delta t$, $p_t$}
\KwResult{$\mathbf{z}_{n+1}$ }

Set $\bm{w} = \left( \bm{I} - \bm{P}\right) \bm{r}\left(t_n\right) - \bm{A}\left(\bm{I}-\bm{P}\right) \mathbf{z}_{n}$ and $\bm{q}_0 = 2 \bm{z}_n$; 

Compute $\bm{q}_{1/p} = \bm{q}_0 + \dfrac{1}{2} \left( \dfrac{\Delta t}{p}\right)^2 \left[2\bm{w} + 2\bm{P}\bm{r}\left(t_n\right) - \bm{A}\bm{P}\bm{q}_0 \right]$;

\For{$m = 1, \dots, p_t -1$ }{
    $\bm{q}_{\left(m+1\right)/p} = 2\bm{q_{m/p}} - 2\bm{q_{\left(m-1\right)/p}} + \left( \dfrac{\Delta t}{p}\right)^2 \left\{2\bm{w} + \bm{P}\left[\bm{r}\left(t_n + m \Delta t\right)+ \bm{r}\left(t_n - m \Delta t\right) \right]  - \bm{A}\bm{P}\bm{q}_{m/p} \right\}$;
}

Compute $\bm{z}_{n+1} = - \bm{z}_{n-1} + \bm{q}_1$

\caption{Second order leap-frog algorithm}
\end{algorithm}\label{alg:leapfrog}

Herein, diagonal entries of the selection matrix $\bm{P}$ are set to unity for all nodes belonging to cut elements, while parameter $p_t$ from \autoref{alg:leapfrog} is chosen to yield a time step that is smaller than the smallest critical time step among all cut elements. Even though the overhead introduced by the method should be small, excessively large numbers of fine time steps, can render the method inefficient. Therefore, combining the method with additional measures for limiting the critical time step, as described in the beginning of the present subsection, is necessary to maintain efficiency.

%% file: figures/time_integration/dtcrit_2d_sp4.tikz
\begin{tikzpicture}
    \begin{axis}[
        xmin = 0,
        xmax = 1,
        ymin = 0,
        ymax = 1.5,
        xtick = {0, 0.172673, 0.5, 0.827327, 1},
        xticklabels = {$0$,$0.17$,$0.5$,$0.83$, $1$},
        ytick = {0.2, 0.4, 0.6, 0.8, 1, 1.2, 1.4},
        %xmode = log,
        %ymode = log,
        xlabel = {$ \frac{\Delta l_{x}}{h}$},
        ylabel = {$ \frac{\Delta t_{crit}}{\Delta t_{crit, 0}}  $},
        xticklabel style = {font =\fontsize{\figureFontSize pt}{10pt}\selectfont},
        yticklabel style = {font=\fontsize{\figureFontSize pt}{10pt}\selectfont},
        xlabel style = {font =\fontsize{\figureFontSize pt}{\figureFontSize pt}\selectfont},
        ylabel style = {font=\fontsize{\figureFontSize pt}{\figureFontSize pt}\selectfont},
        grid = both,
        width=0.4\textwidth,
        height=0.3\textwidth,
        legend style={at={(0,1)},anchor=south west,nodes={font=\fontsize{\figureFontSize pt}{\figureFontSize pt}\selectfont}}
        %legend pos = north west
        ]
\addplot[style = dashed, color = \uncutColor, mark = triangle*, line width=0.75pt] coordinates{
        (0.01, 0.2641964906601005)
        (0.043168291161502836, 0.30684882734242275)
        (0.08633658232300567, 0.3644742298444194)
        (0.12950487348450856, 0.4231807354448334)
        (0.1726731646460114, 0.4807323802542928)
        (0.25450487348450856, 0.5785716432223844)
        (0.3363365823230057, 0.657184255799655)
        (0.41816829116150284, 0.7232654511789176)
        (0.5, 0.7851535374218374)
        (0.5818317088384971, 0.8449490706036075)
        (0.6636634176769942, 0.9010716837733997)
        (0.7454951265154914, 0.9523727024645292)
        (0.8273268353539885, 1.0001042184079956)
        (0.8704951265154914, 1.0244675760722146)
        (0.9136634176769942, 1.0483214104608218)
        (0.9568317088384971, 1.0710016950514645)
        (0.99, 1.0824979810087736)
	};
    %\addlegendentry{SCM scaled}
    
\addplot[style = dashed, color = \cutHRZColor, mark =\explicitMarker*, line width=\plotLineWidth pt] coordinates{
        (0.01, 0.050409560585432334)
        (0.043168291161502836, 0.24085204094688964)
        (0.08633658232300567, 0.4519353359023781)
        (0.12950487348450856, 0.5113216738129834)
        (0.1726731646460114, 0.4281213612588271)
        (0.25450487348450856, 0.5169777239854435)
        (0.3363365823230057, 0.8582022456053375)
        (0.41816829116150284, 0.9922254026186931)
        (0.5, 0.8674889506959529)
        (0.5818317088384971, 0.8515317740459306)
        (0.6636634176769942, 1.2506073678882352)
        (0.7454951265154914, 1.3300682154118961)
        (0.8273268353539885, 1.2125299975371757)
        (0.8704951265154914, 1.0845497488816287)
        (0.9136634176769942, 1.2722596580403895)
        (0.9568317088384971, 1.4476290950016186)
        (0.99, 1.445873789020323)
	};
	%\addlegendentry{SCM HRZ}

	\addplot[color = \cutOptimizedColor, mark =square, line width=\plotLineWidth pt] coordinates{
         (0.01, 0.027761109644918)
        (0.043168291161502836, 0.03660651873136646)
        (0.08633658232300567, 0.04963727590832755)
        (0.12950487348450856, 0.05643844786748583)
        (0.1726731646460114, 0.05383062510180194)
        (0.25450487348450856, 0.04540399720453005)
        (0.3363365823230057, 0.1144890801442808)
        (0.41816829116150284, 0.12116979332001743)
        (0.5, 0.10832185296124484)
        (0.5818317088384971, 0.0848450647588377)
        (0.6636634176769942, 0.2748032305312954)
        (0.7454951265154914, 0.2600775537531307)
        (0.8273268353539885, 0.21906215380479999)
        (0.8704951265154914, 0.1580600511973773)
        (0.9136634176769942, 0.1106287062605793)
        (0.9568317088384971, 0.9302407944828126)
        (0.99, 1.0651878056779185)
	};
	%\addlegendentry{SCM fitted $\epsilon =0.01$}
   
    \addplot[color = \cutOptimizedColor, mark =o, line width=\plotLineWidth pt] coordinates{
        (0.01, 0.08240179379687637)
        (0.043168291161502836, 0.11375016202603336)
        (0.08633658232300567, 0.15508764532106697)
        (0.12950487348450856, 0.1773915331601402)
        (0.1726731646460114, 0.16980436473597948)
        (0.25450487348450856, 0.1432773776758344)
        (0.3363365823230057, 0.3490409914323668)
        (0.41816829116150284, 0.37724675735474994)
        (0.5, 0.341114768083311)
        (0.5818317088384971, 0.2681184709423851)
        (0.6636634176769942, 0.4756098925685184)
        (0.7454951265154914, 0.66204990417773)
        (0.8273268353539885, 0.6730154646134665)
        (0.8704951265154914, 0.4983553489983455)
        (0.9136634176769942, 0.3487686839663039)
        (0.9568317088384971, 0.9327174927169536)
        (0.99, 1.0647082810846435)
    };
	%\addlegendentry{SCM fitted $\epsilon = 0.1$}
	
	    \addplot[color = \cutOptimizedColor, mark =triangle, line width=\plotLineWidth pt] coordinates{
        (0.01, 0.10440530778007959)
        (0.043168291161502836, 0.15759560907114584)
        (0.08633658232300567, 0.2165361836786295)
        (0.12950487348450856, 0.24933678825271074)
        (0.1726731646460114, 0.2396658234472961)
        (0.25450487348450856, 0.20253659439645538)
        (0.3363365823230057, 0.47143956817784)
        (0.41816829116150284, 0.509768169030474)
        (0.5, 0.48072682502977165)
        (0.5818317088384971, 0.37883525037577875)
        (0.6636634176769942, 0.4508166576479736)
        (0.7454951265154914, 0.7202177974246696)
        (0.8273268353539885, 0.8128159630045381)
        (0.8704951265154914, 0.702304061418395)
        (0.9136634176769942, 0.49245743569380035)
        (0.9568317088384971, 0.9424160250827545)
        (0.99, 1.0647281239621795)
    };
	%\addlegendentry{SCM fitted $\epsilon = 0.2$}
    \end{axis}
    
\end{tikzpicture}

%% file: figures/time_integration/dtcrit_2d_sp5.tikz
\begin{tikzpicture}
    \begin{axis}[
        xmin = 0,
        xmax = 1,
        ymin = 0,
        ymax = 1.6,
         xtick = {0, 0.117472, 0.357384, 0.642616, 0.882528, 1},
        xticklabels = {$0$,$0.12$,$0.38$,$0.64$, $0.88$, $1$},
        ytick = {0.2, 0.4, 0.6, 0.8, 1, 1.2, 1.4},
        %xmode = log,
        %ymode = log,
        xlabel = {$ \frac{\Delta l_{x}}{h}$},
        ylabel = {$ \frac{\Delta t_{crit}}{\Delta t_{crit, 0}}  $},
        xticklabel style = {font =\fontsize{\figureFontSize pt}{10pt}\selectfont},
        yticklabel style = {font=\fontsize{\figureFontSize pt}{10pt}\selectfont},
        xlabel style = {font =\fontsize{\figureFontSize pt}{\figureFontSize pt}\selectfont},
        ylabel style = {font=\fontsize{\figureFontSize pt}{\figureFontSize pt}\selectfont},
        grid = both,
        width=0.4\textwidth,
        height=0.3\textwidth,
        %legend style={at={(0,1)},anchor=north west,nodes={font=\fontsize{\figureFontSize pt}{\figureFontSize pt}\selectfont}}
%        legend pos = south east
        ]
    
    \addplot[style = dashed, color = \uncutColor, mark = triangle*, line width=0.75pt] coordinates{
        	(0.01, 0.21684633626146452)
        (0.02936808450881684, 0.24592711014060922)
        (0.05873616901763368, 0.2926219136707477)
        (0.08810425352645052, 0.34046164827619735)
        (0.11747233803526735, 0.3871520187899513)
        (0.17745031396636984, 0.47169423734907506)
        (0.23742828989747233, 0.5385947505588479)
        (0.2974062658285749, 0.5950666314668002)
        (0.3573842417596773, 0.6481988386021055)
        (0.42869212087983866, 0.7085379410289157)
        (0.5, 0.7638411197738123)
        (0.5713078791201613, 0.8139153291243363)
        (0.6426157582403226, 0.8610437934298609)
        (0.702593734171425, 0.8995813883157872)
        (0.7625717101025276, 0.9368245429042058)
        (0.82254968603363, 0.9721380251118953)
        (0.8825276619647324, 1.0056521259459308)
        (0.9118957464735493, 1.021667630155514)
        (0.9412638309823662, 1.0375037310953075)
        (0.9706319154911831, 1.0528782321021055)
        (0.99, 1.0616196058329295)
	};
    %\addlegendentry{scaled}
    
    \addplot[style = dashed, color = \cutHRZColor, mark =\explicitMarker*, line width=\plotLineWidth pt] coordinates{
        (0.01, 0.06702971167225315)
        (0.02936808450881684, 0.21328449495378385)
        (0.05873616901763368, 0.39668882393094895)
        (0.08810425352645052, 0.43804836146967874)
        (0.11747233803526735, 0.36613729924836214)
        (0.17745031396636984, 0.45735434441967626)
        (0.23742828989747233, 0.7565565351386381)
        (0.2974062658285749, 0.8857215904503256)
        (0.3573842417596773, 0.7354770378062047)
        (0.42869212087983866, 0.7724118339378567)
        (0.5, 1.1549074581565353)
        (0.5713078791201613, 1.183023230690076)
        (0.6426157582403226, 1.0507403539698799)
        (0.702593734171425, 1.0637844207020128)
        (0.7625717101025276, 1.4241869855443006)
        (0.82254968603363, 1.5157310285672865)
        (0.8825276619647324, 1.4159017875036064)
        (0.9118957464735493, 1.2217899285384948)
        (0.9412638309823662, 1.3858263629923957)
        (0.9706319154911831, 1.587809321683712)
        (0.99, 1.5935721037749129)
	};
	%\addlegendentry{cut HRZ}

	\addplot[color = \cutOptimizedColor, mark =square, line width=\plotLineWidth pt] coordinates{
        (0.01, 0.019779629227760085)
        (0.02936808450881684, 0.025332313460234224)
        (0.05873616901763368, 0.034448606404106685)
        (0.08810425352645052, 0.03914656683587978)
        (0.11747233803526735, 0.03765905792891239)
        (0.17745031396636984, 0.032677610619222)
        (0.23742828989747233, 0.07505196987985321)
        (0.2974062658285749, 0.08015888400923583)
        (0.3573842417596773, 0.07375045510804996)
        (0.42869212087983866, 0.06030046604996584)
        (0.5, 0.13220591873845353)
        (0.5713078791201613, 0.12657100569134594)
        (0.6426157582403226, 0.11578363612858328)
        (0.702593734171425, 0.0898682447254796)
        (0.7625717101025276, 0.3971147714326075)
        (0.82254968603363, 0.25852295409620585)
        (0.8825276619647324, 0.2084146452177811)
        (0.9118957464735493, 0.14792617376965006)
        (0.9412638309823662, 0.10260424547933171)
        (0.9706319154911831, 0.9452962566971258)
        (0.99, 1.0449800145324908)
	};
	%\addlegendentry{cut optimized $\epsilon = 0.01$}
	
	\addplot[color = \cutOptimizedColor, mark =o, line width=\plotLineWidth pt] coordinates{
        (0.01, 0.0613107190602356)
        (0.02936808450881684, 0.07918517025506669)
        (0.05873616901763368, 0.1082737362097926)
        (0.08810425352645052, 0.12344639372492967)
        (0.11747233803526735, 0.11895691448051386)
        (0.17745031396636984, 0.10263943615282284)
        (0.23742828989747233, 0.2331535838620877)
        (0.2974062658285749, 0.2513466867427393)
        (0.3573842417596773, 0.23244265987321883)
        (0.42869212087983866, 0.19057002221268238)
        (0.5, 0.37306406940779197)
        (0.5713078791201613, 0.3805684750592517)
        (0.6426157582403226, 0.36376698389277623)
        (0.702593734171425, 0.2817003308340774)
        (0.7625717101025276, 0.4613947239458877)
        (0.82254968603363, 0.6379417870017586)
        (0.8825276619647324, 0.6484842366439519)
        (0.9118957464735493, 0.4654688492450875)
        (0.9412638309823662, 0.32103380685548016)
        (0.9706319154911831, 0.9469077788806117)
        (0.99, 1.0463262983615667)
	};
	%\addlegendentry{cut optimized $\epsilon = 0.1$}
	
		\addplot[color = \cutOptimizedColor, mark =triangle, line width=\plotLineWidth pt] coordinates{
        (0.01, 0.08435696162369051)
        (0.02936808450881684, 0.110836927692566)
        (0.05873616901763368, 0.15215936031766905)
        (0.08810425352645052, 0.17407158533894918)
        (0.11747233803526735, 0.168068077129766)
        (0.17745031396636984, 0.14504285000403183)
        (0.23742828989747233, 0.325085266892302)
        (0.2974062658285749, 0.35325597392161634)
        (0.3573842417596773, 0.32815856295632484)
        (0.42869212087983866, 0.2692771339442799)
        (0.5, 0.40409191502604247)
        (0.5713078791201613, 0.49263055494767505)
        (0.6426157582403226, 0.5097241151872098)
        (0.702593734171425, 0.3979719118589007)
        (0.7625717101025276, 0.3591791611350356)
        (0.82254968603363, 0.6673890353114847)
        (0.8825276619647324, 0.7916584483936104)
        (0.9118957464735493, 0.6563330820167802)
        (0.9412638309823662, 0.453263149197781)
        (0.9706319154911831, 0.9379938335301844)
        (0.99, 1.0489942756034316)
	};
	%\addlegendentry{cut optimized $\epsilon = 0.2$}
	
    \end{axis}
    
\end{tikzpicture}

%% file: figures/time_integration/dtcrit_2d_sp6.tikz
\begin{tikzpicture}
    \begin{axis}[
        xmin = 0,
        xmax = 1,
        ymin = 0,
        ymax = 1.71,
        xtick = {0, 0.0848881, 0.265576, 0.5, 0.734424, 0.915112, 1},
        xticklabels = {$0$,$0.08$,$0.27$,$0.5$,$0.73$, $0.92$, $1$},
        ytick = {0.2, 0.4, 0.6, 0.8, 1, 1.2, 1.4, 1.6},
        %xmode = log,
        %ymode = log,
        xlabel = {$ \frac{\Delta l_{x}}{h}$},
        ylabel = {$ \frac{\Delta t_{crit}}{\Delta t_{crit, 0}}  $},
        xticklabel style = {font =\fontsize{\figureFontSize pt}{10pt}\selectfont},
        yticklabel style = {font=\fontsize{\figureFontSize pt}{10pt}\selectfont},
        xlabel style = {font =\fontsize{\figureFontSize pt}{\figureFontSize pt}\selectfont},
        ylabel style = {font=\fontsize{\figureFontSize pt}{\figureFontSize pt}\selectfont},
        grid = both,
        width=0.4\textwidth,
        height=0.3\textwidth,
        legend style={at={(0,1)},anchor=north west,nodes={font=\fontsize{\figureFontSize pt}{\figureFontSize pt}\selectfont}}
%        legend pos = south east
        ]
    
    \addplot[style = dashed,color = \uncutColor, mark = triangle*, line width=0.75pt] coordinates{
    	(0.01, 0.18558955058700893)
        (0.021222012965179116, 0.20546290938330988)
        (0.042444025930358176, 0.24496850399620745)
        (0.06366603889553724, 0.2853456153723236)
        (0.08488805186071635, 0.3246542970933922)
        (0.13005993971169794, 0.39833623408484364)
        (0.17523182756267958, 0.45612100205633077)
        (0.22040371541366122, 0.5049960214644077)
        (0.2655756032646428, 0.5510661547885888)
        (0.3241817024484821, 0.6077691423122629)
        (0.3827878016323214, 0.6590538243960168)
        (0.4413939008161607, 0.705274636237089)
        (0.5, 0.7488959020769004)
        (0.5586060991838392, 0.7910539533841316)
        (0.6172121983676785, 0.831131315286843)
        (0.6758182975515177, 0.8687043484782977)
        (0.734424396735357, 0.9045503721860472)
        (0.7795962845863386, 0.9315823978087142)
        (0.8247681724373201, 0.9580474803206672)
        (0.8699400602883017, 0.9835923726512503)
        (0.9151119481392833, 1.008142675514088)
        (0.9363339611044625, 1.01944493252305)
        (0.9575559740696417, 1.0306530081069816)
        (0.9787779870348208, 1.0416275135089792)
        (0.99, 1.0470139630126696)
	};
    %\addlegendentry{scaled}
    
    \addplot[style = dashed, color = \cutHRZColor, mark =\explicitMarker*, line width=\plotLineWidth pt] coordinates{
        (0.01, 0.08596368556926297)
        (0.021222012965179116, 0.19297505998103887)
        (0.042444025930358176, 0.35647897012406754)
        (0.06366603889553724, 0.38748142919622564)
        (0.08488805186071635, 0.323806333080902)
        (0.13005993971169794, 0.41161021182849483)
        (0.17523182756267958, 0.6770396475835733)
        (0.22040371541366122, 0.777308776805883)
        (0.2655756032646428, 0.6467194328212782)
        (0.3241817024484821, 0.7006502132948019)
        (0.3827878016323214, 1.0476648012369765)
        (0.4413939008161607, 1.0786668049678005)
        (0.5, 0.948458914131757)
        (0.5586060991838392, 0.9984215662910266)
        (0.6172121983676785, 1.3105812349568104)
        (0.6758182975515177, 1.345526011593895)
        (0.734424396735357, 1.1917592733035853)
        (0.7795962845863386, 1.2419916683831485)
        (0.8247681724373201, 1.571856974414859)
        (0.8699400602883017, 1.6554219709800033)
        (0.9151119481392833, 1.5729061798996853)
        (0.9363339611044625, 1.3387624108201719)
        (0.9575559740696417, 1.4864333041601616)
        (0.9787779870348208, 1.6949830334180584)
        (0.99, 1.7122866195693398)
	};
	%\addlegendentry{cut HRZ}

	\addplot[color = \cutOptimizedColor, mark =square, line width=\plotLineWidth pt] coordinates{
	    (0.01, 0.015769867274950156)
        (0.021222012965179116, 0.018932642600802543)
        (0.042444025930358176, 0.025828614569628542)
        (0.06366603889553724, 0.029358990000771294)
        (0.08488805186071635, 0.028394835337211943)
        (0.13005993971169794, 0.025956139558194178)
        (0.17523182756267958, 0.0555041317687386)
        (0.22040371541366122, 0.05914905911832782)
        (0.2655756032646428, 0.055274148388751686)
        (0.3241817024484821, 0.04708246546045487)
        (0.3827878016323214, 0.09443424698471438)
        (0.4413939008161607, 0.09149769412658172)
        (0.5, 0.08554829857862133)
        (0.5586060991838392, 0.07152253613814441)
        (0.6172121983676785, 0.14586403544116183)
        (0.6758182975515177, 0.12595278552125563)
        (0.734424396735357, 0.11788607077597198)
        (0.7795962845863386, 0.09216827125168695)
        (0.8247681724373201, 0.49960516594700166)
        (0.8699400602883017, 0.25415400217929823)
        (0.9151119481392833, 0.20332290444637893)
        (0.9363339611044625, 0.14057589837950643)
        (0.9575559740696417, 0.10756406980214657)
        (0.9787779870348208, 0.8412183251203265)
        (0.99, 1.0526533609003168)
	};
	%\addlegendentry{cut optimized $\epsilon = 0.01$}
	
		\addplot[color = \cutOptimizedColor, mark =o, line width=\plotLineWidth pt] coordinates{
	   (0.01, 0.04839276897517653)
        (0.021222012965179116, 0.05933479799214358)
        (0.042444025930358176, 0.08133056785882072)
        (0.06366603889553724, 0.09263372462613569)
        (0.08488805186071635, 0.08963583523757901)
        (0.13005993971169794, 0.0783313077210253)
        (0.17523182756267958, 0.17228427665488452)
        (0.22040371541366122, 0.1862575165358497)
        (0.2655756032646428, 0.174688991052408)
        (0.3241817024484821, 0.1464366644609963)
        (0.3827878016323214, 0.2832915403242642)
        (0.4413939008161607, 0.2834701257465399)
        (0.5, 0.2667227292275553)
        (0.5586060991838392, 0.21545398973499838)
        (0.6172121983676785, 0.3446531290629821)
        (0.6758182975515177, 0.37354446722257034)
        (0.734424396735357, 0.3673474045012824)
        (0.7795962845863386, 0.2828310321260958)
        (0.8247681724373201, 0.44267292595316693)
        (0.8699400602883017, 0.6169760897402284)
        (0.9151119481392833, 0.6215589197178601)
        (0.9363339611044625, 0.4365120364056424)
        (0.9575559740696417, 0.3010955713912787)
        (0.9787779870348208, 0.8249441973384708)
        (0.99, 1.0427387213674018)
	};
	%\addlegendentry{cut optimized $\epsilon = 0.1$}
	
		\addplot[color = \cutOptimizedColor, mark =triangle, line width=\plotLineWidth pt] coordinates{
	  (0.01, 0.06725408539195045)
        (0.021222012965179116, 0.08328930946378878)
        (0.042444025930358176, 0.11455089064956671)
        (0.06366603889553724, 0.1307711753908448)
        (0.08488805186071635, 0.12668398889998195)
        (0.13005993971169794, 0.11082774250150948)
        (0.17523182756267958, 0.2416222751341314)
        (0.22040371541366122, 0.26262867211284874)
        (0.2655756032646428, 0.24661787957746462)
        (0.3241817024484821, 0.20700942761415106)
        (0.3827878016323214, 0.38269862914695324)
        (0.4413939008161607, 0.38677737671479173)
        (0.5, 0.37649325711962184)
        (0.5586060991838392, 0.3043398341918418)
        (0.6172121983676785, 0.2867672010192163)
        (0.6758182975515177, 0.4488444037613987)
        (0.734424396735357, 0.4906686587850502)
        (0.7795962845863386, 0.39970119504980156)
        (0.8247681724373201, 0.36943726262695065)
        (0.8699400602883017, 0.6561625848423153)
        (0.9151119481392833, 0.7703491139987335)
        (0.9363339611044625, 0.6155669211127044)
        (0.9575559740696417, 0.4223240016238854)
        (0.9787779870348208, 0.8069839753856365)
        (0.99, 1.029427885602862)
	};
	%\addlegendentry{cut optimized $\epsilon = 0.2$}
	
    \end{axis}
    
\end{tikzpicture}

%% file: figures/time_integration/dtcrit_2d_sp7.tikz
\begin{tikzpicture}
    \begin{axis}[
        xmin = 0,
        xmax = 1,
        ymin = 0,
        ymax = 1.81,
        xtick = {0, 0.0641299, 0.20415, 0.39535, 0.60465, 0.79585, 0.93587, 1},
        xticklabels = {$0$, $.06$,  $0.20$,  $0.40$,  $0.60$,  $0.80$,   $0.94$, $1$},
        ytick = {0.2, 0.4, 0.6, 0.8, 1, 1.2, 1.4, 1.6, 1.8},
        %xmode = log,
        %ymode = log,
        xlabel = {$ \frac{\Delta l_{x}}{h}$},
        ylabel = {$ \frac{\Delta t_{crit}}{\Delta t_{crit, 0}}  $},
        xticklabel style = {font =\fontsize{\figureFontSize pt}{10pt}\selectfont},
        yticklabel style = {font=\fontsize{\figureFontSize pt}{10pt}\selectfont},
        xlabel style = {font =\fontsize{\figureFontSize pt}{\figureFontSize pt}\selectfont},
        ylabel style = {font=\fontsize{\figureFontSize pt}{\figureFontSize pt}\selectfont},
        grid = both,
        width=0.4\textwidth,
        height=0.3\textwidth,
        %legend style={at={(0,1)},anchor=north west,nodes={font=\fontsize{\figureFontSize pt}{\figureFontSize pt}\selectfont}}
%        legend pos = south east
        ]
    
    \addplot[style = dashed, color = \uncutColor, mark = triangle*, line width=0.75pt] coordinates{
	(0.01, 0.1643446902802986)
    (0.016032481436299317, 0.17669231862008575)
    (0.03206496287259858, 0.21093035445131342)
    (0.04809744430889784, 0.24587100162683187)
    (0.06412992574519716, 0.2798338678595087)
    (0.09913492162975529, 0.3448519690829615)
    (0.13413991751431342, 0.3955503590914911)
    (0.1691449133988715, 0.4384818286487406)
    (0.20414990928342963, 0.4789892322747765)
    (0.2519500297247621, 0.5312353897085141)
    (0.29975015016609463, 0.5781042997874597)
    (0.34755027060742716, 0.6202324337012076)
    (0.39535039104875963, 0.6600314066620341)
    (0.44767519552437984, 0.7019543397515776)
    (0.5, 0.7414095418073933)
    (0.5523248044756203, 0.7781873583455002)
    (0.6046496089512405, 0.8133089248941638)
    (0.6524497293925725, 0.8445800298425186)
    (0.7002498498339046, 0.874828290803224)
    (0.7480499702752367, 0.9037263977536646)
    (0.7958500907165686, 0.9315712328674298)
    (0.8308550866011275, 0.9516150132288559)
    (0.8658600824856865, 0.9713797673893663)
    (0.9008650783702453, 0.9906543141558783)
    (0.9358700742548042, 1.0093288872389465)
    (0.9519025556911032, 1.0177350901005657)
    (0.967935037127402, 1.026086504071451)
    (0.983967518563701, 1.0342997630147936)
    (0.99, 1.0372519146798453)
	};
    %\addlegendentry{scaled}
    
    \addplot[style = dashed, color = \cutHRZColor, mark =\explicitMarker*, line width=\plotLineWidth pt] coordinates{
        (0.01, 0.10734633167430932)
        (0.016032481436299317, 0.17784536654527192)
        (0.03206496287259858, 0.3268750274139485)
        (0.04809744430889784, 0.3514460308834401)
        (0.06412992574519716, 0.29373561699141293)
        (0.09913492162975529, 0.37710245490348476)
        (0.13413991751431342, 0.6169824989839098)
        (0.1691449133988715, 0.6984352735135785)
        (0.20414990928342963, 0.5831763884160913)
        (0.2519500297247621, 0.6428067193161577)
        (0.29975015016609463, 0.9582606284121683)
        (0.34755027060742716, 1.0035846094185246)
        (0.39535039104875963, 0.8829975708751366)
        (0.44767519552437984, 0.9279859427416274)
        (0.5, 1.209866331598534)
        (0.5523248044756203, 1.2071360194195846)
        (0.6046496089512405, 1.0634506349327797)
        (0.6524497293925725, 1.1549865133872583)
        (0.7002498498339046, 1.441755107103566)
        (0.7480499702752367, 1.4873115412936073)
        (0.7958500907165686, 1.3225973127616126)
        (0.8308550866011275, 1.3893349739624017)
        (0.8658600824856865, 1.7135957791256566)
        (0.9008650783702453, 1.7735272581902677)
        (0.9358700742548042, 1.7253788264598071)
        (0.9519025556911032, 1.4457440935110772)
        (0.967935037127402, 1.5807441078220885)
        (0.983967518563701, 1.8080171288413405)
        (0.99, 1.8225275541785075)
	};
	%\addlegendentry{cut HRZ}

	\addplot[color = \cutOptimizedColor, mark =square, line width=\plotLineWidth pt] coordinates{
            (0.01, 0.013002046815904701)
            (0.016032481436299317, 0.014919674829442536)
            (0.03206496287259858, 0.020394682268802405)
            (0.04809744430889784, 0.0231851592179193)
            (0.06412992574519716, 0.02247429682627246)
            (0.09913492162975529, 0.023821154911598082)
            (0.13413991751431342, 0.043221000143977985)
            (0.1691449133988715, 0.04731575300184401)
            (0.20414990928342963, 0.04456779467789281)
            (0.2519500297247621, 0.03763203110402739)
            (0.29975015016609463, 0.07355907853918932)
            (0.34755027060742716, 0.07575955015915455)
            (0.39535039104875963, 0.0684997271547575)
            (0.44767519552437984, 0.05582993895979905)
            (0.5, 0.11000084066831119)
            (0.5523248044756203, 0.09858163144716763)
            (0.6046496089512405, 0.09115105813353705)
            (0.6524497293925725, 0.07674298352558956)
            (0.7002498498339046, 0.14907226962714024)
            (0.7480499702752367, 0.12820137210101343)
            (0.7958500907165686, 0.11818325623886686)
            (0.8308550866011275, 0.09017075014088845)
            (0.8658600824856865, 0.45887329138883537)
            (0.9008650783702453, 0.2895514846152276)
            (0.9358700742548042, 0.19435313161216308)
            (0.9519025556911032, 0.13605867980839204)
            (0.967935037127402, 0.18445354017864754)
            (0.983967518563701, 0.8521974458306126)
            (0.99, 0.9531178518052245)
	};
	%\addlegendentry{cut optimized $\epsilon = 0.01$}
	
	\addplot[color = \cutOptimizedColor, mark =o, line width=\plotLineWidth pt] coordinates{
           (0.01, 0.04060026666503842)
            (0.016032481436299317, 0.04687916864794127)
            (0.03206496287259858, 0.06438757454697322)
            (0.04809744430889784, 0.0732012845220463)
            (0.06412992574519716, 0.07107975978962183)
            (0.09913492162975529, 0.06322712762586849)
            (0.13413991751431342, 0.13517257301940186)
            (0.1691449133988715, 0.14655347385158476)
            (0.20414990928342963, 0.13835452198552292)
            (0.2519500297247621, 0.11831517462016707)
            (0.29975015016609463, 0.22048502441526036)
            (0.34755027060742716, 0.22583118832105942)
            (0.39535039104875963, 0.21041947331945068)
            (0.44767519552437984, 0.17528972910521629)
            (0.5, 0.29891089818764693)
            (0.5523248044756203, 0.2834940674945242)
            (0.6046496089512405, 0.276767768699441)
            (0.6524497293925725, 0.2283487915233142)
            (0.7002498498339046, 0.25372637737916887)
            (0.7480499702752367, 0.36392193605296613)
            (0.7958500907165686, 0.3582081179418159)
            (0.8308550866011275, 0.27958122761351534)
            (0.8658600824856865, 0.397547413966063)
            (0.9008650783702453, 0.6437801483695264)
            (0.9358700742548042, 0.5930196418387741)
            (0.9519025556911032, 0.4128977351197851)
            (0.967935037127402, 0.3183009240777115)
            (0.983967518563701, 0.8498073029421732)
            (0.99, 0.9482358876196575)
	};
	%\addlegendentry{cut optimized $\epsilon = 0.1$}
	
	\addplot[color = \cutOptimizedColor, mark =triangle, line width=\plotLineWidth pt] coordinates{
           (0.01, 0.05704229723883213)
            (0.016032481436299317, 0.0660146959811153)
            (0.03206496287259858, 0.0907811940288651)
            (0.04809744430889784, 0.1034102271011659)
            (0.06412992574519716, 0.10042106467986658)
            (0.09913492162975529, 0.08855783177524833)
            (0.13413991751431342, 0.19024724506635765)
            (0.1691449133988715, 0.20671051348896843)
            (0.20414990928342963, 0.19538364128400473)
            (0.2519500297247621, 0.16679070400356266)
            (0.29975015016609463, 0.30779907914515686)
            (0.34755027060742716, 0.31331137852317203)
            (0.39535039104875963, 0.29712848083142007)
            (0.44767519552437984, 0.24722684472874265)
            (0.5, 0.27494998838318924)
            (0.5523248044756203, 0.3830692080176971)
            (0.6046496089512405, 0.38164433649398033)
            (0.6524497293925725, 0.32232736083565755)
            (0.7002498498339046, 0.2699018204739101)
            (0.7480499702752367, 0.4368018846188169)
            (0.7958500907165686, 0.47984046871380853)
            (0.8308550866011275, 0.39489206628796414)
            (0.8658600824856865, 0.39670476042574854)
            (0.9008650783702453, 0.727251883858937)
            (0.9358700742548042, 0.7725054334444621)
            (0.9519025556911032, 0.5822231502986553)
            (0.967935037127402, 0.39662295485729)
            (0.983967518563701, 0.8476412971855359)
            (0.99, 0.9442962602550607)
	};
	%\addlegendentry{cut optimized $\epsilon = 0.2$}
    \end{axis}
    
\end{tikzpicture}

%% file: figures/time_integration/legend.tikz
\begin{tikzpicture}
\begin{customlegend}[legend columns=3,legend style={align=left,draw=none,column sep=2ex},
        legend entries={SCM scaled ,
                        SCM HRZ ,
                        SCM fitted $\epsilon =0.01$,
                        SCM fitted $\epsilon =0.1$,
                        SCM fitted $\epsilon =0.2$,
                        }]
        \addlegendimage{style = dashed, color = \uncutColor, mark = triangle*, line width=0.75pt}
        \addlegendimage{style = dashed, color = \cutHRZColor, mark =\explicitMarker*, line width=\plotLineWidth pt} 
        \addlegendimage{color = \cutOptimizedColor, mark =square, line width=\plotLineWidth pt}
        \addlegendimage{color = \cutOptimizedColor, mark =o, line width=\plotLineWidth pt}
        \addlegendimage{color = \cutOptimizedColor, mark =triangle, line width=\plotLineWidth pt}
        \end{customlegend}
\end{tikzpicture}

%% file: 4_1_example2D.tex
\section{Numerical examples}\label{sec:numerical examples}
In this section, the novel method is applied to the solution of three problems of increasing complexity. In the first example, the data presented in \autoref{fig:element_critical_dt} is complemented by studying the accuracy of all the aforementioned lumping methods by means of a 2D example. Lamb waves are then modeled with the 3D version of the method in \autoref{sec:ex2}, and \autoref{sec:ex3}, where its performance is compared to the mesh-conforming SEM.

\subsection{Plane bar with straight cut}\label{sec:ex1}

\begin{figure}[]
    \centering
    \includegraphics[width=1\linewidth]{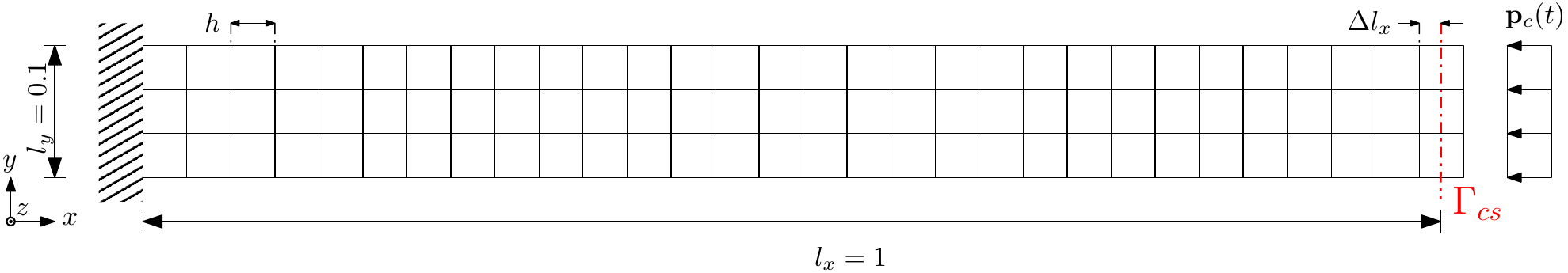}
    \caption{2D benchmark example with fixed left boundary, uniform planar loading $\mathbf{p}_c(t)$ acting on cut interface, $l_x = 1, ly = 0.1, l_z = 1$}
    \label{fig:example1}
\end{figure}

The 2D mesh illustrated in \autoref{fig:example1} represents a bar of unit width and thickness $l_y = 0.1$, with fixed boundary conditions on its left end. For this example, we assume academic material properties ($E = 1$, $\nu = 0$, $\rho = 1$) and plane strain conditions. The benchmark consists in letting a mesh of length $l_x + h - \Delta l_x$ be cut at the interface $\Gamma_{cs}$, represented by the plane $x = l_x$. A distributed load $\mathbf{p}_c(t) = -\mathbf{e}_x p(t)$ acts uniformly on $\Gamma_{cs}$ and is modulated by the Hann window:
\begin{equation}
    p(t) = p \sin{(\omega t) } \sin^2\left({\frac{\omega t}{2n}}\right), \ \ \ t \in \left[0; \frac{n}{f} \right]
    \label{eq:hanning}
\end{equation}
where $p = 10^{6}$ is the load amplitude and $\omega = 2 \pi f$ is the angular frequency. $n = 5$ represents the number of cycles within one pulse, which, for a frequency of $f = 20 \ [Hz]$, leads to an excitation window of $n / f = 0.25 \ [s]$. In this setup, inspired by a similar benchmark in \cite{kumar2018enriched}, the analytical expression for the velocity in a rod can be used to validate numerical results over $\Omega$:
\begin{align} \label{eq:velSolution}
&\dot{u}_x(x, t) = \frac{cp}{E} \sin^2 \left( \frac{\omega \ell}{2 c n} \right) \sin\left(\frac{\omega \ell }{c}\right),
\hspace{15pt}x \in \left[l_x - ct; \ l_x - c\left(t+\frac{n}{f}\right)\right],\\
&\nonumber c = \sqrt{\frac{E}{\rho}},\\
&\nonumber \ell = x + ct - l_x
\end{align}
For all simulations of this example, a finer than necessary time discretization ($\Delta t = 10^{-5} [s]$) is used, as the accuracy of space discretization is of interest. The accuracy is assessed by computing the $L2$ error norm of the velocity field at time $t = 0.4 \ [s]$:
\begin{equation} \label{eq:L2_velocity}
\epsilon_{h,\Omega} = \sqrt{ \frac{ \int\limits_{\Omega} \lVert \dot{u}_{h,e}(t) - \dot{u}_{ref,e}(t)\rVert^2 d\Omega}{\int\limits_{\Omega}  \lVert \dot{u}_{ref,e}^2(t) \rVert \ d\Omega} }
\end{equation}
%%%%%%%%%%%%%%%%%%%%%%%%%%%%%%%%%%%%%%%%%%%%%%%%%%%%%%
%sp5
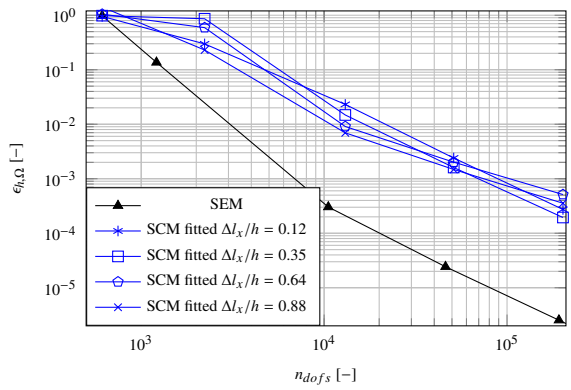
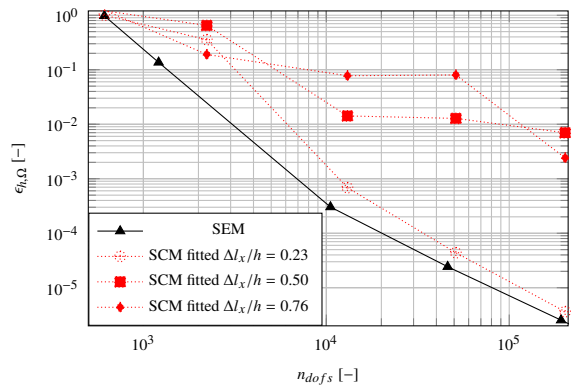
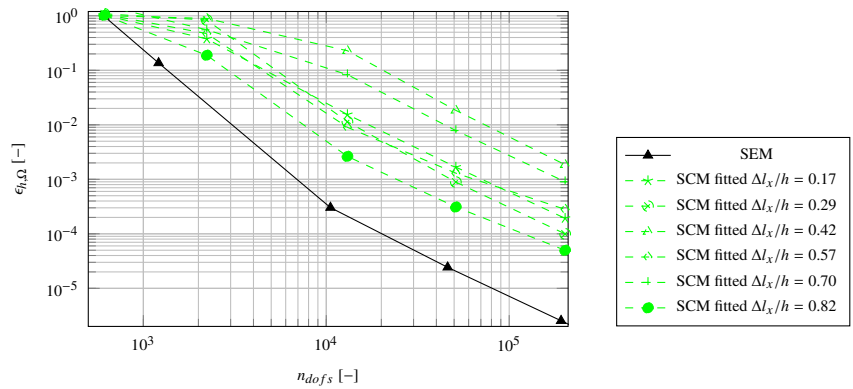
\begin{figure}[]
\centering
\subfigure[Cut aligned with nodes]{
    \input{figures/ex1/by_cut/conv_align.tikz}
}
\subfigure[Cut bisecting distance between nodes]{
    \input{figures/ex1/by_cut/conv_half.tikz}
}
\subfigure[Cut at quarter distance between nodes]{
    \input{figures/ex1/by_cut/conv_quarter.tikz}
}
\caption{L2 error norm for a mesh of elements $\mathcal{N}_{5,5}$ cut at different ratios $\Delta l_x / h$. Cf \autoref{fig:unit_square_cut}. When the cut matches a column of nodes, consistent convergence is observed. When the boundary bisects the node distance, inconsistent performance and elevated errors occur. At the quarter points good convergence can be achieved, although at more variable levels of accuracy.}
\label{fig:ex1_sp5_conv}
\end{figure}
%%%%%%%%%%%%%%%%%%%%%%%%%%%%%%%%%%%%%%%%%%%%%%%%%%%%%%
where $\lVert \cdot \rVert$ is the Euclidean norm, $\dot{u}_{h,e}(t)$ is the computed element velocity field, and $\dot{u}_{ref,h}$ is evaluated according to \autoref{eq:velSolution}. Initially, the ratio $\Delta l_x / h$ is varied similarly to the studies of \autoref{fig:element_critical_dt} and the proposed method (``SCM fitted'') is applied to elements of a fixed polynomial degree ($\mathcal{N}_{5,5}$) for a set of increasingly refined meshes. The $L2$ convergence for each cut configuration can be seen in \autoref{fig:ex1_sp5_conv}. In \autoref{fig:ex1_sp5_conv}(a), $\Gamma_{cs}$ is aligned with a column of element nodes, in \autoref{fig:ex1_sp5_conv}(b), the interface bisects the distance between nodes,  in \autoref{fig:ex1_sp5_conv}(b) $\Gamma_{cs}$ lies at $1/4$ or $3/4$ of the distance between nodes. The convergence of the SEM with a conformal mesh is also shown, which, for this example, represents an upper bound for the accuracy of the SCM. We can observe that the performance of the proposed method strongly varies when the interface bisects the node spacing, while a more consistent behavior is observed as the interface moves near the nodes. That said, with exception of a few outliers ($\Delta_x / h \in \{0.5, 0.76\}$) these results can be considered satisfactory.

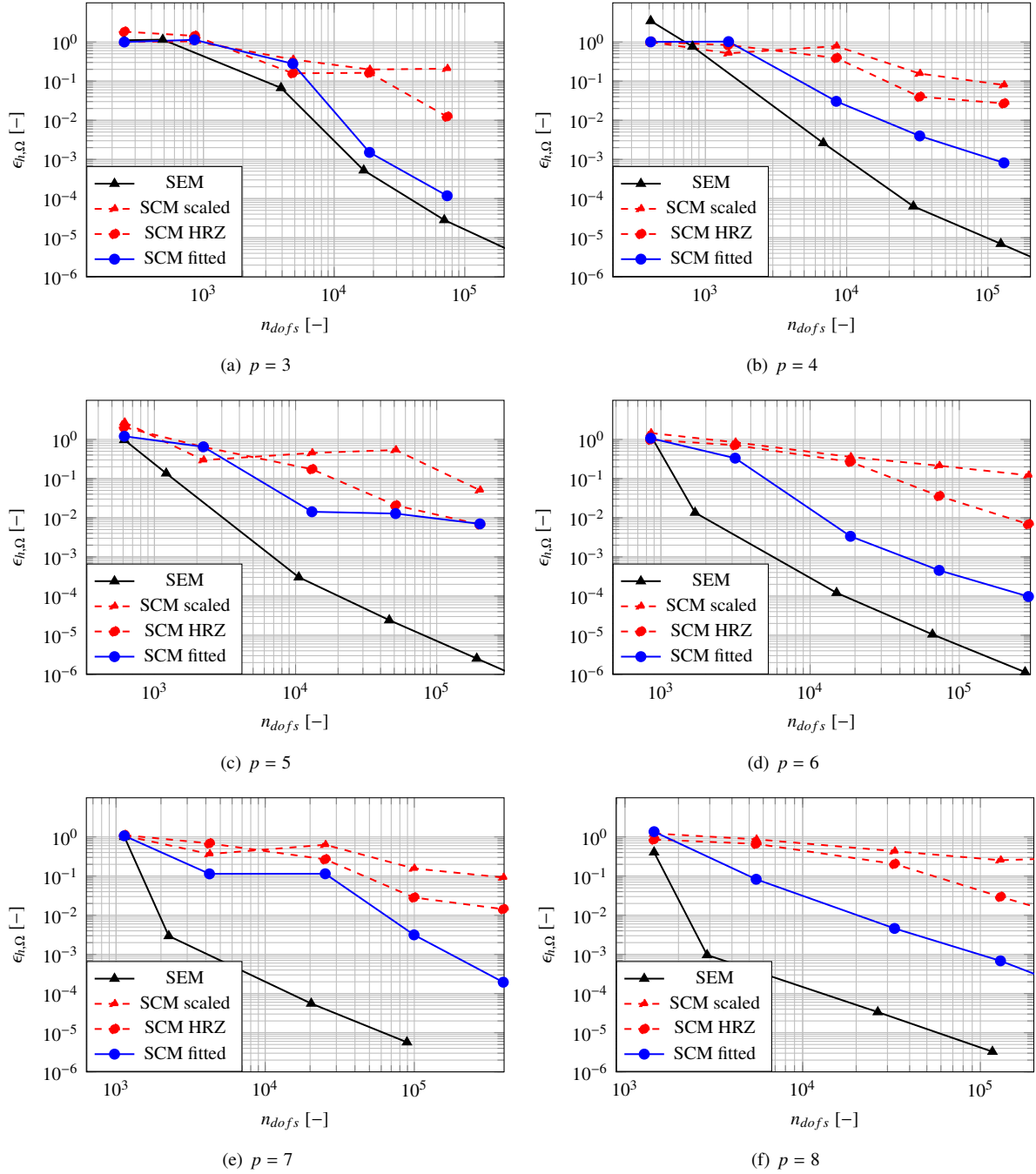
\begin{figure}[]
%\begin{tabular}{C{.5\textwidth}C{.5\textwidth}}
\subfigure [$p = 3$] {
    \input{figures/ex1/f50/sp3_conv.tikz}
}
\subfigure [$p = 4$] {
    \input{figures/ex1/f50/sp4_conv.tikz}
}\\
\subfigure [$p = 5$] {
    \input{figures/ex1/f50/sp5_conv.tikz}
}
\subfigure [$p = 6$] {
    \input{figures/ex1/f50/sp6_conv.tikz}
}\\
\subfigure [$p = 7$] {
    \input{figures/ex1/f50/sp7_conv.tikz}
}
\subfigure [$p = 8$] {
    \input{figures/ex1/f50/sp8_conv.tikz}
}\\
%\end{tabular}
\caption{Error norm convergence of the velocity profile with $h$ refinement of SE meshes with different polynomial degrees $t = 0.4 [s]$ for $\Delta l_x / h = 0.5 [-]$}
\label{fig:ex1_poly}
\end{figure}

Next, we study the effect of varying polynomial degree on the novel procedure as well as on the lumping techniques ``1'' and ``2'' by Joulaian et al. \cite{joulaian2014finite}, which we label ``SCM scaled'' and ``SCM HRZ'', respectively. For all instances of the SCM, the physical domain is modeled according to \autoref{sec:LSM}, so that only the effect of lumping might emerge. %With that in mind, a variationally consistent solution of the SCM is computed with the implicit version of the Newmark algorithm \cite{newmark1959method}.
The benchmark is repeated for a cut configuration with $\Delta_x / h = 0.5$ and elements $\mathcal{N}_{p,p}, p \in \{3,4,...,8 \}$. The results are reported in \autoref{fig:ex1_poly}. While the HRZ method delivers an improvement with respect to the scaled version by including an integration rule based on the cut configuration, the fitting procedure exploits the same principle but can further enhance accuracy as it aims at minimizing lumping error by design. Compared to the SEM, lumping procedures seem to incur a higher loss of accuracy as the polynomial order increases. This can be attributed to the fact that cut elements of high order encompass more nodes and, thus, lumping error affects a larger portion of the domain. At lower orders, this effect is restricted due to comparatively smaller elements with respect to the model's number of DOFs. For elements of odd order, the interface bisects the distance between nodes (see, e.g. \autoref{fig:element_critical_dt}) representing one of the worst case scenarios highlighted in \autoref{fig:ex1_sp5_conv}. However, for even orders, and conceded a loss of accuracy, the novel method shows convergence rates comparable to the conformal SEM.

%f50

%% file: figures/ex1/by_cut/conv_align.tikz
\begin{tikzpicture}
    \begin{axis}[
        xmin = 500,
        xmax = 2.1e5,
        ymin = 2e-6,
        ymax = 1.2,
        xtick = {10^3,10^4, 10^5},
        xticklabels = {$10^3$,$10^4$,$10^5$,},
        ytick = { 1e-5, 1e-4, 1e-3, 1e-2, 1e-1, 1},
        xmode = log,
        ymode = log,
        xlabel = {$n_{dofs} \ [-]$},
        ylabel = {$\epsilon_{h,\Omega} \ [-]$},
        xticklabel style = {font =\fontsize{\bigFigureFontSize pt}{10pt}\selectfont},
        yticklabel style = {font=\fontsize{\bigFigureFontSize pt}{10pt}\selectfont},
        xlabel style = {font =\fontsize{\bigFigureFontSize pt}{\bigFigureFontSize pt}\selectfont},
        ylabel style = {font=\fontsize{\bigFigureFontSize pt}{\bigFigureFontSize pt}\selectfont},
        grid = both,
        width= 0.48 \textwidth,
        height= 0.35 \textwidth,
        legend style={at={(0,0)},anchor=south west,nodes={font=\fontsize{\bigFigureFontSize pt}{\bigFigureFontSize pt}\selectfont}}
        ]
        \addplot [ color = \cutConsistentColor, style = solid, mark =triangle*] table [x=numDofs, y=L2ErrorNorm, col sep=semicolon]
         {figures/ex1/by_cut/sp5_conv/planeBarUncut_sp_5_te-6Conv.csv};
        \addlegendentry{SEM}
        \addplot [color = blue, style = solid, mark = asterisk ] table [x=numDofs, y=L2ErrorNorm, col sep=semicolon] {figures/ex1/by_cut/sp5_conv/planeBeamCutOpt_sp5_f11_te-5Conv.csv};
        \addlegendentry{SCM fitted $\Delta l_x / h = 0.12$}
        
        \addplot [color = blue, style = solid, mark = square] table [x=numDofs, y=L2ErrorNorm, col sep=semicolon] {figures/ex1/by_cut/sp5_conv/planeBeamCutOpt_sp5_f35_te-5Conv.csv};
        \addlegendentry{SCM fitted $\Delta l_x / h = 0.35$}
        
        \addplot [color = blue, style = solid, mark = pentagon] table [x=numDofs, y=L2ErrorNorm, col sep=semicolon] {figures/ex1/by_cut/sp5_conv/planeBeamCutOpt_sp5_f64_te-5Conv.csv};
        \addlegendentry{SCM fitted $\Delta l_x / h = 0.64$}
        \addplot [color = blue, style = solid, mark = x] table [x=numDofs, y=L2ErrorNorm, col sep=semicolon] {figures/ex1/by_cut/sp5_conv/planeBeamCutOpt_sp5_f88_te-5Conv.csv};
        \addlegendentry{SCM fitted $\Delta l_x / h = 0.88$}
    
    \end{axis}
\end{tikzpicture}

%% file: figures/ex1/by_cut/conv_half.tikz
\begin{tikzpicture}
    \begin{axis}[
        xmin = 500,
        xmax = 2.1e5,
        ymin = 2e-6,
        ymax = 1.2,
        xtick = {10^3,10^4, 10^5},
        xticklabels = {$10^3$,$10^4$,$10^5$,},
        ytick = { 1e-5, 1e-4, 1e-3, 1e-2, 1e-1, 1},
        xmode = log,
        ymode = log,
        xlabel = {$n_{dofs} \ [-]$},
        ylabel = {$\epsilon_{h,\Omega} \ [-]$},
        xticklabel style = {font =\fontsize{\bigFigureFontSize pt}{10pt}\selectfont},
        yticklabel style = {font=\fontsize{\bigFigureFontSize pt}{10pt}\selectfont},
        xlabel style = {font =\fontsize{\bigFigureFontSize pt}{\bigFigureFontSize pt}\selectfont},
        ylabel style = {font=\fontsize{\bigFigureFontSize pt}{\bigFigureFontSize pt}\selectfont},
        grid = both,
        width= 0.48 \textwidth,
        height= 0.35 \textwidth,
        legend style={at={(0,0)},anchor=south west,nodes={font=\fontsize{\bigFigureFontSize pt}{\bigFigureFontSize pt}\selectfont}}
        ]
        \addplot [ color = \cutConsistentColor, style = solid, mark =triangle*] table [x=numDofs, y=L2ErrorNorm, col sep=semicolon]
         {figures/ex1/by_cut/sp5_conv/planeBarUncut_sp_5_te-6Conv.csv};
        \addlegendentry{SEM}
        \addplot [ color = red, style = densely dotted, mark = oplus] table [x=numDofs, y=L2ErrorNorm, col sep=semicolon] {figures/ex1/by_cut/sp5_conv/planeBeamCutOpt_sp5_f23_te-5Conv.csv};
        \addlegendentry{SCM fitted $\Delta l_x / h = 0.23$}
        \addplot [color = red, mark = square*, style = densely dotted] table [x=numDofs, y=L2ErrorNorm, col sep=semicolon] {figures/ex1/by_cut/sp5_conv/spec_p5_f50_te-6_Conv.csv};
        \addlegendentry{SCM fitted $\Delta l_x / h = 0.50$}
        \addplot [color = green, style = dashed, color = red, style = densely dotted, mark = diamond*] table [x=numDofs, y=L2ErrorNorm, col sep=semicolon] {figures/ex1/by_cut/sp5_conv/planeBeamCutOpt_sp5_f76_te-5Conv.csv};
        \addlegendentry{SCM fitted $\Delta l_x / h = 0.76$}
    \end{axis}
\end{tikzpicture}

%% file: figures/ex1/by_cut/conv_quarter.tikz
\begin{tikzpicture}
    \begin{axis}[
        xmin = 500,
        xmax = 2.1e5,
        ymin = 2e-6,
        ymax = 1.2,
        xtick = {10^3,10^4, 10^5},
        xticklabels = {$10^3$,$10^4$,$10^5$,},
        ytick = { 1e-5, 1e-4, 1e-3, 1e-2, 1e-1, 1},
        xmode = log,
        ymode = log,
        xlabel = {$n_{dofs} \ [-]$},
        ylabel = {$\epsilon_{h,\Omega} \ [-]$},
        xticklabel style = {font =\fontsize{\bigFigureFontSize pt}{10pt}\selectfont},
        yticklabel style = {font=\fontsize{\bigFigureFontSize pt}{10pt}\selectfont},
        xlabel style = {font =\fontsize{\bigFigureFontSize pt}{\bigFigureFontSize pt}\selectfont},
        ylabel style = {font=\fontsize{\bigFigureFontSize pt}{\bigFigureFontSize pt}\selectfont},
        grid = both,
        width= 0.48 \textwidth,
        height= 0.35 \textwidth,
        legend style={at={(1.1,0)},anchor=south west,nodes={font=\fontsize{\bigFigureFontSize pt}{\bigFigureFontSize pt}\selectfont}}
        ]
        \addplot [ color = \cutConsistentColor, style = solid, mark =triangle*] table [x=numDofs, y=L2ErrorNorm, col sep=semicolon]
         {figures/ex1/by_cut/sp5_conv/planeBarUncut_sp_5_te-6Conv.csv};
        \addlegendentry{SEM}
        \addplot [color = green, style = dashed, mark = star ] table [x=numDofs, y=L2ErrorNorm, col sep=semicolon] {figures/ex1/by_cut/sp5_conv/planeBeamCutOpt_sp5_f17_te-5Conv.csv};
        \addlegendentry{SCM fitted $\Delta l_x / h = 0.17$}
        \addplot [color = green, style = dashed, mark = otimes] table [x=numDofs, y=L2ErrorNorm, col sep=semicolon] {figures/ex1/by_cut/sp5_conv/planeBeamCutOpt_sp5_f29_te-5Conv.csv};
        \addlegendentry{SCM fitted $\Delta l_x / h = 0.29$}
        \addplot [color = green, style = dashed, mark = triangle] table [x=numDofs, y=L2ErrorNorm, col sep=semicolon] {figures/ex1/by_cut/sp5_conv/planeBeamCutOpt_sp5_f42_te-5Conv.csv};
        \addlegendentry{SCM fitted $\Delta l_x / h = 0.42$}
        \addplot [color = green, style = dashed, mark = diamond] table [x=numDofs, y=L2ErrorNorm, col sep=semicolon] {figures/ex1/by_cut/sp5_conv/planeBeamCutOpt_sp5_f57_te-5Conv.csv};
        \addlegendentry{SCM fitted $\Delta l_x / h = 0.57$}
        \addplot [color = green, style = dashed, mark = +] table [x=numDofs, y=L2ErrorNorm, col sep=semicolon] {figures/ex1/by_cut/sp5_conv/planeBeamCutOpt_sp5_f70_te-5Conv.csv};
        \addlegendentry{SCM fitted $\Delta l_x / h = 0.70$}
        \addplot [color = green, style = dashed, mark = *] table [x=numDofs, y=L2ErrorNorm, col sep=semicolon] {figures/ex1/by_cut/sp5_conv/planeBeamCutOpt_sp5_f82_te-5Conv.csv};
        \addlegendentry{SCM fitted $\Delta l_x / h = 0.82$}
    \end{axis}
\end{tikzpicture}

%% file: figures/ex1/f50/sp3_conv.tikz
\begin{tikzpicture}
    \begin{axis}[
        xmin = 0,
        xmax = 2e5,
        ymin = 1e-6,
        ymax = 10,
        xtick = {10^2,10^3,10^4,10^5},
        xticklabels = {$10^2$,$10^3$,$10^4$,$10^5$},
        ytick = {1e-6, 1e-5, 1e-4, 1e-3, 1e-2, 1e-1, 1},
        xmode = log,
        ymode = log,
        xlabel = {$n_{dofs} \ [-]$},
        ylabel = {$\epsilon_{h,\Omega} \ [-]$},
        xticklabel style = {font =\fontsize{\figureFontSize pt}{10pt}\selectfont},
        yticklabel style = {font=\fontsize{\figureFontSize pt}{10pt}\selectfont},
        xlabel style = {font =\fontsize{\figureFontSize pt}{\figureFontSize pt}\selectfont},
        ylabel style = {font=\fontsize{\figureFontSize pt}{\figureFontSize pt}\selectfont},
        grid = both,
        width=0.48\textwidth,
        height=0.35\textwidth,
        legend style={at={(0,0)},anchor=south west,nodes={font=\fontsize{\figureFontSize pt}{\figureFontSize pt}\selectfont}}
%        legend pos = south east
        ]
        
    \addplot[color = \uncutColor, mark =\conformMarker*, line width=0.75pt] coordinates{
	(248,1.1076156791894507)
	(488,1.1475889819362621)
	(3926,0.06705049948128924)
	(16856,0.000527281790901969)
	(69716,2.8378806006152156e-05)
	(283436, 3.2436511389906503e-06)
	};
    \addlegendentry{SEM}

	\addplot[style = dashed, color = \cutHRZColor, mark = triangle*, line width=\plotLineWidth pt] coordinates{
    %convergence analysis for: "cutScaled"
    (248, 1.0005743648485677)
    (854, 1.0354951154337875)
    (4832, 0.3587122252294364)
    (18662, 0.19835480886563453)
    (73322, 0.20804258614452123)
	};
	\addlegendentry{SCM scaled}
	
	\addplot[style = dashed, color = \cutHRZColor, mark = \explicitMarker*, line width=\plotLineWidth pt] coordinates{
    %convergence analysis for: "cutHRZ"
    (248, 1.8336643028166593)
    (854, 1.4342673401568167)
    (4832, 0.1584404247145158)
    (18662, 0.16293923009684827)
    (73322, 0.012355958923290803)
	};
	\addlegendentry{SCM HRZ}
	
	\addplot[color = \cutOptimizedColor, mark = \explicitMarker*, line width=\plotLineWidth pt] coordinates{
    %convergence analysis for: "cutOptimized_me0.010000"
    (248, 0.9981204833553057)
    (854, 1.1348745675322918)
    (4832, 0.2782104752034135)
    (18662, 0.0015118414408090523)
    (73322, 0.00011784450502416515)
	};
   	\addlegendentry{SCM fitted}
   
    \end{axis}
\end{tikzpicture}

%% file: figures/ex1/f50/sp4_conv.tikz
\begin{tikzpicture}
    \begin{axis}[
        xmin = 0,
        xmax = 2e5,
        ymin = 1e-6,
        ymax = 10,
        xtick = {10^2,10^3,10^4,10^5},
        xticklabels = {$10^2$,$10^3$,$10^4$,$10^5$},
        ytick = {1e-6, 1e-5, 1e-4, 1e-3, 1e-2, 1e-1, 1},
        xmode = log,
        ymode = log,
        xlabel = {$n_{dofs} \ [-]$},
        ylabel = {$\epsilon_{h,\Omega} \ [-]$},
        xticklabel style = {font =\fontsize{\figureFontSize pt}{10pt}\selectfont},
        yticklabel style = {font=\fontsize{\figureFontSize pt}{10pt}\selectfont},
        xlabel style = {font =\fontsize{\figureFontSize pt}{\figureFontSize pt}\selectfont},
        ylabel style = {font=\fontsize{\figureFontSize pt}{\figureFontSize pt}\selectfont},
        grid = both,
        width=0.48\textwidth,
        height=0.35\textwidth,
        legend style={at={(0,0)},anchor=south west,nodes={font=\fontsize{\figureFontSize pt}{\figureFontSize pt}\selectfont}}
%        legend pos = south east
        ]
    
    \addplot[color = \uncutColor, mark =\conformMarker*, line width=0.75pt] coordinates{
		(410,3.44358)
		(810,0.764351)
		(6834,0.00262622)
		(29674,6.34249e-05)
		(123354,7.00582e-06)
		(502714,7.75912e-07)
	};
    \addlegendentry{SEM}
	\addplot[style = dashed, color = \cutHRZColor, mark = triangle*, line width=\plotLineWidth pt] coordinates{
    %convergence analysis for: "cutScaled"
        (410, 1.0011255413961928)
        (1458, 0.5140814051491609)
        (8442, 0.7811815954906599)
        (32882, 0.15571292999714784)
        (129762, 0.07947275486338826)
	};
	\addlegendentry{SCM scaled}
	
    %\addplot[color = \cutConsistentColor, mark =\implicitMarker*, line width=\plotLineWidth pt] coordinates{
	%	(410,1.65282)
	%	(1458,0.360305)
	%	(8442,0.00205)
	%	(32882,0.000140239)
	%	(129762,0.000126647)
	%};
	
	\addplot[style = dashed, color = \cutHRZColor, mark = \explicitMarker*, line width=\plotLineWidth pt] coordinates{
	    (410, 1.00623)
		(1458,0.836194)
		(8442,0.390052)
		(32882,0.039746)
		(129762,0.0269851)
	};
	\addlegendentry{SCM HRZ}
	\addplot[color = \cutOptimizedColor, mark = \explicitMarker*, line width=\plotLineWidth pt] coordinates{
        (410, 1.00736)
		(1458,1.02163)
		(8442,0.0304809)
		(32882,0.00398114)
		(129762,0.000817417)
	};

   	%\addlegendentry{SCM consistent}
   	
   	\addlegendentry{SCM fitted}
   
    \end{axis}
\end{tikzpicture}

%% file: figures/ex1/f50/sp5_conv.tikz
\begin{tikzpicture}
    \begin{axis}[
               xmin = 0,
        xmax = 3e5,
        ymin = 1e-6,
        ymax = 10,
        xtick = {10^2,10^3,10^4,10^5},
        xticklabels = {$10^2$,$10^3$,$10^4$,$10^5$},
        ytick = {1e-6, 1e-5, 1e-4, 1e-3, 1e-2, 1e-1, 1},
        xmode = log,
        ymode = log,
        xlabel = {$n_{dofs} \ [-]$},
        ylabel = {$\epsilon_{h,\Omega} \ [-]$},
        xticklabel style = {font =\fontsize{\figureFontSize pt}{10pt}\selectfont},
        yticklabel style = {font=\fontsize{\figureFontSize pt}{10pt}\selectfont},
        xlabel style = {font =\fontsize{\figureFontSize pt}{\figureFontSize pt}\selectfont},
        ylabel style = {font=\fontsize{\figureFontSize pt}{\figureFontSize pt}\selectfont},
        grid = both,
        width=0.48\textwidth,
        height=0.35\textwidth,
        legend style={at={(0,0)},anchor=south west,nodes={font=\fontsize{\figureFontSize pt}{\figureFontSize pt}\selectfont}}
%        legend pos = south east
        ]
        
    \addplot[color = \uncutColor, mark =\conformMarker*, line width=0.75pt] coordinates{
		(612,0.974708)
		(1212,0.136337)
		(10542,0.000302776)
		(46092,2.42903e-05)
		(192192,2.54081e-06)
		(784392,2.76677e-07)
	};
    \addlegendentry{SEM}
    
   % \addplot[color = \cutConsistentColor, mark =\implicitMarker*, line width=\plotLineWidth pt] coordinates{
	%	(612,2.53552)
	%	(2222,0.131336)
	%	(13052,0.000314897)
	%	(51102,0.000129028)
	%};
	%\addlegendentry{SCM consistent}
	
	\addplot[style = dashed, color = \cutHRZColor, mark = triangle*, line width=\plotLineWidth pt] coordinates{
    %convergence analysis for: "cutScaled"
    (612, 2.7482193383510074)
    (2222, 0.2967073514688995)
    (13052, 0.4515021831348223)
    (51102, 0.5386880995053476)
    (202202, 0.0504728293867545)
	};
	\addlegendentry{SCM scaled}
	
	\addplot[style = dashed, color = \cutHRZColor, mark = \explicitMarker*, line width=\plotLineWidth pt] coordinates{
	    %(612, 1.8484)
		%(2222,0.605867)
		%(13052,0.164164)
		%(51102,0.0239428)
		(612, 2.053250782096266)
        (2222, 0.6430434853777173)
        (13052, 0.17374520706627863)
        (51102, 0.020745682952563387)
        (202202, 0.006741370057243215)
	};
	\addlegendentry{SCM HRZ}
	
	\addplot[color = \cutOptimizedColor, mark = \explicitMarker*, line width=\plotLineWidth pt] coordinates{
        (612, 1.2068588241321787)
        (2222, 0.6482956645655112)
        (13052, 0.014218005618414617)
        (51102, 0.012796789928993767)
        (202202, 0.006975562342754442)
	};
    \addlegendentry{SCM fitted}

    \end{axis}
\end{tikzpicture}

%% file: figures/ex1/f50/sp6_conv.tikz
\begin{tikzpicture}
    \begin{axis}[
        xmin = 0,
        xmax = 3e5,
        ymin = 1e-6,
        ymax = 10,
        xtick = {10^2,10^3,10^4,10^5},
        xticklabels = {$10^2$,$10^3$,$10^4$,$10^5$},
        ytick = {1e-6, 1e-5, 1e-4, 1e-3, 1e-2, 1e-1, 1},
        xmode = log,
        ymode = log,
        xlabel = {$n_{dofs} \ [-]$},
        ylabel = {$\epsilon_{h,\Omega} \ [-]$},
        xticklabel style = {font =\fontsize{\figureFontSize pt}{10pt}\selectfont},
        yticklabel style = {font=\fontsize{\figureFontSize pt}{10pt}\selectfont},
        xlabel style = {font =\fontsize{\figureFontSize pt}{\figureFontSize pt}\selectfont},
        ylabel style = {font=\fontsize{\figureFontSize pt}{\figureFontSize pt}\selectfont},
        grid = both,
        width=0.48\textwidth,
        height=0.35\textwidth,
        legend style={at={(0,0)},anchor=south west,nodes={font=\fontsize{\figureFontSize pt}{\figureFontSize pt}\selectfont}}
%        legend pos = south east
        ]
        
    \addplot[color = \uncutColor, mark =\conformMarker*, line width=0.75pt] coordinates{
		(854,1.26761)
		(1694,0.01352)
		(15050,0.000120933)
		(66110,1.04946e-05)
		(276230,1.14069e-06)
	};
    \addlegendentry{SEM}
    
    \addplot[style = dashed, color = \cutHRZColor, mark = triangle*, line width=\plotLineWidth pt] coordinates{
    %convergence analysis for: "cutScaled"
        (854, 1.4594026561382618)
        (3146, 0.8458214612168666)
        (18662, 0.35809999748085364)
        (73322, 0.21327440757641633)
        (290642, 0.12157623008197203)
	};
	\addlegendentry{SCM scaled}
	
    %\addplot[color = \cutConsistentColor, mark =\implicitMarker*, line width=\plotLineWidth pt] coordinates{
	%	(854,1.08085)
	%	(3146,0.0146982)
	%	(18662,0.000167727)
	%	(73322,0.000126972)
	%};
	%\addlegendentry{SCM consistent}
	
	\addplot[style = dashed, color = \cutHRZColor, mark = \explicitMarker*, line width=\plotLineWidth pt] coordinates{
		%(854,0.98799)
		%(3146,0.699413)
		%(18662,0.242274)
		%73322,0.036136)
		(854, 0.9818355285841567)
        (3146, 0.714381798642481)
        (18662, 0.26766903345550414)
        (73322, 0.035816825653299626)
        (290642, 0.0068478573416346795)
	};
	\addlegendentry{SCM HRZ}
	
	\addplot[color = \cutOptimizedColor, mark = \explicitMarker*, line width=\plotLineWidth pt] coordinates{
        (854, 1.0739345400211062)
        (3146, 0.332800990294011)
        (18662, 0.0033569402514106246)
        (73322, 0.00045325653157038897)
        (290642, 9.740009703018861e-05)
	};
   	\addlegendentry{SCM fitted}
   
    \end{axis}
\end{tikzpicture}

%% file: figures/ex1/f50/sp7_conv.tikz
\begin{tikzpicture}
    \begin{axis}[
        xmin = 0,
        xmax = 4e5,
        ymin = 1e-6,
        ymax = 10,
        xtick = {10^2,10^3,10^4,10^5},
        xticklabels = {$10^2$,$10^3$,$10^4$,$10^5$},
        ytick = {1e-6, 1e-5, 1e-4, 1e-3, 1e-2, 1e-1, 1},
        xmode = log,
        ymode = log,
        xlabel = {$n_{dofs} \ [-]$},
        ylabel = {$\epsilon_{h,\Omega} \ [-]$},
        xticklabel style = {font =\fontsize{\figureFontSize pt}{10pt}\selectfont},
        yticklabel style = {font=\fontsize{\figureFontSize pt}{10pt}\selectfont},
        xlabel style = {font =\fontsize{\figureFontSize pt}{\figureFontSize pt}\selectfont},
        ylabel style = {font=\fontsize{\figureFontSize pt}{\figureFontSize pt}\selectfont},
        grid = both,
        width=0.48\textwidth,
        height=0.35\textwidth,
        legend style={at={(0,0)},anchor=south west,nodes={font=\fontsize{\figureFontSize pt}{\figureFontSize pt}\selectfont}}
%        legend pos = south east
        ]
        
    \addplot[color = \uncutColor, mark =\conformMarker*, line width=0.75pt] coordinates{
		(1136,1.01089)
		(2256,0.00295569)
		(20358,5.56129e-05)
		(89278,5.70878e-06)
	};
    \addlegendentry{SEM}
    
    %\addplot[color = \cutConsistentColor, mark =\implicitMarker*, line width=\plotLineWidth pt] coordinates{
	%	(1136,0.777327)
	%	(4230,0.00345942)
	%	(25272,0.000136322)
	%};
	%\addlegendentry{SCM consistent}
	
	 \addplot[style = dashed, color = \cutHRZColor, mark = triangle*, line width=\plotLineWidth pt] coordinates{
    %convergence analysis for: "cutScaled"
    (1136, 1.0787254369035126)
    (4230, 0.36664291831842816)
    (25272, 0.631416704093931)
    (99542, 0.15684123133798228)
    (395082, 0.09200581747224085)
	};
	\addlegendentry{SCM scaled}
	
	\addplot[style = dashed, color = \cutHRZColor, mark = \explicitMarker*, line width=\plotLineWidth pt] coordinates{
    %convergence analysis for: "cutHRZ"
    (1136, 1.1138631772492058)
    (4230, 0.6838068705561672)
    (25272, 0.26800951831110315)
    (99542, 0.02835623484411551)
    (395082, 0.014320320736698151)
	};
	\addlegendentry{SCM HRZ}
	
	\addplot[color = \cutOptimizedColor, mark = \explicitMarker*, line width=\plotLineWidth pt] coordinates{
    %convergence analysis for: "cutOptimized_me0.010000"
    (1136, 1.0687571017826818)
    (4230, 0.11407475396768414)
    (25272, 0.11464841348425242)
    (99542, 0.003154228511549808)
    (395082, 0.0001949211003812488)
	};
   	\addlegendentry{SCM fitted}
   
    \end{axis}
\end{tikzpicture}

%% file: figures/ex1/f50/sp8_conv.tikz
\begin{tikzpicture}
    \begin{axis}[
        xmin = 0,
        xmax = 2e5,
        ymin = 1e-6,
        ymax = 10,
        xtick = {10^2,10^3,10^4,10^5},
        xticklabels = {$10^2$,$10^3$,$10^4$,$10^5$},
        ytick = {1e-6, 1e-5, 1e-4, 1e-3, 1e-2, 1e-1, 1},
        xmode = log,
        ymode = log,
        xlabel = {$n_{dofs} \ [-]$},
        ylabel = {$\epsilon_{h,\Omega} \ [-]$},
        xticklabel style = {font =\fontsize{\figureFontSize pt}{10pt}\selectfont},
        yticklabel style = {font=\fontsize{\figureFontSize pt}{10pt}\selectfont},
        xlabel style = {font =\fontsize{\figureFontSize pt}{\figureFontSize pt}\selectfont},
        ylabel style = {font=\fontsize{\figureFontSize pt}{\figureFontSize pt}\selectfont},
        grid = both,
        width=0.48\textwidth,
        height=0.35\textwidth,
        legend style={at={(0,0)},anchor=south west,nodes={font=\fontsize{\figureFontSize pt}{\figureFontSize pt}\selectfont}}
%        legend pos = south east
        ]
        
    \addplot[color = \uncutColor, mark =\conformMarker*, line width=0.75pt] coordinates{
		(1458,0.406067)
		(2898,0.00096645)
		(26466,3.36553e-05)
		(116946,3.27902e-06)
	};
    \addlegendentry{SEM}
    
    %\addplot[color = \cutConsistentColor, mark =\implicitMarker*, line width=\plotLineWidth pt] coordinates{
	%	(1458,0.660699)
	%	(5474,0.00132163)
	%	(32882,0.000132154)
	%};
	%\addlegendentry{SCM consistent}
	
	\addplot[style = dashed, color = \cutHRZColor, mark = triangle*, line width=\plotLineWidth pt] coordinates{
    %convergence analysis for: "cutScaled"
    (1458, 1.226871056882062)
    (5474, 0.8701715113050873)
    (32882, 0.43201545296794497)
    (129762, 0.2544431489615419)
    (515522, 0.31021325091413043)
	};
	\addlegendentry{SCM scaled}
	
	\addplot[style = dashed, color = \cutHRZColor, mark = \explicitMarker*, line width=\plotLineWidth pt] coordinates{
    %convergence analysis for: "cutHRZ"
    (1458, 0.8573607318954668)
    (5474, 0.6668915021795192)
    (32882, 0.20415031329240255)
    (129762, 0.029541261804847146)
    (515522, 0.004831049760225338)
	};
	\addlegendentry{SCM HRZ}
	
	\addplot[color = \cutOptimizedColor, mark = \explicitMarker*, line width=\plotLineWidth pt] coordinates{
    %convergence analysis for: "cutOptimized_me0.010000"
    (1458, 1.362373249603172)
    (5474, 0.08294250267559485)
    (32882, 0.00460826065224635)
    (129762, 0.0006842063723988055)
    (515522, 6.164761084241494e-05)
	};
   	\addlegendentry{SCM fitted}
   
    \end{axis}
\end{tikzpicture}

%% file: 4_2_willberg.tex
\newpage
\subsection{Aluminium plate with conic hole}\label{sec:ex2}
\begin{figure}[]
    \centering
    \includegraphics[width=1\linewidth]{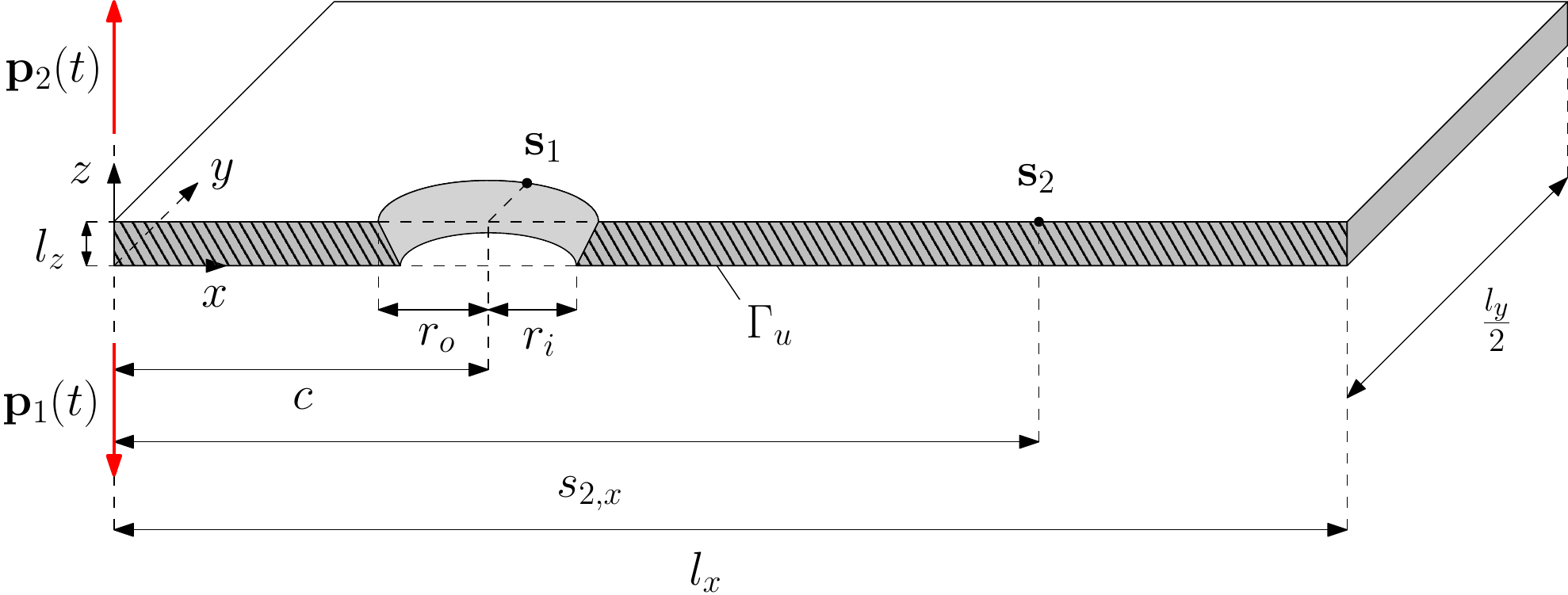}
    \caption{Aluminium plate with dimensions $ l_x = 300 \ [mm]$, $ l_y = 200 \ [mm]$, $l_z = 2 \ [mm]$ and $\mathbf{s}_{2,x} = 200 \ [mm]$. Conic hole at $c = 50 \ [mm], y = 0$ with radiuses $ri = 9 \ [mm], \ r_o = 10 \ [mm]$. Mirrored model with symmetric BC on $\Gamma_u$ and loading $\mathbf{p}_1(t), \mathbf{p}_2(t)$.}
    \label{fig:plateWillberg}
\end{figure}

The $l_z = 2 \ [mm]$ thick, $ l_x = 300 \ [mm]$ long, and $ l_y = 200 \ [mm]$ wide aluminium plate depicted in figure \ref{fig:plateWillberg} features a conic hole with an inner radius $r_i = 9 \ [mm] $ and an outer radius $ r_o = 10 \ [mm] $. A mono-modal excitation of the symmetric mode is provided by the point forces $\mathbf{p}_1(t),\ \mathbf{p}_2(t)$, which act at the bottom and top surfaces of the plate with opposite orientations:
\begin{equation}
\begin{split}
     \mathbf{p}_1(t)& = -\mathbf{e}_z \ p(t) \\
    \mathbf{p}_2(t)& =  \mathbf{e}_z \ p(t)
\end{split}
\label{eq:will_load}
\end{equation}
Where $\mathbf{e}_z$ is the third basis vector of the standard Cartesian system and the load modulation $  p(t) $ is a Hann window (\autoref{eq:hanning}) with an amplitude of $p = 10^7 \ [N]$, a frequency of $f = 175 \ [kHz]$ and $n = 3$ cycles within one pulse, resulting in an excitation window of $n / f \approx 1.71 \cdot 10^{-5} \ [s]$. In this setup, introduced by Willberg et al. \cite{willberg2012comparison}, the excited symmetric mode is converted into an asymmetric mode due to its reflections against the inclined boundary of the conical hole, leading to a non-uniform displacement field along the cross section.

To benchmark the proposed method, the vertical component of the displacement $u_z(t)$ is recorded at the sensors $\mathbf{s}_1 (c, r_o, l_z)$ and $\mathbf{s}_2 (s_{2,x}, 0, l_z)$, with $c = 50 \ [mm],\ s_{2,x} = 200 \ [mm]$ (see \autoref{fig:plateWillberg}). In particular, the time history at $\mathbf{s}_1$  is of great interest, as the successful modeling of mode conversion crucially relies on the discretization of the nearby boundary. As shown in \autoref{fig:plateWillberg}, the computational cost of the analysis can be reduced by mirroring the model along the xz-plane, and applying symmetric boundary conditions (i.e. $\mathbf{u}_y(\globalVec, t) \equiv 0 \ \forall \ \globalVec \in \Gamma_u$). For clarity, we should note that the aforementioned loading is applied on this mirrored model and, therefore, represents only one half of the loading applied on the full plate.

\begin{figure}[b!]
\begin{tabular}{C{.53\textwidth} C{.015\textwidth} C{.40\textwidth}}
\subfigure [Conformal 3D mesh for the SEM with GLL nodes and shape functions of type $\mathcal{N}_{3,3,4}$ (see \autoref{eq:2d3dshapeFunc}).] {
    \resizebox{0.53\textwidth}{!}{%
    \includegraphics{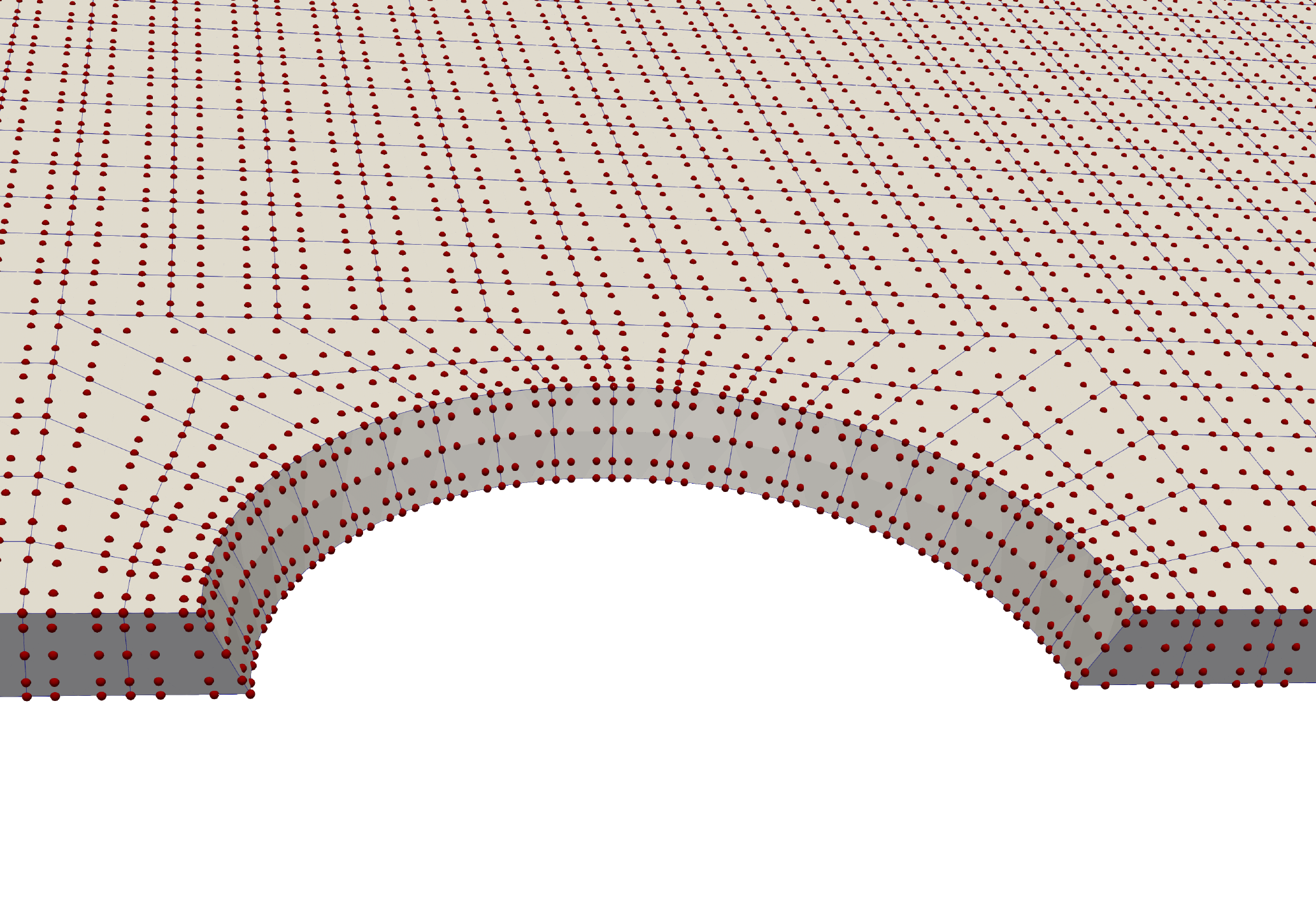}
    }
} & 
  &
\subfigure [Above: Structured GLL-SE grid (nodes are omitted for clarity) independent form the hole. Below: close-up of the local octree mesh and the element partitions. ] {
    \resizebox{0.4\textwidth}{!}{%
    \includegraphics{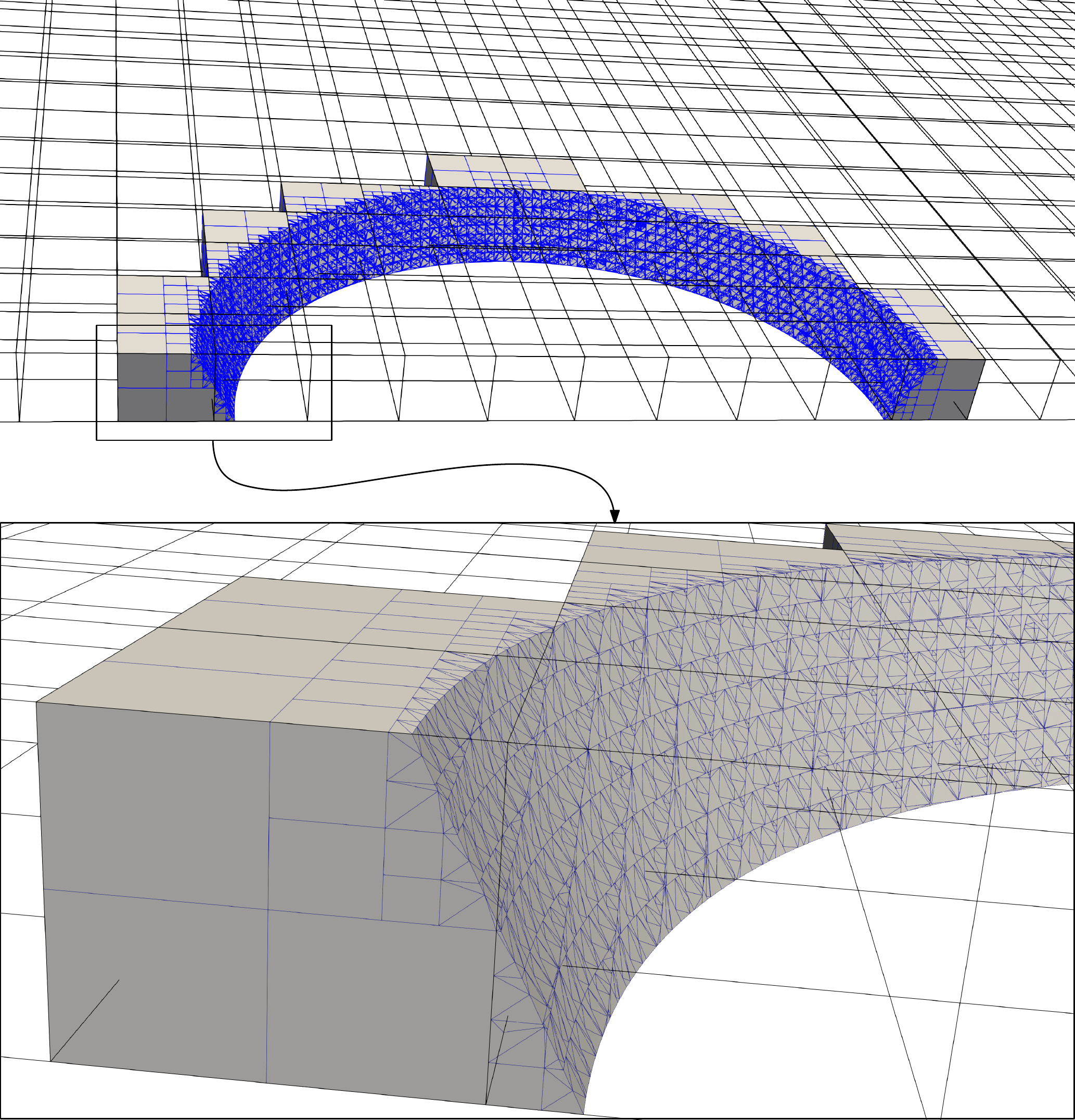}
    }
}
\end{tabular}
\caption{Discretizaion of the conic hole for the example by Willberg \cite{willberg2012comparison}. (a) With the SEM, a conformal mesh customized for the example at hand must be produced, while ensuring good quality of the resulting SE. (b) With the SCM, a structured mesh can be used, the hole being represented implicitly. Elements not intersected by the boundary are standard SE, while elements fully in the void are discarded from the assembly. Cut elements are treated with the proposed version of the SCM (\autoref{sec:LSM} and \autoref{sec:momentFitting}).
}
\label{fig:ex2_meshes}
\end{figure}

A fourth order approximation is chosen in the vertical direction to accurately represent mode conversion. In the x and y directions we use a polynomial order of $p = q = 3$, which has been shown to provide the most efficient use of memory storage with respect to the size of the stiffness matrix \cite{willberg2012comparison}. The problem is solved with the proposed variant of the SCM as well as with the SEM using a conforming mesh. In \autoref{fig:ex2_meshes}, the differences between these two spatial discretization strategies are highlighted. The conformal mesh shown in \autoref{fig:ex2_meshes}a is produced by generating a cylindrical hole via extrusion of a 2D mesh obtained with Gmsh's \emph{transfinite} method \cite{geuzaine2009gmsh}, and performing mesh morphing according to linear elasticity to render the hole conical. On the other hand, with the SCM (\autoref{fig:ex2_meshes}b) a structured mesh can be employed, as the hole is defined by an implicit function.

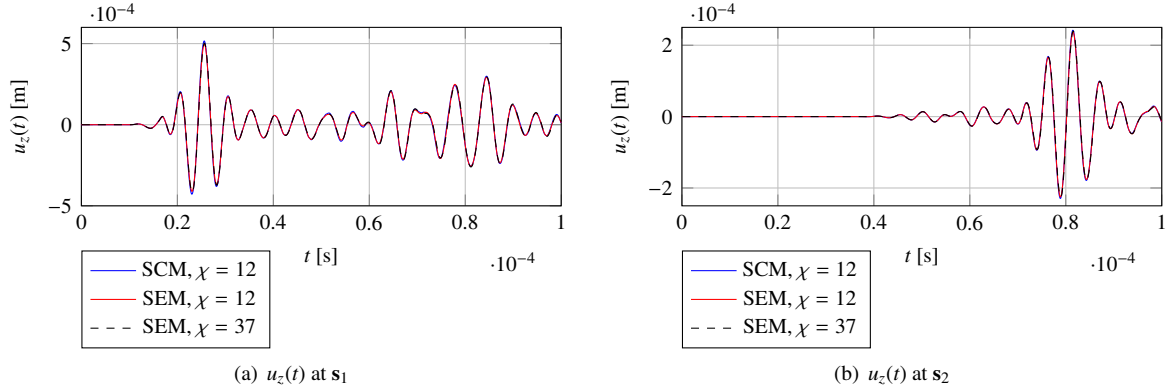
\begin{figure}[]
%\begin{tabular}{C{.49\textwidth} C{.02\textwidth} C{.49\textwidth}}
\subfigure [$u_{z}(t)$ at $\mathbf{s}_1$] {
    \input{figures/ex2/s1_displ.tikz}
} 
\subfigure [$u_{z}(t)$ at $\mathbf{s}_2$] {
    \input{figures/ex2/S2_displ.tikz}
}
%\end{tabular}
\caption{Comparison of time histories computed with SEM and the SCM. In both cases $\chi = 12$ nodes per wavelength are used. The time discretization closely satisfies the CFL condition of the respective mesh (i.e. $\Delta t \approx \Delta t_{crit})$. The reference solution is computed with the SEM with $\chi = 37$ and $\Delta t = 2.5 \cdot 10^{-10}$.}
\label{fig:ex2_timehist}
\end{figure}

A preliminary study was performed with an element size of $h = 2.29 \ [mm]$. This delivers $\chi = 12 $ nodes per wavelength, which is the recommended mesh refinement for the chosen polynomial degrees  \cite{willberg2012comparison}. In order to minimize the cost of time integration, the time step is chosen as to closely fulfill the CFL condition of the respective mesh.
For the SEM, the Central Difference Method (CDM) with a time step of $\Delta t = 2.5 \cdot 10^{-8} \ [s]$ is used, while for the SCM the frog-leap algorithm (see \autoref{sec:leapfrog}) with a global time step of $\Delta t = 4 \cdot 10^{-8} \ [s]$ and a local time refinement ratio of $ p_t = 16 $ is employed. This difference is due to the fact that, in the second case, the global time step only depends on the uncut portion of the structured mesh, as the refinement ratio can be adjusted to comply with the critical time step of the cut elements.
The corresponding time histories of the vertical displacements at the sensors are given in  \autoref{fig:ex2_timehist}. For comparison, a reference solution with roughly  $7.1\cdot 10^6$ DOFs and a time step of $\Delta t = 2.5 \cdot 10^{-10} \ [s]$ is computed with the SEM. One can observe that the fitted SCM and the SEM are in very good agreement with the reference, and match the solutions provided in the literature by the degree to which comparison of the time histories is possible. This suggests that the proposed spatial and time discretizations are adequate, and that the error introduced by moment fitting is acceptable.

\begin{figure}[]
%\begin{tabular}{C{.49\textwidth} C{.02\textwidth} C{.49\textwidth}}
\subfigure [L2 error norm convergence for $\mathbf{u}_z(t)$ at $\mathbf{s}_1$] {
    \input{figures/ex2/by_dof/s1_L2.tikz}
    \label{fig:ex2_s1_L2_dofs}
}
\subfigure [L2 error norm convergence for $\mathbf{u}_z(t)$ at $\mathbf{s}_2$] {
    \input{figures/ex2/by_dof/s2_L2.tikz}
    \label{fig:ex2_s2_L2_dofs}
}\\
\subfigure [L2 error norm convergence for $\mathbf{u}_z(t)$ at $\mathbf{s}_1$.] {
    \input{figures/ex2/s1_L2_time.tikz}
     \label{fig:ex2_s1_L2_time}
}
\subfigure [L2 error norm convergence for $\mathbf{u}_z(t)$ at $\mathbf{s}_2$.] {
    \input{figures/ex2/s2_L2_time.tikz}
    \label{fig:ex2_s2_L2_time}
}
%\end{tabular}
\caption{L2 error norm convergence of time histories at sensors $\mathbf{s}_1$ (left) and $\mathbf{s}_2$ (right)}
\label{fig:ex2_conv}
\end{figure}
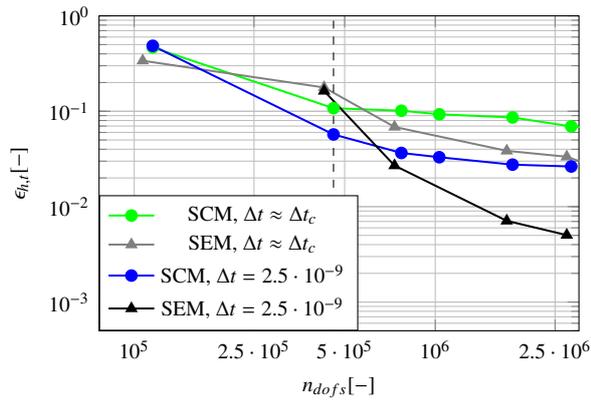
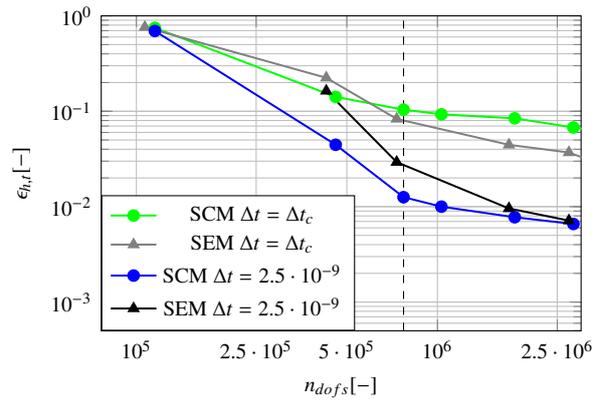
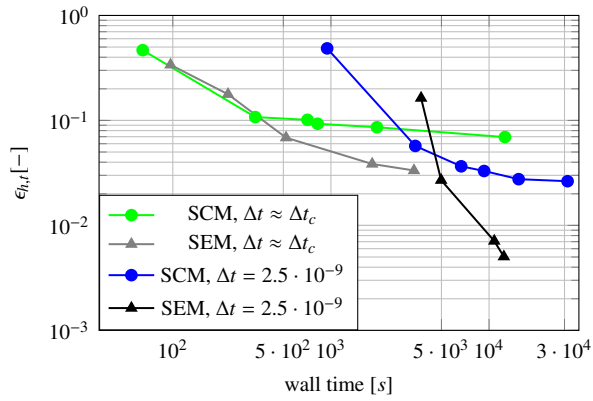
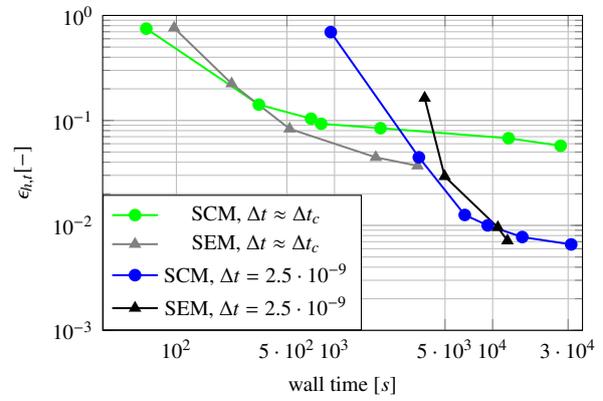

Next, the accuracy of both approaches is assessed by computing the convergence of the L2 error norm of the time histories at the sensors with h-refinement:
\begin{equation} \label{eq:L2_timeHist}
\epsilon_{h,t} = \sqrt{ \frac{\sum\limits_{i = 0}^{n_t} \left[u_h(i\cdot\dt) - u_{ref}(i\cdot\dt)\right]^2 }{\sum\limits_{i = 0}^{n_t} u_{ref}^2(i\cdot\dt)} }
\end{equation}
Where $\epsilon_{h,t}$ is the L2 error norm in time, $ u_h(t) $ is the solution to be evaluated, consisting of $ n_t $ is time steps, and $u_{ref}(t)$ is the aforementioned reference solution. The simulations marked by $\Delta t \approx \Delta t_{c}$ are performed while closely fulfilling the CFL condition, while results obtained with a fixed integration step ($\Delta t = 2.5 \cdot 10^{-9} [s]$) across all models are also provided. The results are plotted with respect to the number of Degrees of Freedom (DOFs) in figures  \autoref{fig:ex2_s1_L2_dofs} and \autoref{fig:ex2_s2_L2_dofs}, which correspond to $ \mathbf{s}_1$ and $ \mathbf{s}_2$, respectively. 

For the cases with $\Delta t \approx \Delta t_{c}$, the SEM performs better than the SCM because, as previously shown, a larger (global) time step can be used with the SCM, leading to a higher time discretization error. To better evaluate the efficacy of space discretization, the simulations with a fixed, lower, time step are considered next. For  $\mathbf{s}_2$, representing a generic point of the mesh, the SCM performs comparably or better than the SEM, while this is not the case for $\mathbf{s}_1$, which represents a worst-case scenario for the proposed method. In both cases, an inflection point is encountered (marked by dashed lines) after which the convergence of the SCM is reduced. This can be indicative of the fact that, before this point, the error introduced by moment fitting is comparable or smaller than the one of the SE discretization, and thus the SCM can leverage its structured mesh to outperform the slightly distorted conforming mesh used with the SEM. This explanation is also consistent with the fact that the loss of accuracy occurs sooner and is more pronounced in proximity of the hole ($\mathbf{s}_1$), for which, at finer meshes, the SEM performs better.

Figures \autoref{fig:ex2_s1_L2_time} and \autoref{fig:ex2_s1_L2_time} display the error $\epsilon_{h,t}$ as a function of the time employed by the dynamic solver to integrate the solution in time, which, given enough memory storage, represents the bulk of the computational cost of an elastodynamic analysis. We implemented both the CDM and the frog-leap solver using openMP parallelization \cite{openmp1998} via the linear algebra library Eigen \cite{guennebaud2010eigen} and solved all problems using 8 threads. These plots confirm that, for coarse meshes, it is beneficial to use the maximum allowed time step, as the error is driven by the spatial discretization, while, for finer meshes, this strategy severely limits the accuracy of the analysis. Compared to figures \autoref{fig:ex2_s1_L2_dofs} and \autoref{fig:ex2_s2_L2_dofs}, the performance of the SEM is improved by the remarkable effectiveness and simplicity of the CDM, while the more complicated leap-frog algorithm might incur some computational overhead and cache misses, and therefore its implementation is more challenging to optimize. Nevertheless, by looking at figure \autoref{fig:ex2_s2_L2_time}, one could also note than this algorithm enables to handle the very fine time step required for cut elements while offering a computation time that is comparable to the CDM.

%%%%%%%%%%%%%%%%%%%%%%%
%time histories
%%%%%%%%%%%%%%%%%%%%%%%

%% file: figures/ex2/s1_displ.tikz
\begin{tikzpicture}
[spy using outlines=
	{magnification=1, anchor = center, connect spies}]
    \begin{axis}[
        name = ex2_s1,
        xmin = 0,
        xmax = 1e-4,
        ymin = -5e-4,
        ymax = 6e-4,
        xlabel = {$t$ [s]},
        ylabel = {$u_{z}(t)$ [m]},
        xticklabel style = {font =\fontsize{\figureFontSize pt}{10pt}\selectfont},
        yticklabel style = {font=\fontsize{\figureFontSize pt}{10pt}\selectfont},
        xlabel style = {font =\fontsize{\figureFontSize pt}{\figureFontSize}\selectfont},
        ylabel style = {font=\fontsize{\figureFontSize pt}{\figureFontSize}\selectfont},
        no markers,
        grid = both,
        width=0.48\textwidth,
        height=0.24\textwidth,
        legend pos=south west,
        legend style={
                    at = {(0,-0.25)},
                    anchor=north west,
                    nodes={font=\fontsize{\figureFontSize pt}{\figureFontSize}\selectfont}
                    }
        ]
        
        \addplot table [x=time, y=sensor_2, col sep=semicolon] {figures/ex2/cut_132.csv};
        \addplot table [x=time, y=sensor_2, col sep=semicolon] {figures/ex2/expl_132.csv};
        \addplot [dashed, color=black] table [x=time, y=sensor_2, col sep=semicolon] {figures/ex2/expl_401_short.csv};

        \addlegendentry{SCM, $\chi = 12$}
        \addlegendentry{SEM, $\chi = 12$ }
        \addlegendentry{SEM, $\chi = 37$}

\end{axis}
\end{tikzpicture}

%% file: Moment fitted cut spectral elements for explicit analysis of guided wave propagation/figures/ex3/timehist/S2_displ.tikz
\begin{tikzpicture}
[spy using outlines=
	{magnification=1, anchor = center, connect spies}]
    \begin{axis}[
        name = ex2_s1,
        xmin = 0,
        xmax = 1e-4,
        ymin = -3e-12,
        ymax = 3e-12,
        xlabel = {time [s]},
        ylabel = {$u_{1,z}$ [m]},
        xticklabel style = {font =\fontsize{\figureFontSize pt}{10pt}\selectfont},
        yticklabel style = {font=\fontsize{\figureFontSize pt}{10pt}\selectfont},
        xlabel style = {font =\fontsize{\figureFontSize pt}{\figureFontSize}\selectfont},
        ylabel style = {font=\fontsize{\figureFontSize pt}{\figureFontSize}\selectfont},
        no markers,
        grid = both,
        width=0.48\textwidth,
        height=0.24\textwidth,
        legend pos=south west,
        legend style={
                    at = {(0,-0.35)},
                    anchor=north west,
                    nodes={font=\fontsize{\figureFontSize pt}{\figureFontSize}\selectfont}
                    }
        ]
        
        % \addplot [color = green] table [x=time, y=sensor_5, col sep=semicolon] {figures/ex3/timehist/cut76_sh334_dtcrit.csv};
        %\addlegendentry{SCM, $\chi = 16$ }
        
         \addplot table [x=time, y=sensor_5, col sep=semicolon] {figures/ex3/timehist/cut101_sh334_dtcrit.csv};
        \addlegendentry{SCM, $\chi = 21$ }
        
        \addplot table [x=time, y=sensor_5, col sep=semicolon] {figures/ex3/timehist/cut201_sh334_dtcrit.csv};
        \addlegendentry{SCM, $\chi = 41$ }
        
        \addplot table [x=time, y=sensor_5, col sep=semicolon] {figures/ex3/timehist/expl_h19e-4_O2_sh334_dtcrit.csv};
        \addlegendentry{SEM, $\chi \approx 41$}
        
        \addplot table [x=time, y=sensor_5, col sep=semicolon] {figures/ex3/timehist/reduced_h8e-4_O2_sh334.csv};
        \addlegendentry{SEM,  $\chi \approx 101$}

\end{axis}
\end{tikzpicture}

%% file: figures/ex2/by_dof/s1_L2.tikz
\begin{tikzpicture}
    \begin{axis}[
        xmin = 0,
        xmax = 3e6,
        ymin = 5e-4,
        ymax = 1,
        xtick = {10^5, 2.5*10^5, 5*10^5, 10^6, 2.5*10^6},
        xticklabels = {$10^5$,$2.5 \cdot 10^5$, $5 \cdot 10^5$,$10^6$,$2.5 \cdot 10^6$},
        xmode = log,
        ymode = log,
        xlabel = {$n_{dofs} [-]$},
        ylabel = {$\epsilon_{h,t} [-]$},
        xticklabel style = {font =\fontsize{\figureFontSize pt}{10pt}\selectfont},
        yticklabel style = {font=\fontsize{\figureFontSize pt}{10pt}\selectfont},
        xlabel style = {font =\fontsize{\figureFontSize pt}{\figureFontSize pt}\selectfont},
        ylabel style = {font=\fontsize{\figureFontSize pt}{\figureFontSize pt}\selectfont},
        grid = both,
        width=0.48\textwidth,
        height=0.35\textwidth,
        legend style={at={(0,0)},anchor=south west,nodes={font=\fontsize{\figureFontSize pt}{\figureFontSize pt}\selectfont}}
%        legend pos = south east
        ]
    \addplot[color = \cutOptimizedColorb, mark =\explicitMarker*, line width=0.75pt] coordinates{
        %cut_min_w1e-1_kill1e-3/dt_is_dtcrit: sensor 1 error conv of L2 by ndof
        (115215,4.671273e-01)        %perforated_solidPlateWillbergStrSym51_sh334
        (459195,1.076071e-01)        %perforated_solidPlateWillbergStrSym101_sh334
        (771810,1.013030e-01)        %perforated_solidPlateWillbergStrSym132_sh334
        (1031055,9.297349e-02)        %perforated_solidPlateWillbergStrSym151_sh334
        (1806000,8.610332e-02)        %perforated_solidPlateWillbergStrSym201_sh334
        (2828355,6.938696e-02)        %perforated_solidPlateWillbergStrSym251_sh334
        (4079085,6.410019e-02)        %perforated_solidPlateWillbergStrSym301_sh334
	};
    \addlegendentry{SCM, $\Delta t \approx \Delta t_{c}$}
    
    \addplot[color = \uncutColorb, mark =\conformMarker*, line width=\plotLineWidth pt] coordinates{
        %expl/dt_is_dtcrit: sensor 1 error conv of L2 by ndof
        (106710,3.386918e-01)        %morphed_solidPlateWillbergStrSymCyl51_O2_sh334
        (427290,1.775610e-01)        %morphed_solidPlateWillbergStrSymCyl101_O2_sh334
        (732390,6.851490e-02)        %morphed_solidPlateWillbergStrSymCyl132_O2_sh334
        (1727835,3.847980e-02)        %morphed_solidPlateWillbergStrSymCyl201_O2_sh334
        (2733225,3.336194e-02)        %morphed_solidPlateWillbergStrSymCyl251_O2_sh334
        (3977520,2.199634e-02)        %morphed_solidPlateWillbergStrSymCyl301_O2_sh334
        %(7103535,2.275379e-02)        %morphed_solidPlateWillbergStrSymCyl401_O2_sh334
	};
    \addlegendentry{SEM, $\Delta t \approx \Delta t_{c}$}
    
    \addplot[color = \cutOptimizedColor, mark =\explicitMarker*, line width=0.75pt] coordinates{
        (115215,4.856207e-01)        %perforated_solidPlateWillbergStrSym51_sh334
        (459195,5.714948e-02)        %perforated_solidPlateWillbergStrSym101_sh334
        (771810,3.655594e-02)        %perforated_solidPlateWillbergStrSym132_sh334
        (1031055,3.298123e-02)        %perforated_solidPlateWillbergStrSym151_sh334
        (1806000,2.756921e-02)        %perforated_solidPlateWillbergStrSym201_sh334
        (2828355,2.636406e-02)        %perforated_solidPlateWillbergStrSym251_sh334
	};
    \addlegendentry{SCM, $\Delta t = 2.5 \cdot 10^{-9}$}
    
    \addplot[color = \uncutColor, mark =\conformMarker*, line width=\plotLineWidth pt] coordinates{
        (427290,1.633026e-01)        %morphed_solidPlateWillbergStrSymCyl101_O2_sh334
        (732390,2.697889e-02)        %morphed_solidPlateWillbergStrSymCyl132_O2_sh334
        (1727835,7.101148e-03)        %morphed_solidPlateWillbergStrSymCyl201_O2_sh334
        (2733225,5.036780e-03)        %morphed_solidPlateWillbergStrSymCyl251_O2_sh334
        %(3977520,4.630423e-03)        %morphed_solidPlateWillbergStrSymCyl301_O2_sh334
	};
    \addlegendentry{SEM, $\Delta t = 2.5 \cdot 10^{-9}$}
    
    \addplot[dashed, mark=none, black] coordinates {(459195,5e-4) (459195,1)};
    
    \end{axis}
\end{tikzpicture}

%% file: figures/ex2/by_dof/s2_L2.tikz
\begin{tikzpicture}
    \begin{axis}[
        xmin = 0,
        xmax = 3e6,
        ymin = 5e-4,
        ymax = 1,
        xtick = {10^5, 2.5*10^5, 5*10^5, 10^6, 2.5*10^6},
        xticklabels = {$10^5$,$2.5 \cdot 10^5$, $5 \cdot 10^5$,$10^6$,$2.5 \cdot 10^6$},
        xmode = log,
        ymode = log,
        xlabel = {$n_{dofs} [-]$},
        ylabel = {$\epsilon_{h,t} [-]$},
        xticklabel style = {font =\fontsize{\figureFontSize pt}{10pt}\selectfont},
        yticklabel style = {font=\fontsize{\figureFontSize pt}{10pt}\selectfont},
        xlabel style = {font =\fontsize{\figureFontSize pt}{\figureFontSize pt}\selectfont},
        ylabel style = {font=\fontsize{\figureFontSize pt}{\figureFontSize pt}\selectfont},
        grid = both,
        width=0.48\textwidth,
        height=0.35\textwidth,
        legend style={at={(0,0)},anchor=south west,nodes={font=\fontsize{\figureFontSize pt}{\figureFontSize pt}\selectfont}}
%        legend pos = south east
        ]
            \addplot[color = \cutOptimizedColorb, mark =\explicitMarker*, line width=0.75pt] coordinates{
        %cut_min_w1e-1_kill1e-3/dt_is_dtcrit: sensor 2 error conv of L2 by ndof
        %(19140,3.016599e+00)        %solidPlateWillbergStrSym21_sh334
        (115215,7.492915e-01)        %perforated_solidPlateWillbergStrSym51_sh334
        (459195,1.414332e-01)        %perforated_solidPlateWillbergStrSym101_sh334
        (771810,1.039666e-01)        %perforated_solidPlateWillbergStrSym132_sh334
        (1031055,9.289755e-02)        %perforated_solidPlateWillbergStrSym151_sh334
        (1806000,8.430959e-02)        %perforated_solidPlateWillbergStrSym201_sh334
        (2828355,6.765287e-02)        %perforated_solidPlateWillbergStrSym251_sh334
        (4079085,5.739852e-02)        %perforated_solidPlateWillbergStrSym301_sh334
	};
    \addlegendentry{SCM $\Delta t = \Delta t_{c}$}
    
    \addplot[color = \uncutColorb, mark =\conformMarker*, line width=\plotLineWidth pt] coordinates{
        %expl/dt_is_dtcrit: sensor 2 error conv of L2 by ndof
        (106710,7.613629e-01)        %morphed_solidPlateWillbergStrSymCyl51_O2_sh334
        (427290,2.242084e-01)        %morphed_solidPlateWillbergStrSymCyl101_O2_sh334
        (732390,8.315060e-02)        %morphed_solidPlateWillbergStrSymCyl132_O2_sh334
        (1727835,4.449440e-02)        %morphed_solidPlateWillbergStrSymCyl201_O2_sh334
        (2733225,3.698472e-02)        %morphed_solidPlateWillbergStrSymCyl251_O2_sh334
        (3977520,2.386830e-02)        %morphed_solidPlateWillbergStrSymCyl301_O2_sh334
        (7103535,2.355977e-02)        %morphed_solidPlateWillbergStrSymCyl401_O2_sh334
	};
    \addlegendentry{SEM $\Delta t = \Delta t_{c}$}
    
    \addplot[color = \cutOptimizedColor, mark =\explicitMarker*, line width=0.75pt] coordinates{
        (115215,6.944212e-01)        %perforated_solidPlateWillbergStrSym51_sh334
        (459195,4.455072e-02)        %perforated_solidPlateWillbergStrSym101_sh334
        (771810,1.257766e-02)        %perforated_solidPlateWillbergStrSym132_sh334
        (1031055,1.000072e-02)        %perforated_solidPlateWillbergStrSym151_sh334
        (1806000,7.748154e-03)        %perforated_solidPlateWillbergStrSym201_sh334
        (2828355,6.587350e-03)        %perforated_solidPlateWillbergStrSym251_sh334
        %(4079085,6.871763e-03)        %perforated_solidPlateWillbergStrSym301_sh334
	};
    \addlegendentry{SCM $\Delta t = 2.5 \cdot 10^{-9}$}
    
    \addplot[color = \uncutColor, mark =\conformMarker*, line width=\plotLineWidth pt] coordinates{
        (427290,1.634893e-01)        %morphed_solidPlateWillbergStrSymCyl101_O2_sh334
        (732390,2.929614e-02)        %morphed_solidPlateWillbergStrSymCyl132_O2_sh334
        (1727835,9.572522e-03)        %morphed_solidPlateWillbergStrSymCyl201_O2_sh334
        (2733225,7.148873e-03)        %morphed_solidPlateWillbergStrSymCyl251_O2_sh334
        %(3977520,5.986474e-03)        %morphed_solidPlateWillbergStrSymCyl301_O2_sh334
	};
    \addlegendentry{SEM $\Delta t = 2.5 \cdot 10^{-9}$}
    
    \addplot[dashed, mark=none, black] coordinates {(771810,5e-4) (771810,1)};
    
    \end{axis}
\end{tikzpicture}

%% file: figures/ex2/s1_L2_time.tikz
\begin{tikzpicture}
    \begin{axis}[
        xmin = 0,
        xmax = 3.7e4,
        ymin = 1e-3,
        ymax = 1,
        xtick = {1e2,5e2, 1e3, 5e3, 1e4, 3e4},
        xticklabels = {$10^2$,$5 \cdot 10^2$, $10^3$, $5 \cdot 10^3$, $10^4$, $3 \cdot 10^4$},
        xmode = log,
        ymode = log,
        xlabel = {wall time $[s]$},
        ylabel = {$\epsilon_{h,t} [-]$},
        xticklabel style = {font =\fontsize{\figureFontSize pt}{10pt}\selectfont},
        yticklabel style = {font=\fontsize{\figureFontSize pt}{10pt}\selectfont},
        xlabel style = {font =\fontsize{\figureFontSize pt}{\figureFontSize pt}\selectfont},
        ylabel style = {font=\fontsize{\figureFontSize pt}{\figureFontSize pt}\selectfont},
        grid = both,
        width=0.48\textwidth,
        height=0.35\textwidth,
        legend style={at={(0,0)},anchor=south west,nodes={font=\fontsize{\figureFontSize pt}{\figureFontSize pt}\selectfont}}
%        legend pos = south east
        ]

     \addplot[color = \cutOptimizedColorb, mark =\explicitMarker*, line width=0.75pt] coordinates{
		%cut_min_w1e-1_kill1e-3/dt_is_dtcrit: sensor 1 error conv of L2 by wall time
		%(9.712922e+00,7.617976e+00)        %solidPlateWillbergStrSym21_sh334
		(6.450508e+01,4.671273e-01)        %perforated_solidPlateWillbergStrSym51_sh334
        (3.323725e+02,1.076071e-01)        %perforated_solidPlateWillbergStrSym101_sh334
        (7.109505e+02,1.013030e-01)        %perforated_solidPlateWillbergStrSym132_sh334
        (8.240353e+02,9.297349e-02)        %perforated_solidPlateWillbergStrSym151_sh334
        (1.954218e+03,8.610332e-02)        %perforated_solidPlateWillbergStrSym201_sh334
        (1.260240e+04,6.938696e-02)        %perforated_solidPlateWillbergStrSym251_sh334
        %(2.690841e+04,6.410019e-02)        %perforated_solidPlateWillbergStrSym301_sh334
	};
    \addlegendentry{SCM, $\Delta t \approx \Delta t_{c}$}
    
    \addplot[color = \uncutColorb, mark =\conformMarker*, line width=\plotLineWidth pt] coordinates{
        %expl/dt_is_dtcrit: sensor 1 error conv of L2 by wall time
        (9.673130e+01,3.386918e-01)        %morphed_solidPlateWillbergStrSymCyl51_O2_sh334
        (2.234510e+02,1.775610e-01)        %morphed_solidPlateWillbergStrSymCyl101_O2_sh334
        (5.201053e+02,6.851490e-02)        %morphed_solidPlateWillbergStrSymCyl132_O2_sh334
        (1.823283e+03,3.847980e-02)        %morphed_solidPlateWillbergStrSymCyl201_O2_sh334
        (3.350779e+03,3.336194e-02)        %morphed_solidPlateWillbergStrSymCyl251_O2_sh334
        %(2.929210e+04,2.199634e-02)        %morphed_solidPlateWillbergStrSymCyl301_O2_sh334
        %(1.307890e+04,2.275379e-02)        %morphed_solidPlateWillbergStrSymCyl401_O2_sh334
        
	};
    
    \addlegendentry{SEM, $\Delta t \approx \Delta t_{c}$}
    
     \addplot[color = \cutOptimizedColor, mark =\explicitMarker*, line width=0.75pt] coordinates{
		%cut_min_w1e-1_kill1e-3/dt25e-10: sensor 1 error conv of L2 by wall time
        (9.463517e+02,4.856207e-01)        %perforated_solidPlateWillbergStrSym51_sh334
        (3.416917e+03,5.714948e-02)        %perforated_solidPlateWillbergStrSym101_sh334
        (6.665817e+03,3.655594e-02)        %perforated_solidPlateWillbergStrSym132_sh334
        (9.306488e+03,3.298123e-02)        %perforated_solidPlateWillbergStrSym151_sh334
        (1.538705e+04,2.756921e-02)        %perforated_solidPlateWillbergStrSym201_sh334
        (3.136903e+04,2.636406e-02)        %perforated_solidPlateWillbergStrSym251_sh334
        %(3.614951e+04,2.644066e-02)        %perforated_solidPlateWillbergStrSym301_sh334
	};
    \addlegendentry{SCM, $\Delta t = 2.5 \cdot 10^{-9}$}
    
    \addplot[color = \uncutColor, mark =\conformMarker*, line width=\plotLineWidth pt] coordinates{
        %expl/dt25e-10: sensor 1 error conv of L2 by wall time
        (3.717419e+03,1.633026e-01)        %morphed_solidPlateWillbergStrSymCyl101_O2_sh334
        (4.976566e+03,2.697889e-02)        %morphed_solidPlateWillbergStrSymCyl132_O2_sh334
        (1.078657e+04,7.101148e-03)        %morphed_solidPlateWillbergStrSymCyl201_O2_sh334
        (1.240475e+04,5.036780e-03)        %morphed_solidPlateWillbergStrSymCyl251_O2_sh334
        %(1.627855e+04,4.630423e-03)        %morphed_solidPlateWillbergStrSymCyl301_O2_sh334
	};
    
    \addlegendentry{SEM, $\Delta t = 2.5 \cdot 10^{-9}$}
    
    \end{axis}
\end{tikzpicture}

%% file: figures/ex2/s2_L2_time.tikz
\begin{tikzpicture}
    \begin{axis}[
        xmin = 0,
        xmax = 3.7e4,
        ymin = 1e-3,
        ymax = 1,
        xtick = {1e2,5e2, 1e3, 5e3, 1e4, 3e4},
        xticklabels = {$10^2$,$5 \cdot 10^2$, $10^3$, $5 \cdot 10^3$, $10^4$, $3 \cdot 10^4$},
        xmode = log,
        ymode = log,
        xlabel = {wall time $[s]$},
        ylabel = {$\epsilon_{h,t} [-]$},
        xticklabel style = {font =\fontsize{\figureFontSize pt}{10pt}\selectfont},
        yticklabel style = {font=\fontsize{\figureFontSize pt}{10pt}\selectfont},
        xlabel style = {font =\fontsize{\figureFontSize pt}{\figureFontSize pt}\selectfont},
        ylabel style = {font=\fontsize{\figureFontSize pt}{\figureFontSize pt}\selectfont},
        grid = both,
        width=0.48\textwidth,
        height=0.35\textwidth,
        legend style={at={(0,0)},anchor=south west,nodes={font=\fontsize{\figureFontSize pt}{\figureFontSize pt}\selectfont}}
%        legend pos = south east
        ]
        
        \addplot[color = \cutOptimizedColorb, mark =\explicitMarker*, line width=0.75pt] coordinates{
		%cut_min_w1e-1_kill1e-3/dt_is_dtcrit: sensor 2 error conv of L2 by wall time
		%(9.712922e+00,3.016599e+00)        %solidPlateWillbergStrSym21_sh334
        (6.450508e+01,7.492915e-01)        %perforated_solidPlateWillbergStrSym51_sh334
        (3.323725e+02,1.414332e-01)        %perforated_solidPlateWillbergStrSym101_sh334
        (7.109505e+02,1.039666e-01)        %perforated_solidPlateWillbergStrSym132_sh334
        (8.240353e+02,9.289755e-02)        %perforated_solidPlateWillbergStrSym151_sh334
        (1.954218e+03,8.430959e-02)        %perforated_solidPlateWillbergStrSym201_sh334
        (1.260240e+04,6.765287e-02)        %perforated_solidPlateWillbergStrSym251_sh334
        (2.690841e+04,5.739852e-02)        %perforated_solidPlateWillbergStrSym301_sh334
	};
    
    \addplot[color = \uncutColorb, mark =\conformMarker*, line width=\plotLineWidth pt] coordinates{
        %expl/dt_is_dtcrit: sensor 2 error conv of L2 by wall time
        (9.673130e+01,7.613629e-01)        %morphed_solidPlateWillbergStrSymCyl51_O2_sh334
        (2.234510e+02,2.242084e-01)        %morphed_solidPlateWillbergStrSymCyl101_O2_sh334
        (5.201053e+02,8.315060e-02)        %morphed_solidPlateWillbergStrSymCyl132_O2_sh334
        (1.823283e+03,4.449440e-02)        %morphed_solidPlateWillbergStrSymCyl201_O2_sh334
        (3.350779e+03,3.698472e-02)        %morphed_solidPlateWillbergStrSymCyl251_O2_sh334
        %(2.929210e+04,2.386830e-02)        %morphed_solidPlateWillbergStrSymCyl301_O2_sh334
        %(1.307890e+04,2.355977e-02)        %morphed_solidPlateWillbergStrSymCyl401_O2_sh334

	};
    \addlegendentry{SCM, $\Delta t \approx \Delta t_{c}$}
    \addlegendentry{SEM, $\Delta t \approx \Delta t_{c}$}
    
    \addplot[color = \cutOptimizedColor, mark =\explicitMarker*, line width=0.75pt] coordinates{
		%cut_min_w1e-1_kill1e-3/dt25e-10: sensor 2 error conv of L2 by wall time
        (9.463517e+02,6.944212e-01)        %perforated_solidPlateWillbergStrSym51_sh334
        (3.416917e+03,4.455072e-02)        %perforated_solidPlateWillbergStrSym101_sh334
        (6.665817e+03,1.257766e-02)        %perforated_solidPlateWillbergStrSym132_sh334
        (9.306488e+03,1.000072e-02)        %perforated_solidPlateWillbergStrSym151_sh334
        (1.538705e+04,7.748154e-03)        %perforated_solidPlateWillbergStrSym201_sh334
        (3.136903e+04,6.587350e-03)        %perforated_solidPlateWillbergStrSym251_sh334
        %(3.614951e+04,6.871763e-03)        %perforated_solidPlateWillbergStrSym301_sh334
	};
    \addlegendentry{SCM, $\Delta t = 2.5 \cdot 10^{-9}$}
    
    \addplot[color = \uncutColor, mark =\conformMarker*, line width=\plotLineWidth pt] coordinates{
        %expl/dt25e-10: sensor 2 error conv of L2 by wall time
        (3.717419e+03,1.634893e-01)        %morphed_solidPlateWillbergStrSymCyl101_O2_sh334
        (4.976566e+03,2.929614e-02)        %morphed_solidPlateWillbergStrSymCyl132_O2_sh334
        (1.078657e+04,9.572522e-03)        %morphed_solidPlateWillbergStrSymCyl201_O2_sh334
        (1.240475e+04,7.148873e-03)        %morphed_solidPlateWillbergStrSymCyl251_O2_sh334
        %(1.627855e+04,5.986474e-03)        %morphed_solidPlateWillbergStrSymCyl301_O2_sh334
	};
    \addlegendentry{SEM, $\Delta t = 2.5 \cdot 10^{-9}$}

    \end{axis}
\end{tikzpicture}

%% file: 4_3_experiment.tex
\newpage
\subsection{Aluminium plate with rivet holes}\label{sec:ex3}
\begin{figure}[] %model vs picture
\begin{tabular}{C{.5\textwidth}C{.5\textwidth}}
\subfigure {
    \resizebox{0.59\textwidth}{!}{%
    \includegraphics[width=1\linewidth]{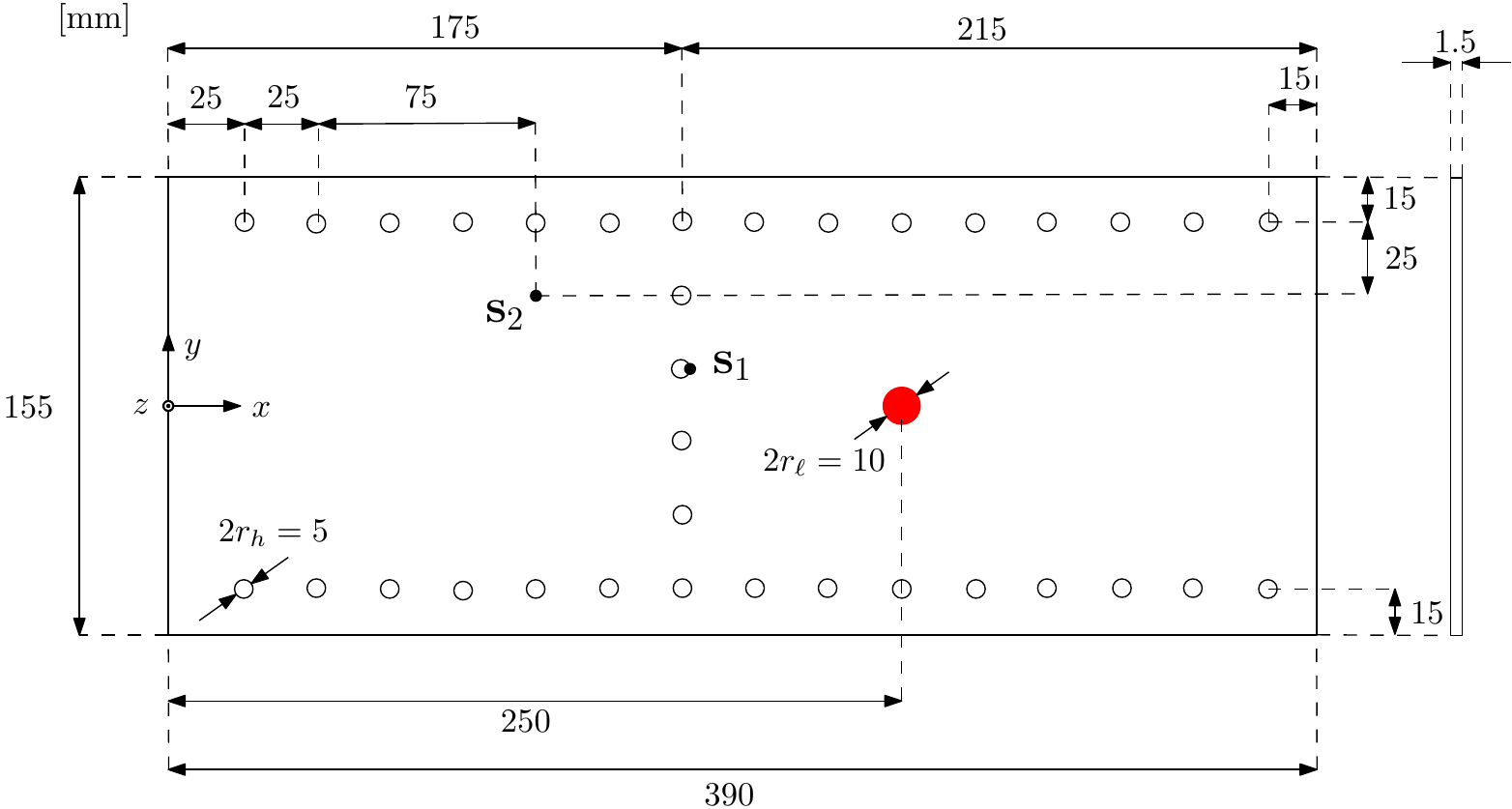}
    }
} & 
\vspace{10mm}
\subfigure  {
    \resizebox{0.39\textwidth}{!}{%
    \includegraphics[width=1\linewidth]{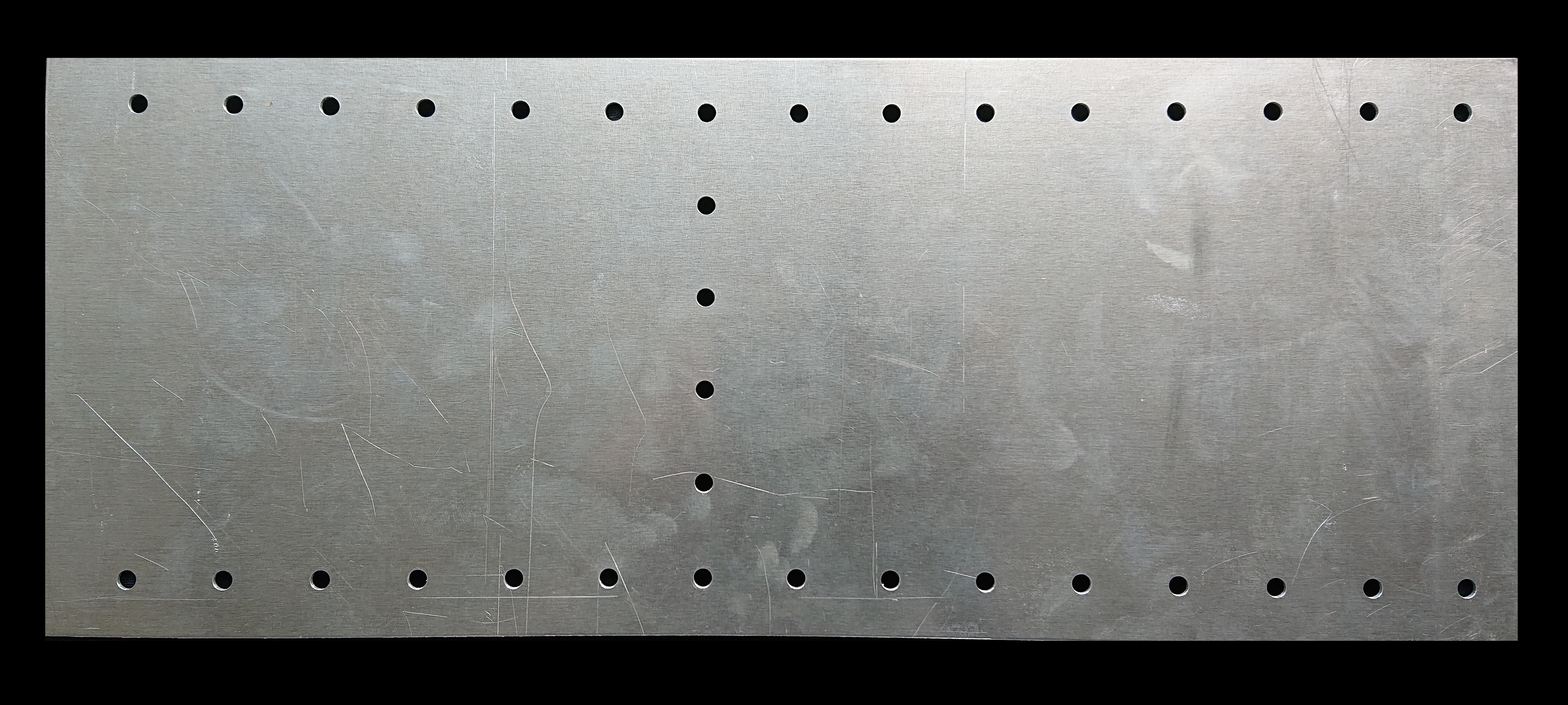}
    }
}
\\
\end{tabular}
\caption{Aluminum plate with rivet holes. Actuators on both sides of the plate are centered ad $(250, 0)$ and modeled by a surface loading area of radius $r_{\ell} = 5 [mm]$.}
\label{fig:ex3_setup}
\end{figure}

In this example we consider the aluminium plate represented in \autoref{fig:ex3_setup}, which is the reproduction of an aerospace panel due for experimental testing.
With respect to the previous example, the plate has increased dimensions (i.e. $l_x = 390 \ [mm], \ l_y = 155 \ [mm]$) and is affected by 34 cylindrical rivet holes with much smaller radii ($r_h = 2.5 \ [mm]$) than the previous conical hole. The configuration of the loading and of all holes are given in \autoref{fig:ex3_setup}, which also highlights the possibility of mirroring the model along the xz-plane, as in the previous example. The modeling of the actuator is improved by introducing circular loading areas with a radius $r_{\ell} = 5 \ [mm]$ on both sides of the plate. Normal out-of-phase surface loading is applied according to \autoref{eq:hanning} and \autoref{eq:will_load} with $\mathbf{p}_1(t), \ \mathbf{p}_2(t)$ acting respectively on the top and bottom surfaces, and $p = 10^4 \ [Pa]$. The frequency is increased to $f = 200 \ [kHz]$ and $n = 5$.
%With a plate thickness of $l_z = 1.5 \ [mm]$, this results to a wave length of $\lambda \approx 7.8 \ [mm]$. The same SE type as before is selected ($\mathcal{N}_{3,3,4}$), leading to a recommended element size of $h = 1.95 \ [mm]$.

To benchmark the quality of the solution with the novel method, the displacement time history at 2 sensors $\mathbf{s}_{1}(177.5, 12.5, 1.5) \ [mm]$ and $\mathbf{s}_{2}(125, 37.5, 1.5) \ [mm]$ is recorded. Similarly to the previous example, these sensor placements enable to evaluate the the accuracy of the method in a ``worst case'' location with respect to the accuracy of spatial discretization ($\mathbf{s}_{1}$), as well as in a sensor location that might be chosen in a real-life applications ($\mathbf{s}_{2}$).

The problem is solved with both, the SEM and the proposed version of the SCM. In the following, we offer some considerations regarding the practical application of both methods.
With the SEM, some limitations quickly arise when representing the domain with a conforming mesh. Aiming for good element quality, the analyst might choose to produce a highly customized structured mesh, which nevertheless would require further human work shall the hole configuration change or a flaw be considered. On the other hand, an unstructured mesh can be generated from more versatile geometrical definitions, however will result in elements of poorer quality and of a broader range of sizes, with negative implications in accuracy of the spatial discretization and in the efficiency of time integration. For the present example, the latter approach was adopted to reflect the requirement for versatility, typical of damage detection applications, as well as the potential interest for even more sophisticated geometries for which high quality, hexaedral, meshes are simply not available. By fixing an upper bound for the element size, we observed that the severity of the aforementioned phenomena strongly varies with the choice of meshing algorithm, which in some cases might even fail to mesh the domain without the use of prismatic elements. We investigated the algorithms offered within the Gmsh library \cite{geuzaine2009gmsh} and selected the BAMG algorithm \cite{hecht1998bamg} for the present example, as it delivered critical time steps that were one or two orders of magnitude larger than the ones obtained with alternative algorithms.

These meshing-related problems are overcome with the SCM, which enables to define each hole at run time by specifying the parameters of the respective signed distance functions. Given any point of the domain as an argument, the level set value is then simply the signed distance to the nearest hole. Moreover, the same approach can be used to represent the loading area (or other boundary conditions). This is rather common in fictitious domain applications \cite{parvizian2007finite}, where the background mesh is often fully immersed in a boundary.

\noindent
\begin{minipage}{0.5\linewidth}
\autoref{fig:ex3_LSM_load} highlights the discretization of the load area with the SCM. The top and bottom surfaces of a $\mathcal{N}_{3,3,4}$ element are represented by connecting the respective nodes with elements of the family $\mathcal{N}_{3,3}$ with $3$ DOFs per node. For the implicitly defined load area, a quadrature rule is provided by a local quadtree mesh and element partitions within the surface elements. For clarity, we should emphasise that these elements do not provide any mass (nor any stiffness) contributions, as they are only employed to discretize the load, and thus no moment fitting is necessary in this case. For a loading of this kind, we use the following adaptation of \autoref{eq:loads}:
\begin{equation}
    \mathbf{f}_e(t) = \int_{\Omega_e } {N^T \mathbf{p}_{s}(t) \ }{d\Omega_e}
    \label{eq:cut_load}
\end{equation}
In \autoref{eq:cut_load}, the distributed traction $\mathbf{p}_{s}(t)$ is integrated over the physical portion $\Omega_e$ of the $\mathcal{N}_{3,3}$ element, delivering the corresponding force vector. Since its DOFs have been chosen to match the ones of the $\mathcal{N}_{3,3,4}$ mesh, $\mathbf{f}_e(t)$ can directly be assembled into the system force vector $\mathbf{f}_s(t)$.
\end{minipage}
\hfill
\begin{minipage}{0.4\linewidth}
\centering
\includegraphics[width=\linewidth]{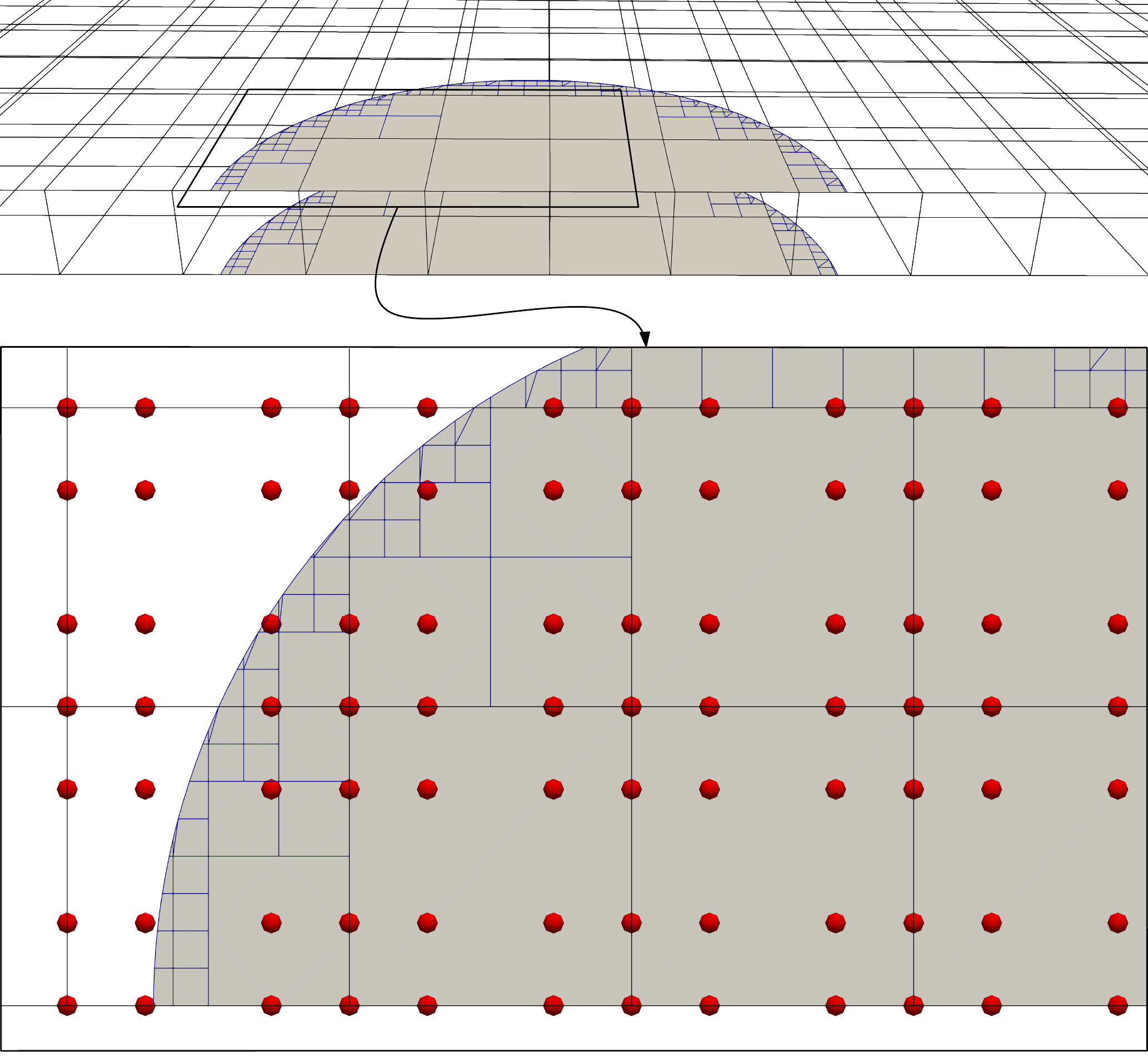}
\captionof{figure}{(Top): GGL-SE grid with top and bottom load surfaces independent form the mesh. (Bottom): Load area discretization with quadtree mesh and element partitions (see \autoref{sec:LSM}). Nodes of $\mathcal{N}_{3,3}$ elements for computation of load vector components.}
\label{fig:ex3_LSM_load}
\end{minipage}

\begin{figure}[htbp]
%\begin{tabular}{C{.49\textwidth} C{.02\textwidth} C{.49\textwidth}}
\subfigure [$u_{z}(t)$ at $\mathbf{s}_1$] {
    \input{figures/ex3/timehist/S1_displ.tikz}

} 
\subfigure [$u_{z}(t)$ at $\mathbf{s}_2$] {
    \input{figures/ex3/timehist/S2_displ.tikz}
}
%\end{tabular}
\caption{Comparison of time histories at for the problem shown in \autoref{fig:ex3_setup}. With the SCM and $\chi = 21$, good results can be obtained, while finer meshes are necessary with the SEM ($\chi = 41$) to achieve conformity of hexahedral elements. Integration in time occurs while closely fulfilling the CFL condition, while for the reference solution ($\chi = 101$) a time step of $\Delta t= 2.5 \cdot 10^{-10}$ has been used.}
\label{fig:ex3_timehist}
\end{figure}
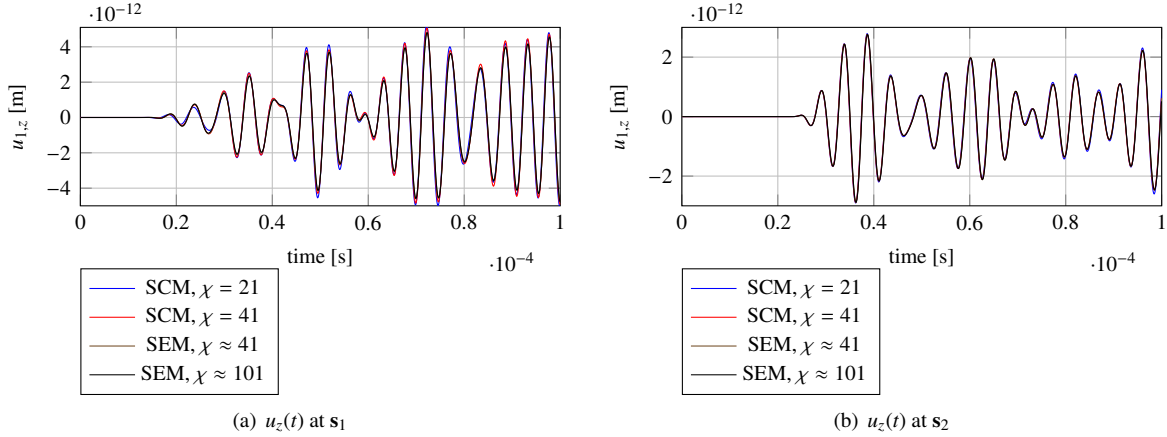

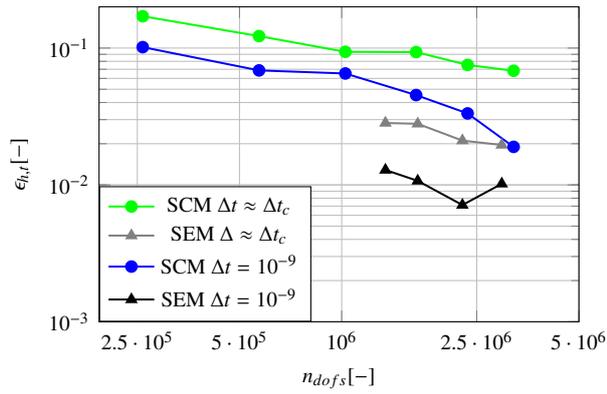
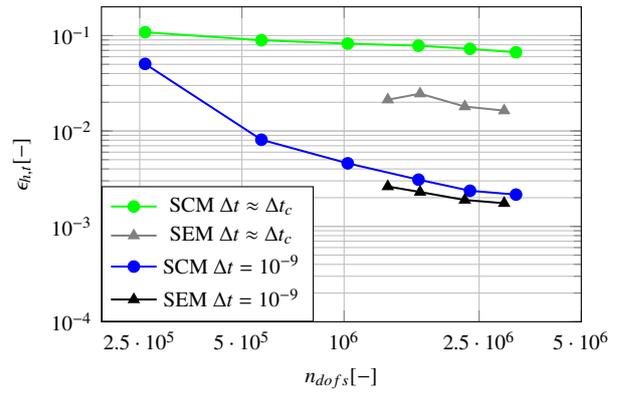
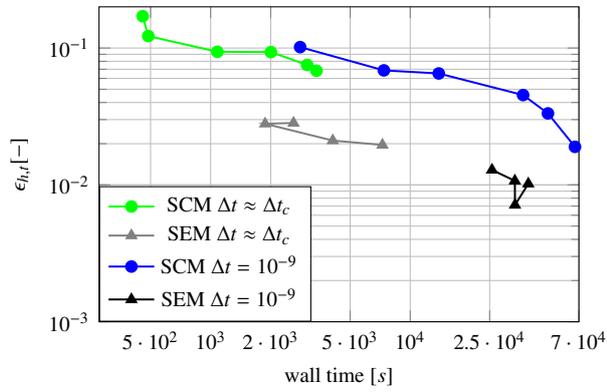
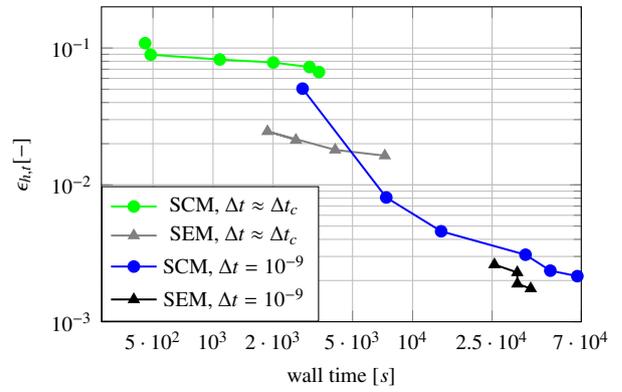
\begin{figure}[]
%\begin{tabular}{C{.49\textwidth} C{.02\textwidth} C{.49\textwidth}}
\subfigure [L2 error norm convergence for $\mathbf{u}_z(t)$ at $\mathbf{s}_1$] {
    \input{figures/ex3/s1_L2.tikz}
}
\subfigure [L2 error norm convergence for $\mathbf{u}_z(t)$ at $\mathbf{s}_2$] {
    \input{figures/ex3/s2_L2.tikz}
}\\
\subfigure [L2 error norm convergence for $\mathbf{u}_z(t)$ at $\mathbf{s}_1$] {
    \input{figures/ex3/s1_L2_time.tikz}
} 
\subfigure [L2 error norm convergence for $\mathbf{u}_z(t)$ at $\mathbf{s}_2$] {
    \input{figures/ex3/s2_L2_time.tikz}
}
%\end{tabular}
\caption{L2 error norm convergence as function of model size and simulation time}
\label{fig:ex3_L2}
\end{figure}

For purely symmetric modes and the given problem parameters, an element size of $h = 7.3 [mm]$ is theoretically recommended. However, this value had to be reduced to $h = 3.9 [mm]$ ($\chi = 21$) to capture the finer modes introduced by the holes. As shown in \autoref{fig:ex3_timehist}, in comparison with a more refined instance of the SCM ($\chi = 41$) and the reference solution obtained with the SEM ($\chi \approx 101$), this choiche leads to acceptable results. For the SEM, due to the aforementioned limitations, the coarsest mesh available has an element size  of $h \approx 1.95 \ [mm]$ (i.e. $\chi \approx 41$). The simulations ``SCM, $\chi = 21$'' and ``SEM, $\chi = 41$'' correspond to the first data points for the respective methods and $\Delta t \approx \Delta t_c$ in \autoref{fig:ex3_L2}, where the accuracy is evaluated with \autoref{eq:L2_timeHist}. In this case, it can be seen in figure \autoref{fig:ex3_L2}(c) that the coarsest conforming mesh leads to a more time consuming simulation with respect to a finer mesh, because its increased distortion leads to a comparatively smaller $\Delta t_c$, which outweighs the difference in model sizes in terms of computational cost. Considering that the quality of the SCM solution with $\chi = 21$ might suffice for the application at hand, its speedup with respect to the fastest SEM solution is roughly $1874 \ [s] / 456 \ [s] = 3.9 \ [-]$. 

By increasing time refinement, the simulations with $\Delta t = 10^{-9} \ [s] $, reveal that at $\mathbf{s}_1$ (\autoref{fig:ex3_L2}(a)) the effectiveness of spatial discretization is limited in both methods. Although the SEM performs better, results are again influenced by the quality of the mesh, to the point where increasing the model size might unpredictably lead to (locally) worst results. On the other hand, consistent improvement can be achieved with the SCM due to the use of an optimal Cartesian mesh. In \autoref{fig:ex3_L2}(b), the results for $\Delta t = 10^{-9} \ [s]$ show that, for a generic point of the domain, the SCM incurs only in a minor loss of accuracy.

\begin{figure}[t!]
%\begin{tabular}{C{.5\textwidth} C{.01\textwidth} C{.5\textwidth}}
\subfigure [$t = 0.1 \cdot 10^{-4}$] {
    \resizebox{0.48\textwidth}{!}{%
    \includegraphics{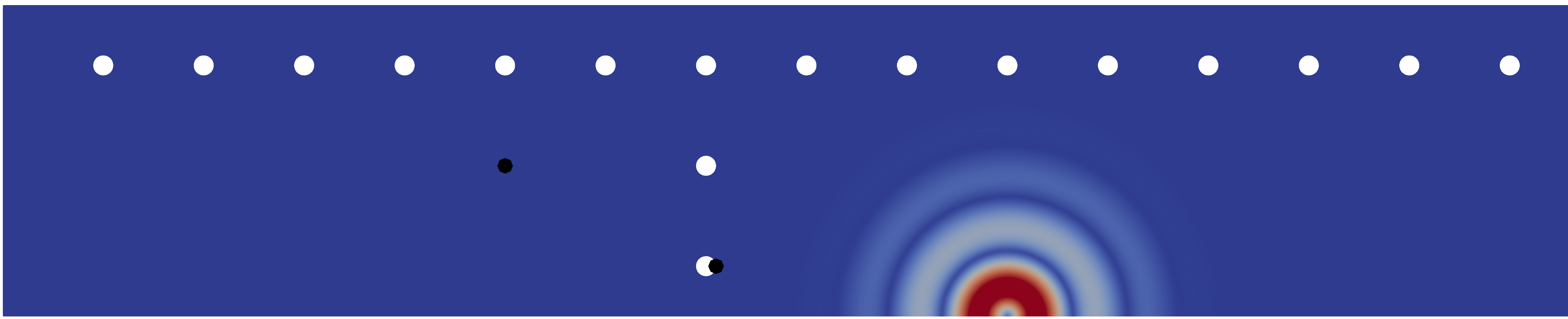}
    }
}
\subfigure [$t = 0.2 \cdot 10^{-4}$] {
    \resizebox{0.48\textwidth}{!}{%
    \includegraphics{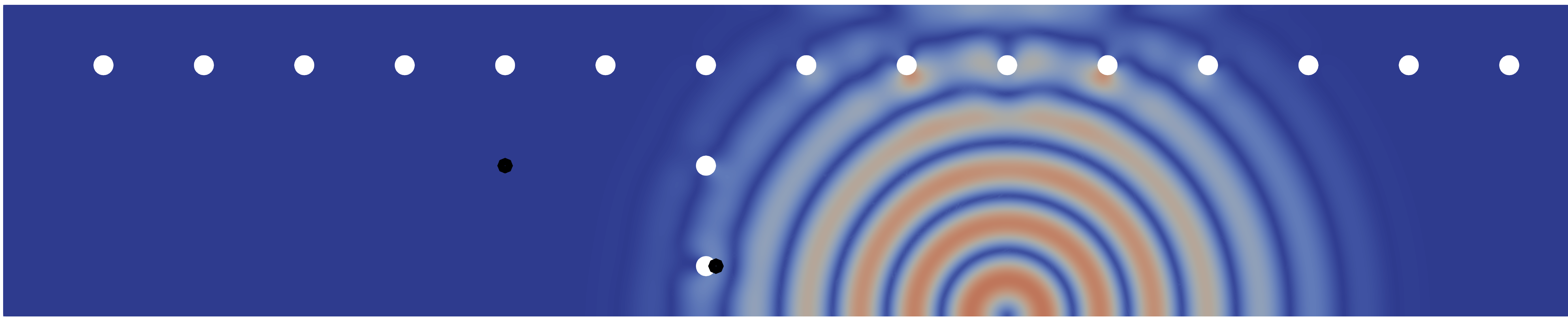}
    }
}\\
\subfigure [$t = 0.3 \cdot 10^{-4}$] {
    \resizebox{0.48\textwidth}{!}{%
    \includegraphics{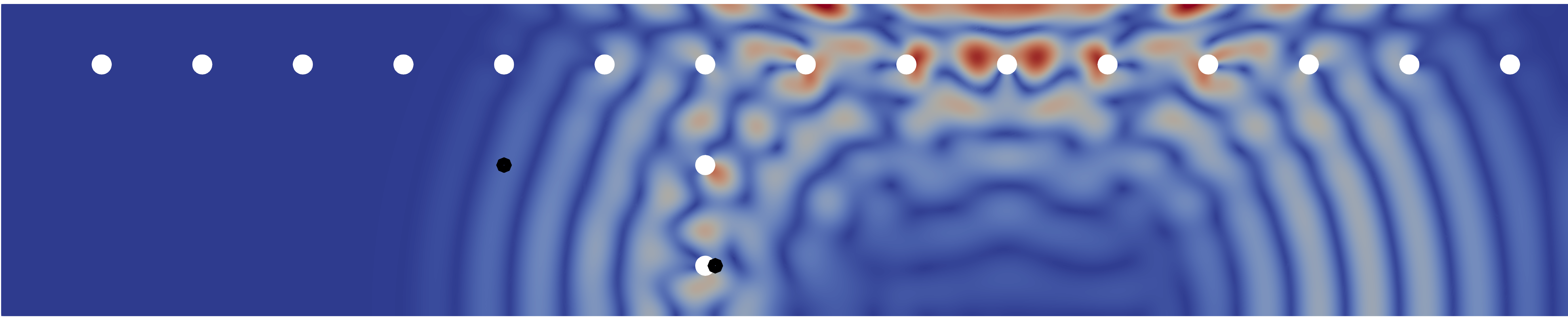}
    }
}
\subfigure [$t = 0.4 \cdot 10^{-4}$] {
    \resizebox{0.48\textwidth}{!}{%
    \includegraphics{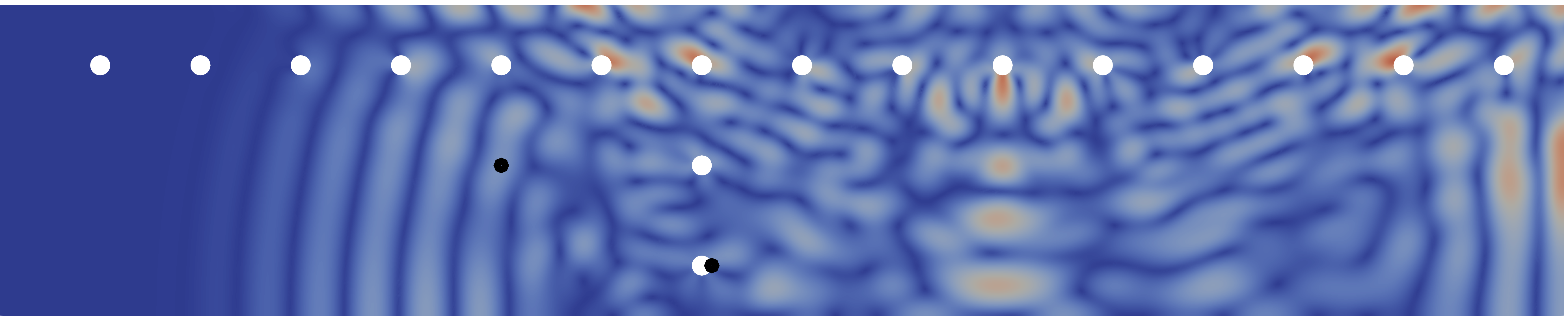}
    }
}
%\end{tabular}
\caption{Four displacement field snapshots for the problem of \autoref{sec:ex3}. Sensor locations are highlighted as black dots. Geometrical dissipation as well as reflections at holes, top and side boundaries can be observed. Due to the quantity and configuration of the holes, a scattering of the initial pulse occurs.}
\label{fig:contours}
\end{figure}

Geometrical dissipation of the pulse can be observed by comparing \autoref{fig:contours}(a) with \autoref{fig:contours}(b), where the top boundary is reached and reflections at the holes are visible. Reflections from the top boundary interact again with the row of holes, leading to the scattered pattern observed in \autoref{fig:contours}(c). At roughly this time ($t = 0.3 \cdot 10^{-4} [s]$) the initial wave packet has reached $\mathbf{s}_2$. In \autoref{fig:contours}(d), reflections from the top boundary reach $\mathbf{s}_1$ and the main wave packet has been reflected at the right edge of the plate.
%time histories
%%%%%%%%%%%%%%%%%%%%%%%%%%%%%%%%%%%%%%%%%%%%%%%%%%%%%%%%%%%%%%%%%%%%%%%%%
%contour plots
%%%%%%%%%%%%%%%%%%%%%%%%%%%%%%%%%%%%%%%%%%%%%%%%%%%%%%%%%%%%%%%%%%%%%%%%%
%%%%%%%%%%%%%%%%%%%%%%%%%%%%%%%%%%%%%%%%%%%%%%%%%%%%%%%%%%%%%%%%%%%%%%%%%

%% file: figures/ex3/timehist/S1_displ.tikz
\begin{tikzpicture}
[spy using outlines=
	{magnification=1, anchor = center, connect spies}]
    \begin{axis}[
        name = ex2_s1,
        xmin = 0,
        xmax = 1e-4,
        ymin = -5e-12,
        ymax = 5.1e-12,
        xlabel = {time [s]},
        ylabel = {$u_{1,z}$ [m]},
        xticklabel style = {font =\fontsize{\figureFontSize pt}{10pt}\selectfont},
        yticklabel style = {font=\fontsize{\figureFontSize pt}{10pt}\selectfont},
        xlabel style = {font =\fontsize{\figureFontSize pt}{\figureFontSize}\selectfont},
        ylabel style = {font=\fontsize{\figureFontSize pt}{\figureFontSize}\selectfont},
        no markers,
        grid = both,
        width=0.48\textwidth,
        height=0.24\textwidth,
        legend pos=south west,
        legend style={
                    at = {(0,-0.35)},
                    anchor=north west,
                    nodes={font=\fontsize{\figureFontSize pt}{\figureFontSize}\selectfont}
                    }
        ]
        
        % \addplot [color = green] table [x=time, y=sensor_2, col sep=semicolon] {figures/ex3/timehist/cut76_sh334_dtcrit.csv};
        %\addlegendentry{SCM, $\chi = 16$ }
        
         \addplot table [x=time, y=sensor_2, col sep=semicolon] {figures/ex3/timehist/cut101_sh334_dtcrit.csv};
        \addlegendentry{SCM, $\chi = 21$ }
        
        \addplot table [x=time, y=sensor_2, col sep=semicolon] {figures/ex3/timehist/cut201_sh334_dtcrit.csv};
        \addlegendentry{SCM, $\chi = 41$ }
        
        \addplot table [x=time, y=sensor_2, col sep=semicolon] {figures/ex3/timehist/expl_h19e-4_O2_sh334_dtcrit.csv};
        \addlegendentry{SEM, $\chi \approx 41$}
        
        \addplot table [x=time, y=sensor_2, col sep=semicolon] {figures/ex3/timehist/reduced_h8e-4_O2_sh334.csv};
        \addlegendentry{SEM,  $\chi \approx 101$}

\end{axis}
\end{tikzpicture}

%% file: figures/ex3/timehist/S2_displ.tikz
\begin{tikzpicture}
[spy using outlines=
	{magnification=1, anchor = center, connect spies}]
    \begin{axis}[
        name = ex2_s1,
        xmin = 0,
        xmax = 1e-4,
        ymin = -3e-12,
        ymax = 3e-12,
        xlabel = {time [s]},
        ylabel = {$u_{1,z}$ [m]},
        xticklabel style = {font =\fontsize{\figureFontSize pt}{10pt}\selectfont},
        yticklabel style = {font=\fontsize{\figureFontSize pt}{10pt}\selectfont},
        xlabel style = {font =\fontsize{\figureFontSize pt}{\figureFontSize}\selectfont},
        ylabel style = {font=\fontsize{\figureFontSize pt}{\figureFontSize}\selectfont},
        no markers,
        grid = both,
        width=0.48\textwidth,
        height=0.24\textwidth,
        legend pos=south west,
        legend style={
                    at = {(0,-0.35)},
                    anchor=north west,
                    nodes={font=\fontsize{\figureFontSize pt}{\figureFontSize}\selectfont}
                    }
        ]
        
        % \addplot [color = green] table [x=time, y=sensor_5, col sep=semicolon] {figures/ex3/timehist/cut76_sh334_dtcrit.csv};
        %\addlegendentry{SCM, $\chi = 16$ }
        
         \addplot table [x=time, y=sensor_5, col sep=semicolon] {figures/ex3/timehist/cut101_sh334_dtcrit.csv};
        \addlegendentry{SCM, $\chi = 21$ }
        
        \addplot table [x=time, y=sensor_5, col sep=semicolon] {figures/ex3/timehist/cut201_sh334_dtcrit.csv};
        \addlegendentry{SCM, $\chi = 41$ }
        
        \addplot table [x=time, y=sensor_5, col sep=semicolon] {figures/ex3/timehist/expl_h19e-4_O2_sh334_dtcrit.csv};
        \addlegendentry{SEM, $\chi \approx 41$}
        
        \addplot table [x=time, y=sensor_5, col sep=semicolon] {figures/ex3/timehist/reduced_h8e-4_O2_sh334.csv};
        \addlegendentry{SEM,  $\chi \approx 101$}

\end{axis}
\end{tikzpicture}

%% file: figures/ex3/s1_L2.tikz
\begin{tikzpicture}
    \begin{axis}[
        xmin = 0,
        xmax = 5e6,
        ymin = 1e-3,
        ymax = 0.2,
        xtick = { 2.5e5, 5e5, 1e6, 2.5e6, 5e6},
        xticklabels = {$2.5\cdot 10^5$, $5\cdot 10^5$, $10^6$, $2.5 \cdot 10^6$, $ 5 \cdot 10^6$},
        xmode = log,
        ymode = log,
        xlabel = {$n_{dofs} [-]$},
        ylabel = {$\epsilon_{h,t} [-]$},
        xticklabel style = {font =\fontsize{\figureFontSize pt}{10pt}\selectfont},
        yticklabel style = {font=\fontsize{\figureFontSize pt}{10pt}\selectfont},
        xlabel style = {font =\fontsize{\figureFontSize pt}{\figureFontSize pt}\selectfont},
        ylabel style = {font=\fontsize{\figureFontSize pt}{\figureFontSize pt}\selectfont},
        grid = both,
        width=0.48\textwidth,
        height=0.35\textwidth,
        legend style={at={(0,0)},anchor=south west,nodes={font=\fontsize{\figureFontSize pt}{\figureFontSize pt}\selectfont}}
%        legend pos = south east
        ]
        
    \addplot[color = \cutOptimizedColorb, mark =\explicitMarker*, line width=0.75pt] coordinates{
        %cut/dt_is_dtcrit: sensor 1 error conv of L2 by ndofs
        (259260,1.708221e-01)        %solidPlateStrSym101_sh334
        (570840,1.223004e-01)        %perforated_solidPlateStrSym151_sh334
        (1025115,9.386458e-02)        %perforated_solidPlateStrSym201_sh334
        (1657110,9.339962e-02)        %perforated_solidPlateStrSym251_sh334
        (2349975,7.526917e-02)        %perforated_solidPlateStrSym301_sh334
        (3210150,6.824648e-02)        %perforated_solidPlateStrSym351_sh334
	};
	\addlegendentry{SCM $\Delta t \approx \Delta t_{c}$}
	
	\addplot[color = \uncutColorb, mark =\conformMarker*, line width=\plotLineWidth pt] coordinates{
        %expl/dt_is_dtcrit: sensor 1 error conv of L2 by ndofs
        (1345350,2.834602e-02)        %solidPlateHoles_sym_h19e-4_O2_sh334
        (1675155,2.796455e-02)        %solidPlateHoles_sym_h175e-5_O2_sh334
        (2268615,2.107740e-02)        %solidPlateHoles_sym_h15e-4_O2_sh334
        (2962650,1.958623e-02)        %solidPlateHoles_sym_h125e-5_O2_sh334
        %(4517490,1.170898e-02)        %solidPlateHoles_sym_h1e-3_O2_sh334
	};
	\addlegendentry{SEM $\Delta \approx \Delta t_{c}$}
	
     \addplot[color = \cutOptimizedColor, mark =\explicitMarker*, line width=0.75pt] coordinates{
        %cut/dt1e-9: sensor 1 error conv of L2 by ndofs
        (259260,1.015637e-01)        %solidPlateStrSym101_sh334
        (570840,6.866942e-02)        %perforated_solidPlateStrSym151_sh334
        (1025115,6.519100e-02)        %perforated_solidPlateStrSym201_sh334
        (1657110,4.533023e-02)        %perforated_solidPlateStrSym251_sh334
        (2349975,3.329602e-02)        %perforated_solidPlateStrSym301_sh334
        (3210150,1.897415e-02)        %perforated_solidPlateStrSym351_sh334
	};
	\addlegendentry{SCM $\Delta t = 10^{-9}$}

	\addplot[color = \uncutColor, mark =\conformMarker*, line width=\plotLineWidth pt] coordinates{
        %expl/dt1e-9: sensor 1 error conv of L2 by ndofs
        (1345350,1.285557e-02)        %solidPlateHoles_sym_h19e-4_O2_sh334
        (1675155,1.070721e-02)        %solidPlateHoles_sym_h175e-5_O2_sh334
        (2268615,7.131228e-03)        %solidPlateHoles_sym_h15e-4_O2_sh334
        (2962650,1.018978e-02)        %solidPlateHoles_sym_h125e-5_O2_sh334

	};
    \addlegendentry{SEM $\Delta t = 10^{-9}$}

    \end{axis}
\end{tikzpicture}

%% file: figures/ex3/s2_L2.tikz
\begin{tikzpicture}
    \begin{axis}[
        xmin = 0,
        xmax = 5e6,
        ymin = 1e-4,
        ymax = 0.2,
        xtick = { 2.5e5, 5e5, 1e6, 2.5e6, 5e6},
        xticklabels = {$2.5\cdot 10^5$, $5\cdot 10^5$, $10^6$, $2.5 \cdot 10^6$, $ 5 \cdot 10^6$},
        xmode = log,
        ymode = log,
        xlabel = {$n_{dofs} [-]$},
        ylabel = {$\epsilon_{h,t} [-]$},
        xticklabel style = {font =\fontsize{\figureFontSize pt}{10pt}\selectfont},
        yticklabel style = {font=\fontsize{\figureFontSize pt}{10pt}\selectfont},
        xlabel style = {font =\fontsize{\figureFontSize pt}{\figureFontSize pt}\selectfont},
        ylabel style = {font=\fontsize{\figureFontSize pt}{\figureFontSize pt}\selectfont},
        grid = both,
        width=0.48\textwidth,
        height=0.35\textwidth,
        legend style={at={(0,0)},anchor=south west,nodes={font=\fontsize{\figureFontSize pt}{\figureFontSize pt}\selectfont}}
%        legend pos = south east
        ]
        
   \addplot[color = \cutOptimizedColorb, mark =\explicitMarker*, line width=0.75pt] coordinates{
        %cut/dt_is_dtcrit: sensor 2 error conv of L2 by ndofs
        (259260,1.084639e-01)        %solidPlateStrSym101_sh334
        (570840,8.936114e-02)        %perforated_solidPlateStrSym151_sh334
        (1025115,8.246246e-02)        %perforated_solidPlateStrSym201_sh334
        (1657110,7.836527e-02)        %perforated_solidPlateStrSym251_sh334
        (2349975,7.263908e-02)        %perforated_solidPlateStrSym301_sh334
        (3210150,6.679602e-02)        %perforated_solidPlateStrSym351_sh334
	};
	\addlegendentry{SCM $\Delta t \approx \Delta t_{c}$}
	
	\addplot[color = \uncutColorb, mark =\conformMarker*, line width=\plotLineWidth pt] coordinates{
        %expl/dt_is_dtcrit: sensor 2 error conv of L2 by ndofs
        (1345350,2.126892e-02)        %solidPlateHoles_sym_h19e-4_O2_sh334
        (1675155,2.460310e-02)        %solidPlateHoles_sym_h175e-5_O2_sh334
        (2268615,1.802795e-02)        %solidPlateHoles_sym_h15e-4_O2_sh334
        (2962650,1.634958e-02)        %solidPlateHoles_sym_h125e-5_O2_sh334
        %(4517490,1.055044e-02)        %solidPlateHoles_sym_h1e-3_O2_sh334
	};
	\addlegendentry{SEM $\Delta t \approx \Delta t_{c}$}
	
     \addplot[color = \cutOptimizedColor, mark =\explicitMarker*, line width=0.75pt] coordinates{
        %cut/dt1e-9: sensor 2 error conv of L2 by ndofs
        (259260,5.050418e-02)        %solidPlateStrSym101_sh334
        (570840,8.090792e-03)        %perforated_solidPlateStrSym151_sh334
        (1025115,4.581005e-03)        %perforated_solidPlateStrSym201_sh334
        (1657110,3.088551e-03)        %perforated_solidPlateStrSym251_sh334
        (2349975,2.362459e-03)        %perforated_solidPlateStrSym301_sh334
        (3210150,2.153475e-03)        %perforated_solidPlateStrSym351_sh334
	};
	\addlegendentry{SCM $\Delta t = 10^{-9}$}

	\addplot[color = \uncutColor, mark =\conformMarker*, line width=\plotLineWidth pt] coordinates{
        %expl/dt1e-9: sensor 2 error conv of L2 by ndofs
        (1345350,2.618550e-03)        %solidPlateHoles_sym_h19e-4_O2_sh334
        (1675155,2.291300e-03)        %solidPlateHoles_sym_h175e-5_O2_sh334
        (2268615,1.891445e-03)        %solidPlateHoles_sym_h15e-4_O2_sh334
        (2962650,1.748211e-03)        %solidPlateHoles_sym_h125e-5_O2_sh334
	};
    \addlegendentry{SEM $\Delta t = 10^{-9}$}

    \end{axis}
\end{tikzpicture}

%% file: figures/ex3/s1_L2_time.tikz
\begin{tikzpicture}
    \begin{axis}[
        xmin = 0,
        xmax = 7e4,
        ymin = 1e-3,
        ymax = 0.2,
        xtick = {5e2, 1e3, 2e3, 5e3, 1e4, 2.5e4, 7e4},
        xticklabels = {$5 \cdot 10^2$, $10^3$, $2 \cdot 10^3$, $5 \cdot 10^3$ ,$10^4$, $2.5 \cdot 10^4$,$7 \cdot 10^4$},
        xmode = log,
        ymode = log,
        xlabel = {wall time $[s]$},
        ylabel = {$\epsilon_{h,t} [-]$},
        xticklabel style = {font =\fontsize{\figureFontSize pt}{10pt}\selectfont},
        yticklabel style = {font=\fontsize{\figureFontSize pt}{10pt}\selectfont},
        xlabel style = {font =\fontsize{\figureFontSize pt}{\figureFontSize pt}\selectfont},
        ylabel style = {font=\fontsize{\figureFontSize pt}{\figureFontSize pt}\selectfont},
        grid = both,
        width=0.48\textwidth,
        height=0.35\textwidth,
        legend style={at={(0,0)},anchor=south west,nodes={font=\fontsize{\figureFontSize pt}{\figureFontSize pt}\selectfont}}
%        legend pos = south east
        ]
        
       \addplot[color = \cutOptimizedColorb, mark =\explicitMarker*, line width=0.75pt] coordinates{
        %cut/dt_is_dtcrit: sensor 1 error conv of L2 by wall_time
        (4.565755e+02,1.708221e-01)        %solidPlateStrSym101_sh334
        (4.870261e+02,1.223004e-01)        %perforated_solidPlateStrSym151_sh334
        (1.080847e+03,9.386458e-02)        %perforated_solidPlateStrSym201_sh334
        (2.003856e+03,9.339962e-02)        %perforated_solidPlateStrSym251_sh334
        (3.044578e+03,7.526917e-02)        %perforated_solidPlateStrSym301_sh334
        (3.391473e+03,6.824648e-02)        %perforated_solidPlateStrSym351_sh334
	};
	\addlegendentry{SCM $\Delta t \approx \Delta t_{c}$}
	
	\addplot[color = \uncutColorb, mark =\conformMarker*, line width=\plotLineWidth pt] coordinates{
        %expl/dt_is_dtcrit: sensor 1 error conv of L2 by wall_time
        (2.602116e+03,2.834602e-02)        %solidPlateHoles_sym_h19e-4_O2_sh334
        (1.874735e+03,2.796455e-02)        %solidPlateHoles_sym_h175e-5_O2_sh334
        (4.091282e+03,2.107740e-02)        %solidPlateHoles_sym_h15e-4_O2_sh334
        (7.256047e+03,1.958623e-02)        %solidPlateHoles_sym_h125e-5_O2_sh334
        %(1.402923e+04,1.170898e-02)        %solidPlateHoles_sym_h1e-3_O2_sh334
        %(4.368216e+04,1.018334e-02)        %solidPlateHoles_sym_h8e-4_O2_sh334
	};
	\addlegendentry{SEM $\Delta t \approx \Delta t_{c}$}
	
     \addplot[color = \cutOptimizedColor, mark =\explicitMarker*, line width=0.75pt] coordinates{
        %cut/dt1e-9: sensor 1 error conv of L2 by wall_time
        (2.809273e+03,1.015637e-01)        %solidPlateStrSym101_sh334
        (7.377172e+03,6.866942e-02)        %perforated_solidPlateStrSym151_sh334
        (1.389889e+04,6.519100e-02)        %perforated_solidPlateStrSym201_sh334
        (3.677576e+04,4.533023e-02)        %perforated_solidPlateStrSym251_sh334
        (4.896877e+04,3.329602e-02)        %perforated_solidPlateStrSym301_sh334
        (6.687691e+04,1.897415e-02)        %perforated_solidPlateStrSym351_sh334
	};
	\addlegendentry{SCM $\Delta t = 10^{-9}$}

	\addplot[color = \uncutColor, mark =\conformMarker*, line width=\plotLineWidth pt] coordinates{
        %expl/dt1e-9: sensor 1 error conv of L2 by wall_time
        (2.569626e+04,1.285557e-02)        %solidPlateHoles_sym_h19e-4_O2_sh334
        (3.343649e+04,1.070721e-02)        %solidPlateHoles_sym_h175e-5_O2_sh334
        (3.348634e+04,7.131222e-03)        %solidPlateHoles_sym_h15e-4_O2_sh334
        (3.904904e+04,1.018978e-02)        %solidPlateHoles_sym_h125e-5_O2_sh334
	};
    \addlegendentry{SEM $\Delta t = 10^{-9}$}
    
   % \addplot[color = \uncutColor, mark =\conformMarker*, line width=\plotLineWidth pt] coordinates{
        %expl/dt1e-9: sensor 1 error conv of L2 by wall_time
        %(2.598813e+04,1.285557e-02)        %solidPlateHoles_sym_h19e-4_O2_sh334
        %(1.819531e+04,1.070721e-02)        %solidPlateHoles_sym_h175e-5_O2_sh334
        %(3.487213e+04,7.131228e-03)        %solidPlateHoles_sym_h15e-4_O2_sh334
        %(3.043573e+04,1.018978e-02)        %solidPlateHoles_sym_h125e-5_O2_sh334
%	};
 %   \addlegendentry{SEM $\Delta t = 10^{-9}$}

    \end{axis}
\end{tikzpicture}

%% file: figures/ex3/s2_L2_time.tikz
\begin{tikzpicture}
    \begin{axis}[
        xmin = 0,
        xmax = 7e4,
        ymin = 1e-3,
        ymax = 0.2,
        xtick = {5e2, 1e3, 2e3, 5e3, 1e4, 2.5e4, 7e4},
        xticklabels = {$5 \cdot 10^2$, $10^3$, $2 \cdot 10^3$, $5 \cdot 10^3$ ,$10^4$, $2.5 \cdot 10^4$,$7 \cdot 10^4$},
        xmode = log,
        ymode = log,
        xlabel = {wall time $[s]$},
        ylabel = {$\epsilon_{h,t} [-]$},
        xticklabel style = {font =\fontsize{\figureFontSize pt}{10pt}\selectfont},
        yticklabel style = {font=\fontsize{\figureFontSize pt}{10pt}\selectfont},
        xlabel style = {font =\fontsize{\figureFontSize pt}{\figureFontSize pt}\selectfont},
        ylabel style = {font=\fontsize{\figureFontSize pt}{\figureFontSize pt}\selectfont},
        grid = both,
        width=0.48\textwidth,
        height=0.35\textwidth,
        legend style={at={(0,0)},anchor=south west,nodes={font=\fontsize{\figureFontSize pt}{\figureFontSize pt}\selectfont}}
%        legend pos = south east
        ]
        
    \addplot[color = \cutOptimizedColorb, mark =\explicitMarker*, line width=0.75pt] coordinates{
        %cut/dt_is_dtcrit: sensor 2 error conv of L2 by wall_time
        (4.565755e+02,1.084639e-01)        %solidPlateStrSym101_sh334
        (4.870261e+02,8.936114e-02)        %perforated_solidPlateStrSym151_sh334
        (1.080847e+03,8.246246e-02)        %perforated_solidPlateStrSym201_sh334
        (2.003856e+03,7.836527e-02)        %perforated_solidPlateStrSym251_sh334
        (3.044578e+03,7.263908e-02)        %perforated_solidPlateStrSym301_sh334
        (3.391473e+03,6.679602e-02)        %perforated_solidPlateStrSym351_sh334
	};
	\addlegendentry{SCM, $\Delta t \approx \Delta t_{c}$}
	
	\addplot[color = \uncutColorb, mark =\conformMarker*, line width=\plotLineWidth pt] coordinates{
        %expl/dt_is_dtcrit: sensor 2 error conv of L2 by wall_time
        (2.602116e+03,2.126892e-02)        %solidPlateHoles_sym_h19e-4_O2_sh334
        (1.874735e+03,2.460310e-02)        %solidPlateHoles_sym_h175e-5_O2_sh334
        (4.091282e+03,1.802795e-02)        %solidPlateHoles_sym_h15e-4_O2_sh334
        (7.256047e+03,1.634958e-02)        %solidPlateHoles_sym_h125e-5_O2_sh334
        %(1.402923e+04,1.055044e-02)        %solidPlateHoles_sym_h1e-3_O2_sh334
        %(4.368216e+04,1.003089e-02)        %solidPlateHoles_sym_h8e-4_O2_sh334
	};
	\addlegendentry{SEM, $\Delta t \approx \Delta t_{c}$}
	
     \addplot[color = \cutOptimizedColor, mark =\explicitMarker*, line width=0.75pt] coordinates{
        %cut/dt1e-9: sensor 2 error conv of L2 by wall_time
        (2.809273e+03,5.050418e-02)        %solidPlateStrSym101_sh334
        (7.377172e+03,8.090792e-03)        %perforated_solidPlateStrSym151_sh334
        (1.389889e+04,4.581005e-03)        %perforated_solidPlateStrSym201_sh334
        (3.677576e+04,3.088551e-03)        %perforated_solidPlateStrSym251_sh334
        (4.896877e+04,2.362459e-03)        %perforated_solidPlateStrSym301_sh334
        (6.687691e+04,2.153475e-03)        %perforated_solidPlateStrSym351_sh334
	};
	\addlegendentry{SCM, $\Delta t = 10^{-9}$}

    \addplot[color = \uncutColor, mark =\conformMarker*, line width=\plotLineWidth pt] coordinates{
        %expl/dt1e-9: sensor 2 error conv of L2 by wall_time
        (2.569626e+04,2.618562e-03)        %solidPlateHoles_sym_h19e-4_O2_sh334
        (3.343649e+04,2.291296e-03)        %solidPlateHoles_sym_h175e-5_O2_sh334
        (3.348634e+04,1.891444e-03)        %solidPlateHoles_sym_h15e-4_O2_sh334
        (3.904904e+04,1.748218e-03)        %solidPlateHoles_sym_h125e-5_O2_sh334
	};
    \addlegendentry{SEM, $\Delta t = 10^{-9}$}
    
	%\addplot[color = \uncutColor, mark =\conformMarker*, line width=\plotLineWidth pt] coordinates{
        %expl/dt1e-9: sensor 2 error conv of L2 by wall_time
    %    (2.598813e+04,2.618550e-03)        %solidPlateHoles_sym_h19e-4_O2_sh334
     %   (1.819531e+04,2.291300e-03)        %solidPlateHoles_sym_h175e-5_O2_sh334
     %   (3.487213e+04,1.891445e-03)        %solidPlateHoles_sym_h15e-4_O2_sh334
     %   (3.043573e+04,1.748211e-03)        %solidPlateHoles_sym_h125e-5_O2_sh334
	%};
    %\addlegendentry{SEM, $\Delta t = 10^{-9}$}

    \end{axis}
\end{tikzpicture}

%% file: 5_conclusions.tex
\section{Conclusions}\label{sec:conclusions}
By leveraging Gauss-Lobatto-Legendre spectral elements, the spectral cell method offers great capabilities to model guided waves and its potential in the context of SHM is enhanced with the versatility offered by decoupled geometrical descriptions. To achieve all these qualities simultaneously, particular attention must be placed on the numerical integration of elements intersected by mesh-independent boundaries. In this paper, a novel moment fitting technique is proposed to restore a diagonal element mass matrix, which is essential for fast simulations. As in many other lumping techniques, this comes at the cost of abandoning the variational formulation. In our procedure, moment fitting equations are relaxed into a quadratic programming problem, which is aimed at minimizing lumping error and can be fine-tuned by means of the parameter $\epsilon$. With the use of widely available libraries for numerical analysis, this subroutine can easily be integrated into existing codes. The library Alglib \cite{bochkanov2011alglib} was used in our case. The additional computational cost introduced in the assembly is offset by the important reduction in Gauss points in the subsequent integration of the mass matrix and its implementation is motivated by the prospect of increasing accuracy with respect to existing mass lumping methods. To integrate the basis monomials over the physical element domain, local hierarchical meshes in combination with standard Lagrangian element partitions were employed. However, no particular restriction is set on the method used for this computation, and thus a number of alternatives might be used: be it pure quad/octrees \cite{duczek2015finite}, high order element partitions based on the blending function method \cite{fries2016higher},  or approaches based on Gauss' divergence theorem \cite{duczek2015efficient, chin2019modeling}, just to name a few.

Although results are promising, the authors feel that further improvements could be achieved by deepening the understanding of the procedure under some key aspects. Firstly, a theoretical justification for the values of the optimization parameter $\epsilon$ could not be found, thus leading to its selection by means of trials. For straight interfaces, and the critical time step not being a concern, the best accuracy was achieved with $\epsilon = 10^{-2}$. For curved interfaces in 3D, we found that an increase to $\epsilon = 0.1$ was beneficial to both accuracy and decreasing critical time step decay, perhaps due to its reduction of distortion-related effects. Secondly, oscillations in the element eigenvalues (\autoref{fig:element_critical_dt}), and, most importantly, in the accuracy of the procedure (\autoref{fig:ex1_sp5_conv}), were observed when an interface is the farthest from a node, and are also challenging to quantify theoretically. Thirdly, although high order elements show the most pronounced decay in critical time step, and despite the increased errors observed in \autoref{fig:ex1_poly}, we suspect that they might be advantageous in this procedure, due to the fact that optimization can occur over larger sets of nodes. To confirm or disprove this intuition, more extended theoretical and numerical investigations will be performed as part of future works.

The main draw-back of the procedure, consisting in a reduction of the critical time step for cut elements, can be alleviated by tuning the parameter $\epsilon$, and, most effectively, by adopting a frog-leap solver. By means of Lamb wave simulations with 3D models, we showed that, although higher spatial discretization errors are introduced by cut elements, this approach performs comparatively or only slightly worse in terms of both, accuracy and computation time, with respect to the state-of-the-art SEM. The SEM and the CDM are straightforward to implement and offer great performance in the simulation of wave propagation. However, their range of application is limited by conforming meshing and the absence of parametric damage descriptions. The proposed approach removes some of these limitations, enabling the use of relatively coarse, structured meshes that, in several cases, provide solutions of sufficient accuracy at a much lower cost than corresponding approaches that rely on conforming meshes.

\section*{Acknowledgements}
This work has received funding from the European Union's Horizon 2020 research and innovation programme under the Marie Sklodowska-Curie grant agreement No. 795917 “SiMAero, Simulation-Driven and On-line Condition Monitoring with Applications to Aerospace.